\documentclass[onecolumn,authoryear]{els-mrw} 

\usepackage{amsmath,amssymb,amsfonts,amsthm,makeidx,graphicx}
\usepackage{txfonts}
\usepackage{helvet}
\usepackage{comment}

\begin{document}

\chapter{Metal-Poor Stars in the Milky Way System}\label{chap1}

\author[1]{Anna Frebel}%

\address[1]{\orgname{Massachusetts Institute of Technology}, \orgdiv{Department of Physics and Kavli Institute for Astrophysics and Space Research}, \orgaddress{77 Massachusetts Avenue, Cambridge, MA 02139, USA}}

\maketitle

\section*{Keywords}
Galactic archaeology, metal-poor stars, Population II stars, CEMP stars, Milky Way stellar halo, s-process, r-process, dwarf galaxies, globular clusters, spectroscopy, chemical abundances, stellar abundances, metallicity

\begin{abstract}[Abstract] 

Ancient, long-lived stars remain present in all components of our home galaxy, the Milky Way. 
Born a few hundred million years after the Big Bang and during a time that marked the very beginning of the chemical evolution, these stars display very low abundances of elements heavier than hydrogen and helium, making them “metal-poor”.  Studying the chemical composition of these stars reveals direct information about the conditions of the early universe because each of them has long preserved the local chemical signature of their individual birth gas clouds in their stellar atmosphere. There are many different types of metal-poor stars, each of them providing information on a different element production history that occurred prior to their own births. Large samples of metal-poor stars enable the reconstruction of nucleosynthesis and the chemical evolution of our Galaxy, early star formation processes, and various aspects of the assembly and evolution of the Milky Way. 

\end{abstract}

\section*{Key points/Learning objectives}

\textbf{CEMP stars} refers to carbon-enhanced metal-poor stars; they are prominent in the Milky Way among the most metal-poor stars\\
\textbf{Dwarf galaxy} is a galaxy much less massive than the Milky Way; they orbit the Milky Way in its outskirts but will be absorbed by our galaxy in the future \\
\textbf{[Fe/H]} stellar iron abundance with respect to that of the Sun which is indicated by the bracket notation “[ ]"; typically used as a proxy for the overall metal abundance or total elemental composition of a star; also referred to as metallicity \\
\textbf{Globular cluster} gravitationally bound, tight collection of stars with up to a million members; some of the oldest objects in the universe\\
\textbf{Halo} refers to the outer part of the Milky Way; it envelopes the galactic disk out to distances beyond 250\,kiloparsec\\
\textbf{Kiloparsec} [kpc] refers to a distance measurement; 1 parsec [pc] is 3.26 lightyears, and 1000 pc = 1 kpc = 3260 lightyears \\
\textbf{Metallicity} refers to the overall metal abundance or total elemental composition of a star \\
\textbf{Metal-poor stars} are stars whose overall chemical composition of metals is at least ten times less than that of the Sun \\
\textbf{Metals} are referred to as all chemical elements heavier than hydrogen and helium \\
\textbf{Milky Way} is our home galaxy\\
\textbf{r-process} refers to the rapid neutron-capture process that produces the heaviest elements in the periodic table\\
\textbf{Spectroscopy} is the technique to determine stellar chemical abundances from the corresponding elemental absorption lines in a high-resolution spectrum\\
\textbf{s-process} refers to the slow neutron-capture process that produces the heaviest elements in the periodic table\\


\section{Introduction} 

After the Big Bang, the universe consisted of just hydrogen, helium and tiny traces of lithium. From this gas, the first stars formed. Given a lack of cooling of the primordial gas, these objects were massive, on the order of 100\,M$_{\odot}$ \citep{bromm02, klessen23}. Given their mass, their lifetimes were correspondingly short, around a few million years, so that they soon exploded as the first supernovae. The first elements heavier than hydrogen and helium, called “metals" in astronomy, were synthesized during their stellar evolution and ejected during these first explosions, thus seeding the universe with more complex atoms.

Since then, all of the heavier elements we know and observe today have been subsequently formed in stars and related explosive events, such as supernovae and neutron star mergers. This continued process of enrichment from different sources is referred to as the chemical evolution. As a consequence, stars born sooner (after the Big Bang) will contain much smaller amounts of metals in their atmospheres than any stars that will form later on. The chemical composition of a star then reflects the integrated element yields of all stellar generations that came before. For example, the Sun's composition reflects the $\sim$8-9 billion years of chemical evolution that occurred prior to its birth. Consequently, and compared to the Sun, long-lived metal-poor stars contain only small amounts of heavier elements, metals, since they formed early in the universe and from gas previously enriched by only very few, or even just one, nucleosynthesis source or site. The current record holder for the most iron-poor star is SMSS J0313-6708, with [Fe/H] $<-7$ \citep{keller07}. This corresponds to of order less than one million kilograms of Fe. 

As such, metal-poor stars offer a rich window into the early universe and the physical and chemical conditions of the early star forming gas and the onset of galaxy formation. This is possible because for 12-13 billion years, these ancient stars have preserved the local chemical composition of the birth gas cloud from which they formed in their atmospheres (e.g., \citealt{fn15, frebel_ji23}). 

A star's lifetime depends on its mass. The Sun's lifespan is 10 billion years while stars with lower masses have longer lifetimes. Stars with about 0.8\,M$_{\odot}$, and currently in the last 10\% of their lives, will have ages of 12-13 billion years. It is these stars that are still shining all across the Milky Way in all its populations and components. They are very efficiently burning just hydrogen to helium in their cores with their outer atmospheres remaining undisturbed ever since their formation billions of years ago. 

By spectroscopically determining the elemental abundance signatures of metal-poor stars in their present state, information about these early gas clouds becomes available and which processes and sites may have provided those early elements. This concept of using local stars in the Milky Way to study the early universe is often referred to as “stellar archaeology". This approach has shown that there are in fact a variety of different types of metal-poor stars, each with their own progenitor history that provides important clues on how element production proceeded in the early universe. 

For the last three decades, both individual stars as well as larger samples of metal-poor stars have enabled enormous progress regarding our understanding of the element nucleosynthesis, associated astrophysical sites, the chemical evolution of the Milky Way, early star formation processes, and various aspects of galaxy evolution. No other approach can provide such direct insights into the early universe, making metal-poor stars ideal probes for carrying out “near-field cosmology" because these relatively easily accessible ancient stars are the local equivalent to the high-redshift universe.


\section{Classifications and definitions of metal-poor stars}\label{class}
 
Historically, as it was recognized that not all stars have the exact same chemical composition or location in the Galaxy, Type I and Type II stars were distinguished in a first attempt to classify and quantify stars across the Milky Way \citep{baade44, chamberlain51}. Today, these groups of stars are referred to as Population I and Population II, with older, more metal-poor Population II making up only a small fraction of the number of the younger, more metal-rich Population I stars. In the early 2000's, this overall concept was extended to include the modern, putative metal-free Population III first stars \citep{brommARAA}. The current definitions are as follows:
\smallskip

\noindent
\textbf{Population I stars:} Depict young metal-rich stars and open clusters located in the gaseous disk of the Milky Way. 

\noindent
\textbf{Population II stars:} Much older, metal-poor stars and ancient globular clusters located primarily in the halo of the Galaxy. The most metal-poor stars are examples of the earliest Population II stars formed.

\noindent
\textbf{Population III stars:} Massive metal-free first stars made from only hydrogen, helium and traces of lithium that lit up the universe for the first time, produced the first metals and quickly exploded as the first supernova, kicking off star and galaxy formation across the universe. 

\smallskip
Besides this broad classification, metal-poor stars are mainly defined and characterized by their iron (Fe) abundances. Often, the Fe abundance is taken as a proxy for the overall chemical composition of the star. This is a reasonable assumption, especially for stars similar to the Sun. The Fe abundance is also often referred to as “metallicity" of the star which can be taken as the overall metal abundance.  

The absolute abundance of an element is then given as 
\vspace{-0.5cm}

\begin{equation}
 \log_{10} \epsilon(X) = \log_{10}(N_{\rm X}/N_{\rm H}) + 12, 
\end{equation}

\noindent for an element X and relative to the abundance of hydrogen (H). $N_{\rm X}$ and $N_{\rm H}$ refers to the number of atoms of X and H. By definition, hydrogen is set to $\log_{10}$ $\epsilon$ (H) = 12.

Stellar elemental abundances are then given as ratios relative to the respective abundances of the Sun. This is depicted by presenting abundance ratios in square brackets “[ ]" (bracket notation). 

Elemental abundances relative to hydrogen are then given as 
\vspace{-0.5cm}

\begin{equation}
 \rm{[X/H]} = \log_{10}(N_{\rm X}/N_{\rm H})_\star - \log_{10}(N_{\rm X}/N_{\rm H})_\odot = \log_{10}\epsilon(X)_\star - \log_{10}\epsilon(X)_\odot
\end{equation}

Using iron as an example, the iron abundance, or metallicity, is given as 
\vspace{-0.5cm}

\begin{equation}
\rm{[Fe/H]} = \log_{10}(N_{\rm Fe}/N_{\rm H}) -  \log_{10}(N_{\rm Fe}/N_{\rm H})_{\star} 
\end{equation}

A star with a metallicity of [Fe/H] = $–$3.0 then has 1/1,000 that of solar iron abundance. Metallicity fundamentally characterizes a star and is the most important classifier. Table~\ref{class_list} reproduces different types of metal-poor stars based on their metallicity, following what was presented in \citet{araa, fn15}. 

\begin{table}[t]
\TBL{\caption{\label{class_list} Metallicity-based classes of metal-poor stars. Based on \citet{frebel18}.}}
{\begin{tabular*}
{\textwidth}{@{\extracolsep{\fill}}@{}lll@{}}
\toprule
Description & Definition & Abbreviation \\
\hline
Super-metal-rich & $\mbox{[Fe/H]} >0.0$ & MR \\
Solar & $\mbox{[Fe/H]} =0.0$ & None\\
Metal-poor & $\mbox{[Fe/H]}<-1.0$ & MP \\
Very metal-poor & $\mbox{[Fe/H]}<-2.0$ & VMP \\
Extremely metal-poor& $\mbox{[Fe/H]}<-3.0$ & EMP \\
Ultra-metal-poor & $\mbox{[Fe/H]}<-4.0$ & UMP \\
Hyper-metal-poor & $\mbox{[Fe/H]}<-5.0$ & HMP \\
Mega-metal-poor & $\mbox{[Fe/H]}<-6.0$ & MMP \\
Septa-metal-poor & $\mbox{[Fe/H]}<-7.0$ & SMP \\
Octa-metal-poor & $\mbox{[Fe/H]}<-8.0$ & OMP \\
\botrule
\end{tabular*}}
{
}%
\end{table}


\smallskip
Abundance ratios involving two elements X and Y (besides hydrogen) are given as 
\vspace{-0.5cm}

\begin{equation}
\rm{[X/Y]} = \log_{10}(N_{\rm X}/N_{\rm Y})_\star - \log_{10}(N_{\rm X}/N_{\rm Y})_\odot
\end{equation}
\vspace{-0.5cm}

Using magnesium and iron as an example of an element abundance ratios lead to 
\vspace{-0.5cm}

\begin{equation}
\rm{[Mg/Fe]} = \log_{10}(N_{\rm Mg}/N_{\rm Fe}) - \log_{10}(N_{\rm Mg}/N_{\rm Fe})_\star  
\end{equation}
\vspace{-0.5cm}

If a star has a Mg abundance of [Mg/Fe] = +0.3, it will have twice as much Mg than Fe compared to what is present in the Sun. Magnesium will be regarded enhanced in the star (ratio is $>0$ and thus supersolar).

A principal caveat with providing stellar abundances relative to the Sun's composition lies in the fact that the Solar abundances themselves are not set in stone. Our general knowledge about stars as well as computational approaches continue to advance. As a consequence, the Solar abundances are occasionally revised. This implies that stellar abundances collectively change when the solar abundances are updated. As such, any stellar abundances should always be quoted along with the respective Solar abundances that were employed to calculate them.


\section{Locations of metal-poor stars in the Milky Way system}\label{loc}

Considering the hierarchical assembly process of a galaxy such as the Milky Way system, i.e. its long-term growth through the accretion of smaller, neighboring galaxies, it can be expected that ancient metal-poor stars should be found across all components, although not in equal amounts. Indeed, data collected over the last $\sim$20 years is finally revealing that metal-poor stars are indeed found within all the stellar populations of the Milky Way and its dwarf satellite galaxies. These include the bulge, the thin disk, the thick disk, the Atari disk, the halo, the ultra-faint dwarf galaxies, the classical dwarf galaxies and the Magellanic Clouds. 

As the Milky Way’s oldest component that formed first, the bulge is supposed to contain the oldest Galactic stars \citep{tumlinson10}. Yet observing stars in the bulge is generally limited due to crowding and extreme reddening because of copious amounts of dust present in the gas near the Galaxy’s central region. Still, stars down to [Fe/H] $\sim -$2.0 are known to exist in the bulge \citep{howes14}. 
While the gas-rich thin disk is generally the most metal-rich component of the Milky Way with most stars having a chemical composition similar to that of the Sun, isolated stars down to [Fe/H] $\sim -$4.0 have been found \citep{mardini22_thin}. These stars are typically identified by kinematic analyses of metal-poor star samples and showing thin disk kinematics along with metal-deficiency.
Kinematic analyses usually distinguish thin and thick disk stars \citep{bensby05}.
As suggested by its name, and with $\sim$2-3\,kpc in height above the plane \citep{Li2017}, the thick disk is vertically more extended than the thin disk. The thick disk is on average also more metal-poor. Metal-poor stars down to [Fe/H] $\sim -$2.0 and lower form a well-populated metal-poor tail of the thick disk. 
The Atari disk is a ancient, more metal-poor version of the thick disk with a somewhat similar kinematic signature. Yet it lags the thick disk in its rotational velocity, and it is vertically more extended \citep{mardini22_atari}. It contains stars with metallicities as low as [Fe/H] $\sim -$5.0. It appears to have an accretion origin and its progenitor likely fell into the early disk of the young Milky Way.

The vast majority of known metal-poor stars are located in the halo of the Galaxy, above and below the disk. Given that the halo has been built up by numerous accretion events, metal-rich stars are fewer than what is found in the bulge and disk populations which are nearly exclusively dominated by younger solar-type stars. This makes finding metal-poor stars in the halo somewhat easier, in principle, albeit still requiring sophisticated search techniques. For several decades, extensive discovery efforts have delivered stars down to [Fe/H] $\sim -$7.0 in the halo \citep{araa, fn15}. Generally, the chance to find a progressively more metal-poor star increases with increasing distance as the outer halo becomes more and more dominated by old, accreted material. Halo stars can show both prograde and retrograde motions, with retrograde motions unambiguously pointing to accretion origins.

The ultra-faint dwarf galaxies are the smallest and faintest galaxies to exist, each with only a few thousand member stars. They are thought to have formed in the early universe and have been orbiting the Milky Way for billions of years. Despite their low total stellar mass of 10$^{4-6}$\,M$_{\odot}$, they are regarded as complete galaxies because they are embedded in a much more massive halo of dark matter, as stellar velocity dispersion measurements have repeatedly shown \citep{simon17}. In comparison, many globular clusters contain a million stars but they are not surrounded by a dark halo since they shown little to no velocity dispersion among their stars. The general challenge when working with dwarf galaxies is to establish membership. New member stars can be identified only if their radial velocity agrees with the systemic velocity (obtained from searching for a velocity peak in a large sample of stars in the direction of the galaxy). An alternative technique to identifying both entire dwarf galaxies as well as any new member stars is by means of the stars' common proper motion compared to the motions of the background stars. 

Ultra-faint dwarf galaxies contain stars with $-$4.0 $<$ [Fe/H] $<$ $-$1.0. Stars with even higher metallicity are typically entirely missing. It is assume that the galaxy’s star formation was quenched early on during processes such as reionization \citep{brown14}. As a result, no additional stars formed and chemical evolution ceased before ever fully operating. Dozens of these ancient systems orbit the Milky Way relatively closely, most of them within 150\,kpc but some out to 250\,kpc. Some of them are in the process of being tidally disrupted while others remain intact. 

The classical dwarf galaxies are more massive (stellar mass of 10$^{6-8}$\,M$_{\odot}$) than the ultra-faint dwarf galaxies and display all characteristic signs of galactic evolution, such as multiple bursts of star formation and an extended chemical evolution. Stars with metallicities as low as [Fe/H] $\sim$ $-$4.0 have been found but the majority of members have metallicities up to and around solar values, reflecting the extended chemical evolution of these systems. At distances of 50-100\,kpc, most of them are visible as stellar overdensities in the sky. With many more members available, membership identification for these systems is straightforward and based on the same criteria and techniques as for the ultra-faint dwarfs. 

Metal-poor stars have also been discovered in the Magellanic Clouds in recent years. Their distance of $\sim$50\,kpc into the outer halo, faintness of their stars, and large amounts of gas and dust limiting access to the central regions has long made it challenging to identify suitable candidates \citep{reggiani21}. Using a new narrowband photometry search technique (see Section~\ref{analysis_section}), stars with [Fe/H] $\sim$ $-$2.0 were found, including several with [Fe/H] $\sim$ $-$3.0 and one with [Fe/H] $\sim$ $-$4.0 \citep{chiti24}. This shows that essentially all galaxies in the Milky Way system contain at least some of the most metal-poor stars. 

Many globular clusters are among the oldest known objects, with ages up to 12\,Gyr \citep{bastian18}. They also tend to be metal-poor, with metallicities reaching down to [Fe/H] $\sim$ $-$2.5. Some clusters contain multiple stellar populations which can result in elemental abundance spreads. Besides typical chemical abundances characteristics in light elements (e.g. the Na-O and Mg-Al anti-correlations), studies of globular cluster stars have long shown a robust metallicity floor of [Fe/H] $\sim$ $-$2.5. However, new studies have revealed several outliers. For example, the long-disrupted cluster C19, now found as a stellar stream across the sky, was found to exhibit stellar abundances of [Fe/H] $\sim$ $-$3.4 \citep{martin22}.


\section{Discovery techniques of metal-poor stars} \label{disc}

Metal-poor stars are  ancient survivors from the early universe, and as such, very rare. In the halo, the number of stars decrease by a factor of ten for each factor of ten decrease in Fe abundance. This implies that extremely metal-poor stars with [Fe/H] $<-$3.5 are a factor 100 less common than metal-poor stars with [Fe/H] $<–$1.5. Accordingly, stars with [Fe/H] $<-$4 to $<-$7 are significantly rarer than even these extremely metal-poor stars. This translates to 1 in every 200,000 stars having [Fe/H] $<–$3.5. For halo stars, this relation roughly holds up to [Fe/H] $<-$1.5 which corresponds to the peak of the (inner) halo metallicity distribution \citep{carollo,araa}. At higher metallicities, the metal-rich stars form the vast majority. See \citet{psss} for a more detailed discussion. 
 
In the bulge, the relative occurrence of metal-poor stars is significantly less than what is found in the halo given the gas-rich nature of the central region that has facilitated star formation and thus produced an overwhelming number of metal-rich stars. Finding metal-poor stars in the disk is a similar challenge, although both the thin and the thick disk appear to contain some metal-poor stars \citep{cordoni21, mardini22_thin} that can be found with a combination of kinematic and metallicity analyses.

Interestingly, in the dwarf satellite galaxies, the situation is somewhat different. Ultra-faint dwarf galaxies exclusively contain metal-poor stars with [Fe/H] $<-$1.0. Their star formation process was quenched during the earliest times due to re-ionization and gas being heated away. As a result, the more metal-rich stars never formed. Instead, the challenge lies in correctly identifying member stars, especially for systems whose systemic radial velocity overlaps with, or is close to, the underlying halo star distribution (see also Section~\ref{loc}). In addition, foreground halo stars can mask as member stars but metallicity and any information on the distance can help to eliminate these cases. In contrast, the classical dwarf galaxies do contain stars at all metallicities which requires both a robust membership and metallicity selection in order to discover stars with e.g., [Fe/H] $<-$2.0. 

To overcome all these difficulties, and to identify metal-poor stars across the Galaxy, a variety of search techniques have been developed over the last three decades. One approach is to select metal-poor candidates based on low- and medium resolution spectroscopy with resolving power $R = \lambda/\Delta \lambda$ of $\sim$400 and $\sim$2,000, respectively. This selection is sometimes aided by kinematic information. Such spectra cover the strongest metal line in any stellar spectrum, the Ca II K line at 3933.6\,{\AA}. From this line, a metallicity can be deduced based on carefully constructed calibrations that relate the strength of the Ca line to a corresponding iron abundance \citep{BeersCaKII, Christlieb:2003}. Alternatively, the same has been done using data that covers the Ca II triplet lines in the near infrared regime \citep{starkenburg10}. More details on important surveys for metal-poor stars, such as the HK survey and the Hamburg/ESO survey, and their many discovery successes are provided in \citep{araa} and \citep{fn15}. 

While spectra unambiguously deliver the best indicator of metallicity in the form of measurable absorption lines, they are time consuming to obtain, even with large telescopes. This becomes an increasing concern for fainter stars with magnitudes beyond g$\gtrsim$16. Fortunately, large numbers of brighter metal-poor stars are present in the halo which could be discovered in this way \citep{frebel_bmps, aoki_cemp_2007, zhao_lamost12}. However, for dwarf galaxies, even medium resolution spectroscopy has already limited feasibility. At tens to hundreds of kiloparsecs away, the very brightest stars typically are at 16-18th magnitude. Sparser systems, such as most of the ultra-faint dwarf galaxies, have stars at 18th magnitude or fainter. These are the brightest red giants and there are only few of them given the short time span of this evolutionary phase. The bulk of the stars in any of these ancient systems are located around the turnoff and the main-sequence. But these stars are several magnitudes fainter than the red giants and not reachable with even low-resolution spectroscopy. 

\begin{figure}[t]
\centering
\includegraphics[width=.64\textwidth]{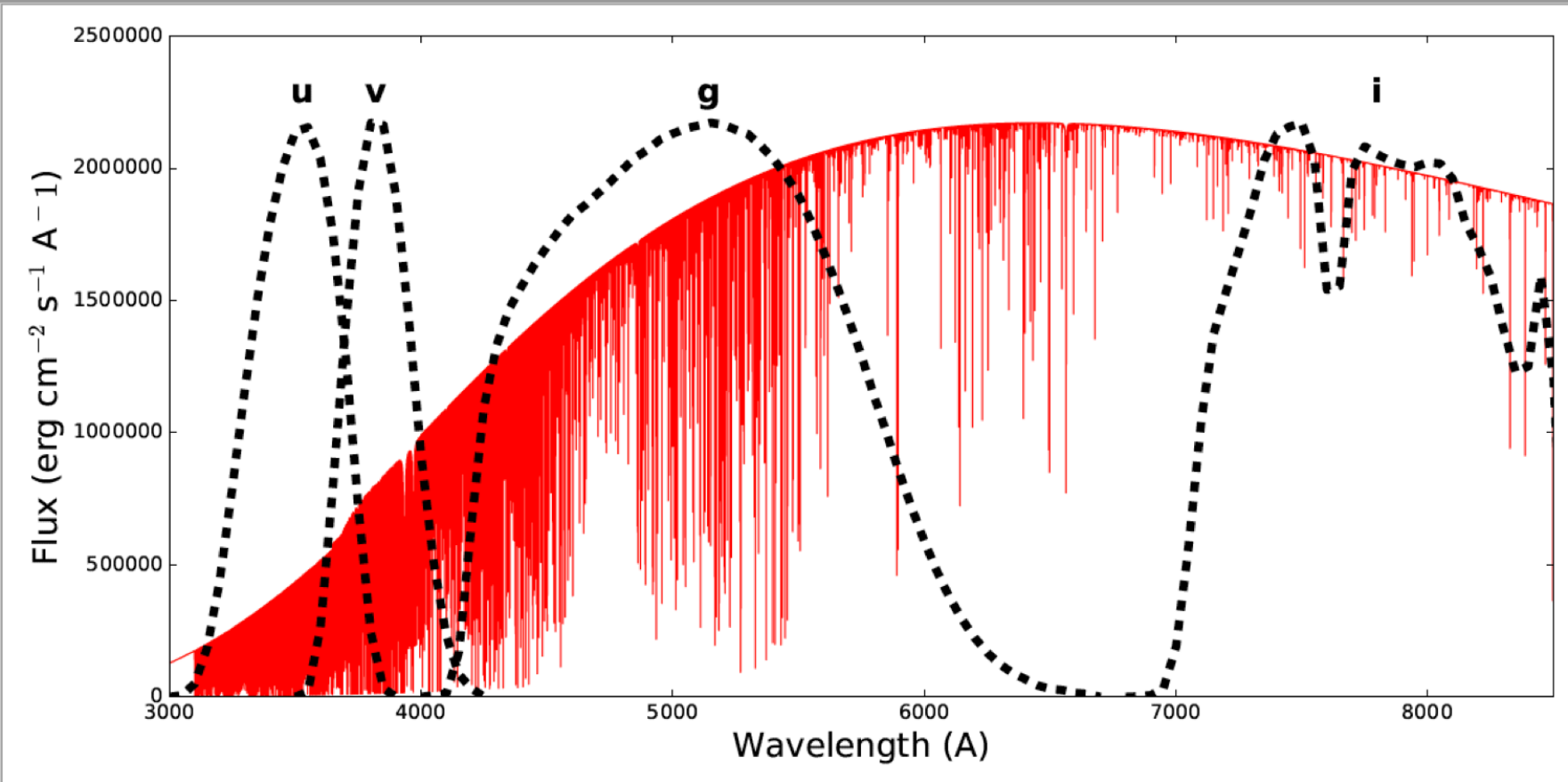}
\includegraphics[width=.35\textwidth]{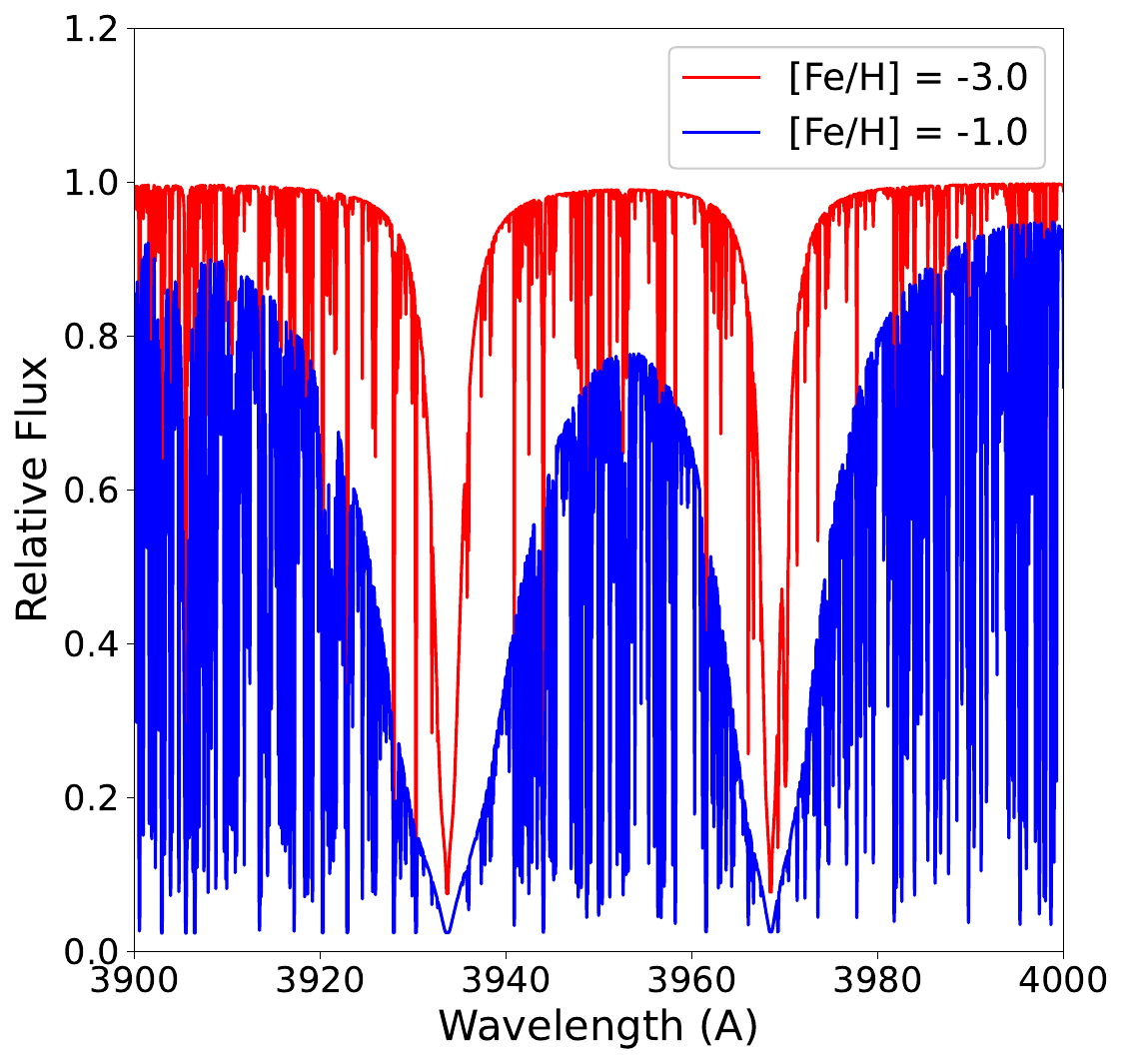}
\caption{Left: Illustration of different filter curves overlaid on a synthetic spectrum of a meatl-poor star with [Fe/H] = $-$3.0. The narrow v filter is centered on the Ca K line at 3933\,{\AA}. Right: Zoom in of the region around the Ca H and K lines (not that Ca H is also blended with H$\epsilon$) which are the two main absorption features in the v filter. Synthetic spectra of stars with with  [Fe/H] = $-$3.0 (red) and [Fe/H] = $-$1.0 (blue) are shown to illustrate that the flux through the v filter is very metallicity sensitive with metal-rich stars having a lower flux due to the increase absorption. Courtesy A. Chiti.}
\label{analysis}
\end{figure}

To overcome these challenges and to study stars fainter than g$\gtrsim$16 and located in the outer halo of the Milky Way as well as the various dwarf galaxies, a new, photometric technique has been developed in recent years that is suitable of obtaining metallicity information from images taken with a narrow-band filter. Typical broadband photometric filters, such as u, g, r, i, z (the SDSS photometric system), cover hundreds of ${\AA}$ and broadly characterize the spectral energy distribution of the star through the respective filter. Spectral lines present within the range covered by the filter blend together; no detailed metallicity information can be gained. Reducing the coverage of the filter to 50 to 100\,{\AA}, and placed over the region that covers the Ca\,II K line, however, provides an avenue for obtaining information on the strength of this line \citep{starkenburg17, chiti20}. In a photometric image, the line strength is recorded as the brightness through the filter. Figure~\ref{analysis} illustrates this concept. 

By definition, metal-poor stars display weaker absorption lines in their spectrum compared to their more metal-rich counterparts. As a result, because of the reduced (lesser) absorption, the flux through the filter is effectively increased compared to that of a more metal-rich star. This principal fortuity is utilized to select metal-poor stars with this technique. However, since line strength and effective temperature of the star are degenerate, at least the temperature needs to be independently determined (such as from broadband photometry, see Section~\ref{analysis_section}) to take this effect into account. Then, in combination with colors from several broadband filters, a metallicity estimate can be obtained (see e.g., \citealt{chiti20} for a detailed description). 

Once metal-poor candidates have been identified with this technique, they still require spectroscopic confirmation, especially to determine detailed chemical abundance measurements of various elements from across the periodic table. After all, absorption lines of most elements in metal-poor stars only become detectable in high-resolution spectra ($R>20,000$) although some elements (e.g. Ca, Mg) can be measured in medium-resolution spectra ($R\sim2,000$ to 4,000).  

The advantage of selecting candidates through narrowband photometry is that the likelihood of the star being metal-poor is dramatically increased. Especially when working with faint stars, telescope time for spectroscopic observations should only be used on high-likelihood targets. For a star with 18th magnitude, even with the largest 6 to 10 meter telescopes, equipped with echelle spectrograph, such as Keck/HIRES, Magellan/MIKE, Subaru/HDS, and VLT/UVES, such spectra take several hours to obtain just to confirm the metallicity from either the Mg triplet lines at 5170\,{\AA} (when obtaining medium resolution spectra) or Fe lines across the spectrum (from high-resolution spectra with $R\sim 20,000$ to 40,000).

Comparisons of the narrowband metallicities with results from high-resolution spectra (in the case of brighter stars) have shown this technique to be about as precise as results from medium-resolution spectra \citep{chiti20, Starkenburg18}. But since much larger samples can be explored from a few images with relatively short exposure times, this is a much more efficient approach. In addition, it has been shown that the discovery rate for stars with [Fe/H] $\lesssim-$2.0 is at an astonishing 85\% level \citep{chiti20}. Some twenty years ago, equivalent yields were 11\% for brighter stars \citep{frebel_bmps} and 20\% for fainter stars \citep{Christlieb:2003}. 

This has made the search for the most metal-poor stars both much easier in terms of the success rate, but also harder as the follow up spectroscopy for any fainter stars takes a lot longer and hardly delivers high quality data from which to determine detailed abundances, if anything at all. In addition, the most iron-poor stars remain extremely rare, and despite new techniques only few have been found. To make further progress in this field, besides placing a narrowband filter over the Ca\,II K line region, other wavelength regions are currently being tested to develop analogous approaches to measuring other elements with strong spectral features such as carbon directly from narrowband images.


\section{Spectroscopy and elemental abundance analysis techniques}\label{analysis_section}

The elemental abundances, and hence the overall chemical composition, of a star are obtained from measuring the strength of the absorption lines of various chemical elements across a stellar spectrum taken with a spectrograph mounted to a telescope. Depending on the nature of the target stars and the scientific objectives, these optical spectra are taken at different resolution levels, ranging from a low (resolving power R = $\lambda/\Delta \lambda \sim 400$), to medium (R $\sim 2,000$), to high resolution (R $> 20,000$) with a given telescope/instrument combination.   The higher the spectral resolution, the more detail and weaker absorption lines become accessible for measurement. 

A star's brightness critically determines whether high-resolution spectroscopy is feasible. A star that is one magnitude fainter than another needs to be observed 2.5 times as long to collect the same amount of photons, i.e. reach the same level of data quality. In the case of high-resolution spectroscopy, this quickly leads to photon starvation, in particular for fainter stars. An alternative is taking medium-resolution spectra but they do not resolve any Fe lines or lines of most other elements. Iron is the element that is usually measured first in high-resolution spectra; its abundance approximates the overall metallicity, and as such, characterizes the star in a major way. Fe lines are numerous and are spread over the entire optical range from 3500 to 6000\,{\AA}. Hence, high-resolution spectra are preferred for a detailed chemical abundance analysis.  

Independent of spectral resolution, strengths of the absorption lines are measured by fitting Gaussian profiles to the line shapes to obtain a so-called equivalent width measurement for each line that reflects its strength. Accuracy of equivalent width measurements depends strongly on the signal-to-noise ratio (S/N) of the data (which also depends on the wavelengths). A S/N of 10-30 of a high-resolution spectrum is regarded low but sufficient for shorter discovery “snapshot" spectra to reveal the principal nature of the star. A very detailed abundance analysis becomes possible when S/N reaches 60-100. 

Elemental abundances are then calculated by modeling the radiative transfer occurring through the stellar atmosphere to obtain the corresponding observed line strengths as a function of elemental abundance. From comparison with the equivalent widths available for each element, the abundance is derived. The resulting abundance uncertainties depend on spectral resolution and S/N of the data but typical values are around 0.05 to 0.10\,dex ($<$26\%) for high-resolution, high-S/N data, and 0.20 to 0.25\,dex for lower S/N spectra. 

The radiative transfer is calculated based on the stellar parameters that consist of the metallicity, effective temperature and surface gravity of the star. Together, these quantities  characterize the stellar atmosphere's physics processes and the emergent flux that makes up the  resulting spectrum \citep{Gray05,Gustafsson08}. This is how a “model atmosphere" is constructed and its output can then be compared to observations. Typically, the assumption of local thermodynamic equilibrium (LTE) is made for these analyses; however, at low metallicity and cooler temperatures, departures from LTE are known to become more significant. Abundance corrections taking into account non-local thermodynamic equilibrium are time consuming to calculate for each individual absorption line and each set of stellar parameters but have been carried out for a number of elements
(e.g. for Fe, see \citealt{bergemann12, lind12, ezzeddine18_H}). In addition, model atmospheres are taken as one-dimensional approximations where the physics processes are calculated for each depth point along the atmosphere. Three-dimensional models have been constructed that describe the dynamic structure of the atmospheres; however they are extremely costly to run and have to be tailored for each individual star and its stellar parameters (e.g., \citealt{behara10, gallagher17, lind24}).

The effective temperature is typically determined from different colors obtained from broad-band photometry such as that of Gaia \citep{gaia21}. They have to be calculated using color-temperature relations \citep{Casagrande14}, taking into account reddening as well as the stellar metallicity. Other methods include taking Fe\,I (neutral Fe) lines and forcing no trend of line abundances with excitation potential of the individual lines. This method is reddening-independent but systematically yields lower temperatures \citep{frebel13} especially for cooler stars. This effect is exacerbated by poor data quality or relatively fewer available weak lines.  

The surface gravity is determined from either placing the star on an isochrone \citep{dcj08} in accordance with its effective temperature, or, if the star is close enough, Gaia's parallax measurements \citep{gaia21} can be used to obtained the distance. Once the distance is known, the gravity can be directly calculated from fundamental relations. Another method is to force no abundance difference of the Fe I (neutral) and Fe II (ionized) average abundances since Fe II is gravity sensitive. 

To complete the stellar parameters, an additional parameter needs to be determined. The microturbulent velocity ensures that both weak and strong lines of a given element (as determined by the atomic properties of the element) yield the same element abundance (on average). This is an artifact of the fact that vast majority of utilized model atmospheres are one-dimensional for practical purposes. More physically realistic state-of-the-art 3D models are computationally much more intensive. 

Output from model atmospheres are “raw" abundances, i.e. the unitless number densities on a logarithmic scale, $\log_{10}$ $\epsilon$ (A), as introduced in Section~\ref{class}. For completeness, it should be noted that these abundances are \textit{elemental} abundances  that represent the contribution of all respective isotopes. Generally, isotopic abundances measurements cannot be measured from stellar spectra, with few exceptions, e.g., C. 

\begin{table}[!h]
\caption{\label{categories} Classes and signatures of metal-poor stars. Based on \citet{frebel18}.}
\begin{tabular}{|l|c|c|c|c@{}}
\hline
Signature & Classes and signatures of metal-poor stars (from \citealt{frebel18})  &Abbreviation\\
\hline
Carbon enhancement & $\mbox{[C/Fe]} > +0.7$ for $\log (L/L_{\odot} ) \le 2.3$ & CEMP\\
& $\mbox{[C/Fe]} \ge [+3.0 - \log(L/L_{\odot} )]$ \mbox{for} $\log(L/L_{\odot}) > 2.3$& CEMP \\
\hline
$\alpha$-element enhancement & $\mbox{[Mg, Si, Ca, Ti/Fe]} \sim +0.4$ & $\alpha$-enhanced \\
\hline
Neutron-capture {normal} & $\mbox{[Ba/Fe]} < 0$ & No\\
\hline
Main $r$-process & $0.3 \le \mbox{[Eu/Fe]} \le +1.0$ and $\mbox{[Ba/Eu]} < 0.0$ & $r$-I \\
& $\mbox{[Eu/Fe]} > +1.0$ and $\mbox{[Ba/Eu]} < 0.0$ & $r$-II \\
\hline
Limited $r$-process & $\mbox{[Eu/Fe]} < 0.3$, 
$\mbox{[Sr/Ba]} > 0.5$, and $\mbox{[Sr/Eu]} > 0.0$ & $r_{\rm{lim}}$\\
\hline
$s$-process & $\mbox{[Ba/Fe]} > +1.0$, $\mbox{[Ba/Eu]} > +0.5$, $\mbox{[Ba/Pb]} > -1.5$ & $s$\\
\hline
$r$- and $s$-processes & $0.0 < \mbox{[Ba/Eu]} <+0.5$ and $-1.0<\mbox{[Ba/Pb]}<-0.5$ &$r+s$\\
\hline
$i$-process & No unambiguous match to neutron-capture element patterns/criteria & $i$\\
\hline
\end{tabular}
\end{table}

\section{Different types of metal-poor stars}\label{types}

In the following, multiple different classes of metal-poor stars are presented to provide a general overview, and to illustrate the large range of science questions that can be addressed with different types metal-poor stars. Table~\ref{categories} summarizes the classes and types. These kinds of stars are all found in the Milky Way's halo. Some of them are also found in globular clusters, stellar streams, and dwarf galaxies, see Sections~\ref{gc} and \ref{dwgal}.  

For easy reference, Table~\ref{top50} provides a representative list of the most interesting and important stars, dwarf galaxies, globular clusters and stellar streams that represent the various groups and classes of stars and systems discussed in this Section.

\subsection{Ordinary metal-poor stars}  

Spectroscopic studies of metal-poor stars carried out over the last four decades have revealed the chemical composition that is typical for most of these stars \citep{psss, frebel15}. 
More than a thousand metal-poor stars have been individually analyzed to date. Many of these stars are captured in the SAGA database \citep{Suda08} and JINAbase \citep{jinabase} for easy access. In addition, massive surveys such as GALAH \citep{DeSilva15} and APOGEE \citep{Majewski17} have contributed thousands of high-resolution survey spectra, although most of then are in the more metal-rich regime.

\begin{figure}[t]
\centering
\includegraphics[width=.9\textwidth]{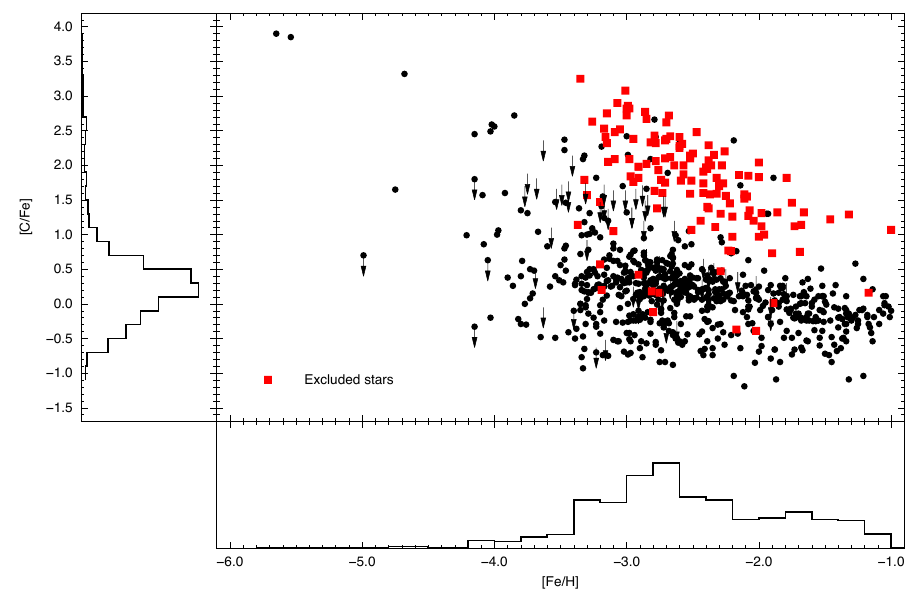}
\caption{[C/Fe] abundance ratios of metal-poor halo star as a function of metallicity [Fe/H]. Black filled dots are ordinary halo stars, and CEMP-no, CEMP-r stars. Arrows indicate upper limits. Red filled squares are CEMP-s and CEMP-i stars whose carbon enhancements have an external origin, see Section~\ref{sproc} for more details. The distributions of each abundance ratio for the ordinary halo stars, and CEMP-no, CEMP-r stars are shown as histograms. Reproduced with permission from \citet{placco14}.}
\label{carbon}
\end{figure}

In the following, selected element abundances and their trends among metal-poor stars are discussed. Carbon abundances among metal-poor stars show a rather large spread reflecting multiple different prior nucleosynthesis channels and associated timelines. This behavior is shown in Figure~\ref{carbon}. The fraction of carbon-rich stars increases with lower [Fe/H], reaching 100\% at the lowest iron abundances. This has made carbon enrichment a major feature of the most metal-poor stars. The nature and features of different types of carbon-rich metal-poor stars are described in more detail in Section~\ref{cemp}). 

\begin{figure}[t]
\centering
\includegraphics[width=.9\textwidth]{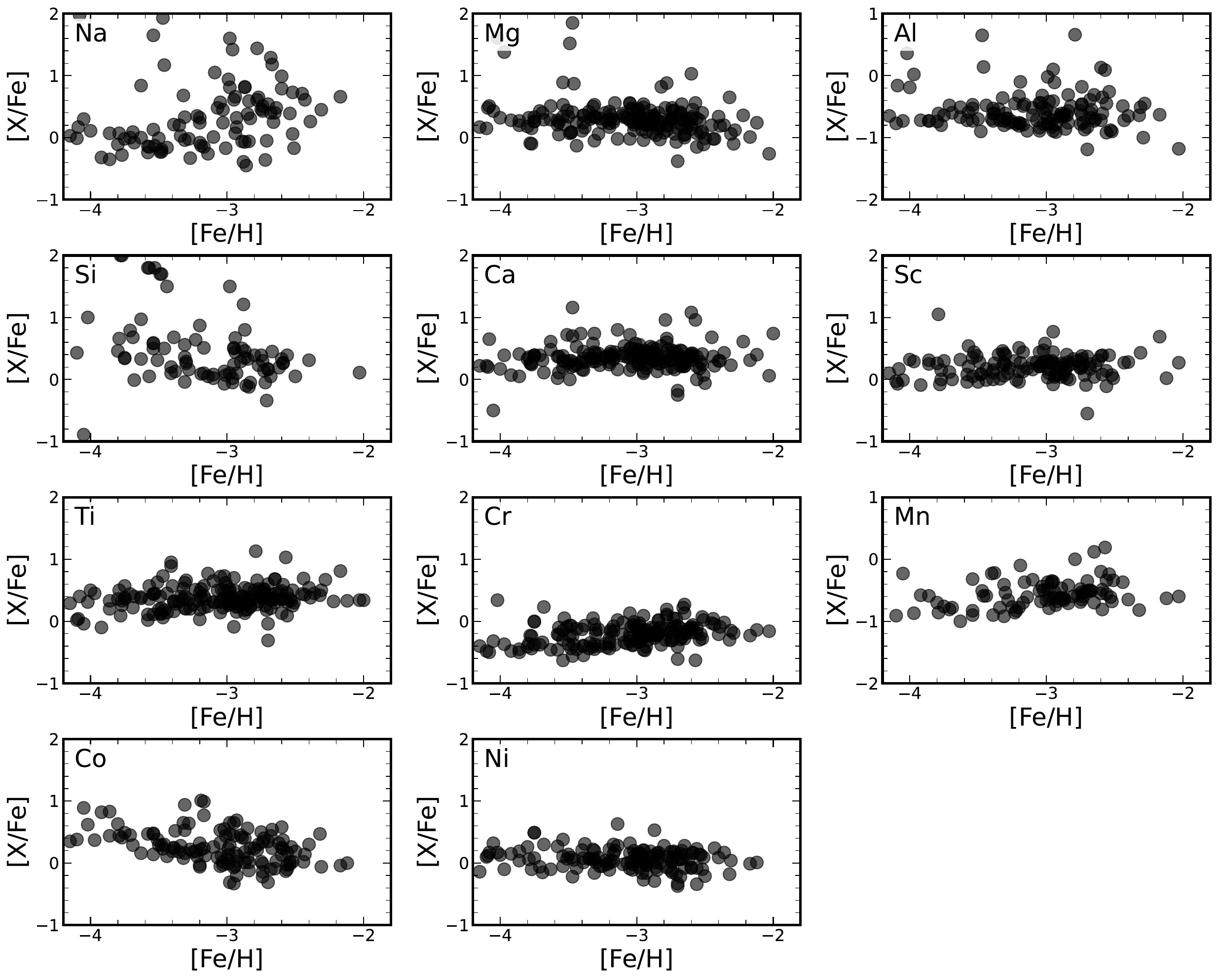}
\caption{Elemental abundance ratios [X/Fe] as a function of metallicity [Fe/H] for various light elements, for a large sample of metal-poor halo stars homogeneously analyzed by \citet{yong13_II}. Note that dwarf galaxy stellar abundances follow these trends extremely well. Courtesy M. Mardini.}
\label{yong}
\end{figure}

Figure~\ref{yong} shows the elemental abundance ratios as a function of [Fe/H] for a large sample of homogeneously analyzed metal-poor halo stars \citep{norris13_I, yong13_II}. Some elemental ratios do vary with [Fe/H], others are stable. Most are at least slightly different than the solar abundance ratios due to the chemical evolution of the Milky Way changing with time until the birth of the Sun 4.6\,Gyr ago.

Sodium (Na) shows a very large scatter among halo stars that appears to be produced somewhat stochastically, at least for halo stars. Note that this is completely different for stars in globular clusters, as described in Section~\ref{gc}.

The $\alpha$-elements magnesium (Mg), calcium (Ca), titanium (Ti) and silicon (Si) are enhanced, typically at the 0.4\,dex level, reflecting an early metal production almost exclusively by core-collapse (type II) supernovae. Only at higher metallicities, around [Fe/H] $\sim-$1.0, the ratios all change, becoming lower and ending up at the solar ratio for [Fe/H] = 0. The turnover reflects the onset of supernovae type Ia that produce largely iron, thus driving down any [X/Fe] ratio towards the solar level. This downturn, or “knee", is generally regarded an important point in time for the evolution of a system since it marks the delay time of the first low-mass stars evolving to the white dwarf stage to then explode as Type Ia. Outliers from the observed trends include “$\alpha$-poor" stars with [$\alpha$/Fe] $\sim$0 \citep{francois20}, or stars with large Mg and Si abundances of [Mg,Si/Fe] $>$1 \citep{aoki_mg}. Note that dwarf galaxies, with their different evolution and timelines, show different behaviors (see Section~\ref{dwgal}.) 

Aluminum (Al) follows a well-defined trend albeit with some scatter, not too different from that of iron-peak elements (Sc, Cr, Mn, Co, Ni). These do not all display the same behavior but generally show well-defined trends. They all track the early metal production of core-collapse supernovae. 

\begin{figure}[t]
\centering
\includegraphics[width=.99\textwidth]{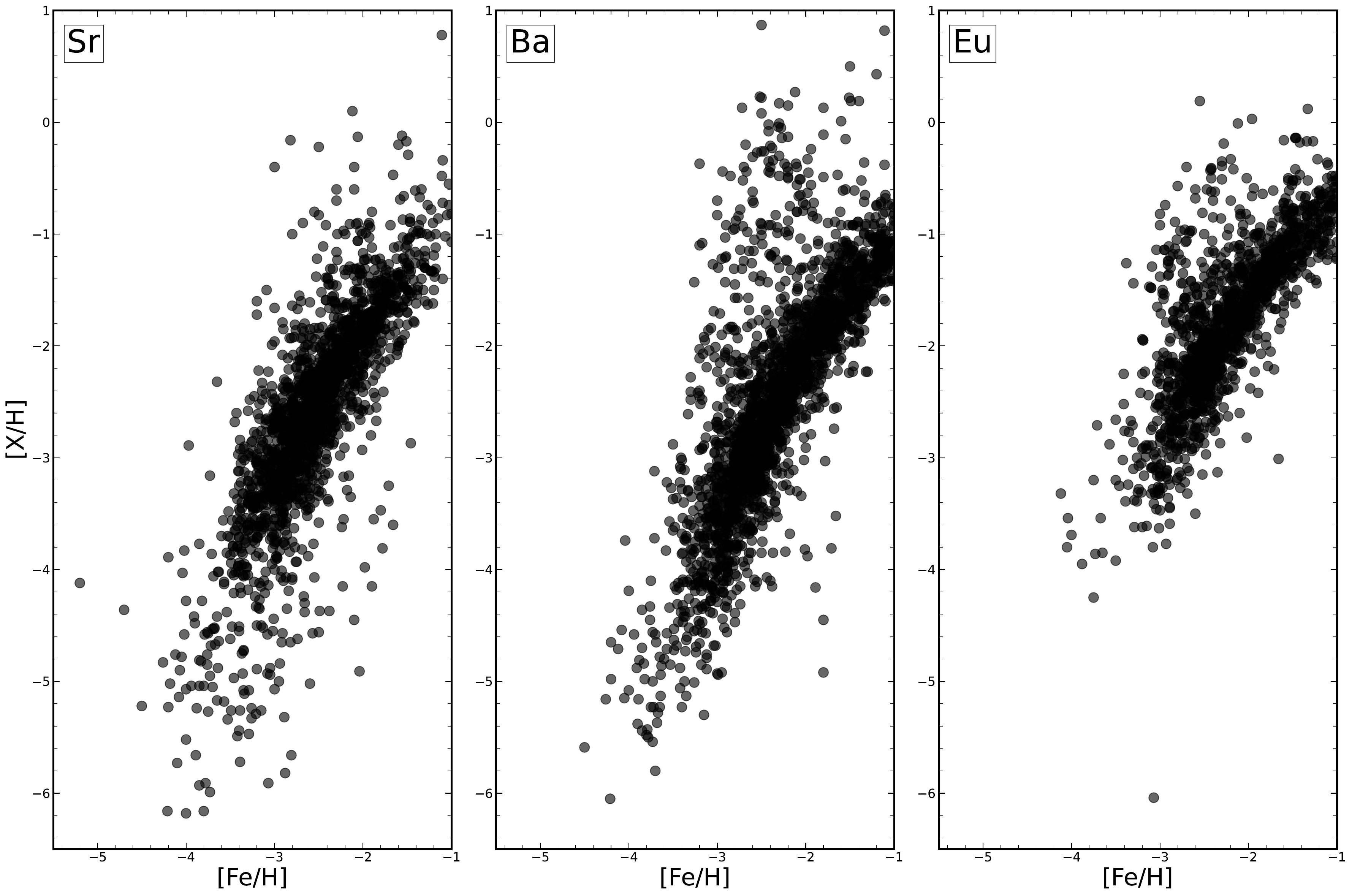}
\caption{Neutron-capture element abundances of metal-poor halo stars for strontium and barium as a function of metallicity [Fe/H]. Note that dwarf galaxy stellar abundances largely follow these trends, with many having extremely low neutron-capture element abundances. Courtesy M. Mardini.}
\label{heavy}
\end{figure}

Neutron-capture elements strontium (Sr) and barium (Ba) show a large spread at the lower [Fe/H], in particular at [Fe/H] $\sim-$3, which then tightens into a more well-defined trend towards higher metallicities, commensurate with a more regulated chemical evolution and the onset of the s-process at [Fe/H] $\sim-$2 (see Section~\ref{sproc}). These trends are shown in Figure~\ref{heavy}. Stars with specific patterns of neutron-capture elements are further discussed in the sections below.

It is important to note that not all elements can easily be measured in stellar spectra, especially in those of metal-poor stars which generally display weak absorption lines in their spectra. As such, nitrogen (N), oxygen (O) are notoriously difficult or impossible to measure. Most neutron-capture elements only become measurable when the star shows a large overabundance in these elements due to specific nucleosynthesis processes that operated prior to the star's birth. 
Other reasons include absorption lines being located in the ultraviolet \citep{roederer12_hubble, roederer20_pb, jacobson15} of infrared \citep{schuler07} wavelength regime for which data is difficult to acquire, or no absorption line being available for measurement in metal-poor stars.

Overall, all these abundance trends have been reproduced with detailed chemical evolution models that take into account the different nucleosynthesis sources and timelines, and the build up of the Milky Way over time \citep{cote19, kobayashi20}. 

\subsection{Most iron-poor stars}

Unlike ordinary metal-poor stars, stars with [Fe/H] $\lesssim-$3.5 usually exhibit a number of abundance ratios that deviate from the regular trends. This likely reflects the more stochastic enrichment events that occurred prior to their births, including those of the first stars before any regular chemical evolution began. Most importantly, these iron-poor stars contain such little iron that they are likely all second-generation stars, born soon after the massive first Population III stars exploded as the first supernovae. A simple example illustrates this point: Mixing 0.1\,M$_{\odot}$ of iron from a first supernova instantaneously and homogeneously into 10$^5$\,M$_{\odot}$ of hydrogen gas results in a next-generation star with [Fe/H] $=-$3.2. Stars with metallicities lower than this value are thus the most likely second-generation stars, the earliest and most extreme Population II stars, and the very first low-mass stars to have formed in the universe. As such, they serve as the best probes of star formation in low-metallicity gas and the associated cooling processes \citep{brommloeb03}. The current record holder for the most iron-poor star is SMSS J0313-6708, with [Fe/H] $<-7$ \citep{keller14}. The Fe abundance is so low that no absorption lines could be detected in the spectrum, resulting in just an upper limit. For the few stars with [Fe/H] $\sim-5$, a few Fe line are often barely detectable. All these stars are regarded second-generation stars and most valuable to stellar archaeology.

Their low Fe abundances makes these stars very attractive for studying the nature and nucleosynthesis yields of the first stars and supernovae that cannot otherwise be obtained. To gain such information, theoretical predictions for the supernova yields are compared with the observed abundance pattern of the iron-poor stars. This reveals details about the first star progenitors, such as their masses, explosion energy and sometimes also the explosion mechanisms. Especially the masses are important for reconstructing the initial mass function (IMF) of Population III; however, these mass estimates are often uncertain or even degenerate between different predictions. 

Regardless, the most iron-poor stars offer an unprecedented view into the early universe and the nucleosynthesis processes that operated then. Certain elements, such as carbon, are extremely overabundant in almost all of these stars (see Section~\ref{cemp}), along with nitrogen and oxygen. The large [C/Fe] ratios are thought to be the result of a lower energy, “faint", first supernovae with significant fallback, i.e. the inner part of the star that contains the iron largely falls back onto the nascent black hole while the outer part, containing the carbon, gets  ejected in the explosion and dispersed in the interstellar medium. Other iron-poor stars show unusually large amounts of magnesium and strontium. Some of the most interesting most iron-poor stars are listed in Table~\ref{top50}.


\clearpage

\begin{table}[h!]
\caption{Astrophysically interesting metal-poor stars}
\begin{tabular*}
{\textwidth}{@{\extracolsep{\fill}}@{}p{2.0cm}p{2.5cm}p{3.0cm}p{4.0cm}p{2.8cm}@{}}
\toprule
\multicolumn{1}{@{}l}{\TCH{Star}} & 
\multicolumn{1}{c}{\TCH{Abundance feature}} &
\multicolumn{1}{c}{\TCH{Classification}} &
\multicolumn{1}{l}{\TCH{Description} }&
\multicolumn{1}{l}{\TCH{Reference\footnote{a}}}\\
\hline\\
\multicolumn{5}{c}{Metal-poor halo stars}\\
\hline
CS~22892-052 	&	\mbox{[Fe/H] $=-$3.2}, \mbox{[Eu/Fe] = 1.2} 	&	R-process star, CEMP-r star 	&	First r-process metal-poor star; with Th measurement; carbon enhanced	&	\citet{sneden03}	\\
HE1327$-$2326 	&	\mbox{[Fe/H] = $-$5.4}, \mbox{[C/Fe] = 4.1}, \mbox{[Sr/Fe] = 1.1}	&	\mbox{Most Fe-poor star}, CEMP star	&	Second [Fe/H] $<-$5 star with high CNO abundances	&	\citet{frebel05}	\\
HE0107$-$5240	&	\mbox{[Fe/H] = $-$5.2}, [\mbox{C/Fe] = 4.0}	& \mbox{Most Fe-poor star}, CEMP-no star	&	First [Fe/H] $<-$5 star with high CNO abundances	&	\citet{christliebetal02}	\\
SM0313$-$6708	&	\mbox{[Fe/H] $<-$ 7}	&	\mbox{Most Fe-poor star}, CEMP-no star	&	Current low [Fe/H] record; no Fe lines detected	&	\citet{keller14}	\\
CS31082-001	&	\mbox{[Fe/H] $=-$3.0}, \mbox{[Eu/Fe] = 1.6}	&	R-process star	&	First r-process star with U measurement; not carbon-hanced	&	\citet{cayrel04}	\\
CS 22949-037	&	\mbox{[Fe/H] = $-$3.9}, \mbox{[O,Na,Mg,Si/Fe]} enhanced, \mbox{[Ba/Fe] $<$ 0}	&	CEMP-no star	&	High C, Mg and Si abundances	&	\citet{depagne02}	\\
HD 196944	&	\mbox{[Fe/H] = $-$2.3}, \mbox{[C/Fe] = 1.3}, \mbox{[Ba/Fe] = 1.1}, \mbox{[Eu/Fe] = 0.2} 	&	CEMP-s/s-process star	&	High Pb abundance, CH star	&	\citet{2001vaneck, aoki02}	\\
LP 625$-$44	&	\mbox{[Fe/H]= $-$2.8}, \mbox{[C/Fe] = 2.3}, \mbox{[Ba/Fe] = 2.8}, \mbox{[Eu/Fe] = 1.9}	&	CEMP-i/i-process star	&	Heavy element pattern distinct from r- and s-process pattern	&	\citet{norris97_carbon, aoki_lead2000}	\\	
J1521$-$3538	&	\mbox{[Fe/H] = $-$2.8}, \mbox{[Ba/Fe] = 1.4}, \mbox{[Eu/Fe] = 2.2}	&	R-process star	&	First r-process stars with \mbox{[Eu/Fe] $>$ 2}	&	\citet{cain20} 	\\
J0949$-$1617	&	\mbox{[Fe/H] = $-$2.2}, \mbox{[Ba/Fe] = 1.0}, \mbox{[Eu/Fe] = 0.6}	&	CEMP r+s star	&	First metal-poor star with combined s- and r-process pattern	&	\citet{gull18}	\\
HD 222925	&	\mbox{[Fe/H] = $-$1.5}, \mbox{[Eu/Fe] = 1.3}	&	R-process star	&	Most complete heavy element r-process pattern	&	\citet{roederer22}	\\
HD140283 	&	[Fe/H] = $-$2.5	&	Metal-poor halo star	&	First classic metal-poor star	&	\citet{chamberlain51}	\\
HD122563	&	[Fe/H] = $-$2.7	&	Limited r-process star, halo star	&	First classic metal-poor star; limited r-process star	&	\citet{wallerstein63, honda06}	\\	
G64$-$12	&	[Fe/H] = $-$3.2	&	Metal-poor halo star	&	Classic bright halo star; radial velocity of 442 km/s	&	\citet{carney96}	\\
BD+80 245	&	[Fe/H] = $-$2.0	&	Metal-poor halo star	&	First "$\alpha$-poor, iron-rich" metal-poor star; unusual iron-group abundances	&	\citet{carney_alphapoor}	\\
J1808$-$5104	&	\mbox{Fe/H] = $-$3.9}	&	Thin disk star	&	Metal-poor thin disk star	&	\citet{mardini22_thin}	\\
J10291+17292	&	\mbox{[Fe/H] = $-$5.0}, \mbox{[C/Fe] $<$1.1}	&	Most Fe-poor star, Atari disk star	&	First [Fe/H] $\sim -$5 star, with low carbon abundance	&	\citet{caffau11}	\\
\mbox{J2050$-$6613}&	\mbox{[Fe/H] = $-$4.1}, \mbox{[Sr/Fe] = 1.0}, \mbox{[Ba/Fe] = $-$1.1}	&	Atari disk star	&	High [Sr/Ba] $\sim2$ disk star, limited r-process star	&	\citet{mardini24_atari}	\\

HD101065 ("Przybylski's Star")	&	[Fe/H] $\sim-$2.0	&	Chemically peculiar star	&	Star with internally altered surface abundances; strong heavy element enhancements	&	\citet{przybylski66, shulyak10}	\\
R Andromedae	&	[Fe/H] = $-$1.0	&	Chemically peculiar star	&	Mira variable star with detected Tc; confirmed nucleosynthesis to occur in stars	&	\citet{merrill52}\\
\botrule
\label{top50}
\end{tabular*}
%
\end{table}

\begin{table}[h!]
\renewcommand\thetable{3}
\caption{-- continued}
 \begin{tabular*}
 {\textwidth}{@{\extracolsep{\fill}}@{}p{2.0cm}p{2.5cm}p{3.0cm}p{4.0cm}p{2.8cm}@{}lllll}
 \toprule
 \multicolumn{1}{@{}l}{\TCH{Star}} & 
 \multicolumn{1}{c}{\TCH{Abundance feature}} &
 \multicolumn{1}{c}{\TCH{Classification}} &
 \multicolumn{1}{l}{\TCH{Description} }&
 \multicolumn{1}{l}{\TCH{Reference\footnote{a}}}\\
 \hline\\
 CarIII-1120	&	\mbox{[Fe/H] = $-$3.9}, \mbox{[C/Fe] = 1.7}	&	Most Fe-poor star, CEMP star	&	CEMP star; most Fe-poor star in an ultra-faint dwarf galaxy	&	\citet{ji20_car} 	\\
AS0039	&	\mbox{[Fe/H] = $-$4.1}, [C/Fe] = $-$0.3	&	Most Fe-poor star	&	Carbon-normal star; most Fe-poor star in any dwarf spheroidal galaxy	&	\citet{skuladottir21}	\\
LMC-119	&	\mbox{[Fe/H] = $-$4.2}, \mbox{[C/Fe] $<$ 0.4}	&	Most Fe-poor star	&	Carbon-normal star; most Fe-poor star in a Milky Way satellite	& \citet{chiti24}\\
\hline\\
\multicolumn{5}{c}{Metal-poor ultra-faint dwarf galaxies}\\
 \hline
 Segue 1	&	$<\rm{[Fe/H]} = -$2.7, \mbox{no stars with} \mbox{[Fe/H]$>-$1.0}	&	Ultra-faint dwarf galaxy	&	Best surviving analog of a first galaxy	&	\citet{frebel14}	\\
 Reticulum II	&	$<\rm{[Fe/H]} = -$2.7, \mbox{most stars have} \mbox{[Eu/Fe] $>$ 0.7}	&	Ultra-faint dwarf galaxy	&	R-process enriched metal-poor dwarf galaxy	&	\citet{ji16a}	\\	
 Tucana II	&	$<\rm{[Fe/H]}> = -$2.9, many stars spatially extended	&	Ultra-faint dwarf galaxy	&	Extremely metal-poor assembled dwarf galaxy with a $<$[Fe/H]$>$ = $-$3.1 extended outer component	&	\citet{chiti21_tuc}	\\			
 Bootes I	&	$<\rm{[Fe/H]}> =-$2.3 	&	Ultra-faint dwarf galaxy	&	"Canonical" ultra-faint dwarf galaxy	&	\citet{norris10a}	\\				
 Sculptor	&	$<\rm{[Fe/H]}> =-$1.7 	&	Classical dwarf spheroidal galaxy	&	"Canonical" classical dwarf spheroidal 	&	\citet{hill19}	\\
 \hline\\
 \multicolumn{5}{c}{Metal-poor globular clusters and stellar streams}\\
 \hline
 M92	&	[Fe/H] $\sim-$2.4	&	Globular cluster	&	R-process rich cluster	&	\citet{kirby23}	\\
 C-19	&	[Fe/H] $<-$3.0	&	Stellar stream	&	Most Fe-poor stellar stream	&	\citet{martin22}	\\	
 47 Tuc	&	[Fe/H] $\sim-$0.8	&	Globular cluster	&	Multiple stellar populations	&	\citet{milone12}	\\	
 Omega Cen	&	\mbox{[Fe/H] $\sim-$1.4 to} $-$1.7	&	Globular cluster	&	Massive globular cluster with ultiple stellar populations	&	\citet{milone17}	\\
 M15	&	[Fe/H] $\sim-$2.5	&	Globular cluster	&	R-process-rich globular cluster	&	\citet{kirby20}	\\		
 Phoenix Stream	&	[Fe/H] $\sim-$2.7	&	Stellar stream	&	Stellar stream/debris of a very metal-poor globular cluster 	&	\citet{wan20}	\\
 Sagittarius Stream	&	[Fe/H] $\sim-$1.0	&	Dwarf galaxy stream	&	First known dwarf galaxy stream	&	\citet{majewski03}	\\
 Orphan-Chenab Stream	&	[Fe/H] $\sim-$1.9	&	Dwarf galaxy stream	&	Second known dwarf galaxy stream; strong impact from LMC; connected two streams	&	\citet{koposov23}	\\
 Palomar 5	&	[Fe/H] $\sim-$1.4	&	Globular cluster stream	&	Disrupting globular cluster with prominent tidal tails	&	\citet{koch04}	\\		
 EXT8	&	[Fe/H] $\sim-$3.0	&	Globular cluster	&	Intact globular cluster in Andromeda (most metal-poor cluster known)	&	\citet{larsen20}	\\
 \botrule
 \end{tabular*}
 \end{table}

\clearpage


\subsection{Carbon-enhanced metal-poor stars}\label{cemp}

Carbon-enhanced metal-poor (CEMP) stars exhibit large abundances of carbon relative to iron, i.e. [C/Fe] $>$ +0.7 \citep{aoki07}, which has long been found to be common feature among halo stars \citep{1999rossicarbon}. As can be seen in Figure~\ref{carbon}, significant number of stars show extreme overabundances of [C/Fe] $>$ +2 and higher, e.g., HE~1327$-$2326 with [Fe/H] $=-$5.4 and [C/Fe] $=$+4.1 \citep{HE1327_Nature, Aokihe1327}. In fact, the frequency of CEMP stars in the halo increases with decreasing metallicty \citep{placco14}, from 24\% at [Fe/H] $<-$2.5 to 100\% at [Fe/H] $<-$5, suggesting that carbon played an important role in early star formation processes \citep{dtrans}. Interestingly, some dwarf galaxies lack this large fraction of CEMP stars and show a different behavior, such as Sculptor \citep{chiti18, skuladottir15} and the Large Magellanic Cloud \citep{chiti24}.

In order to explain extreme stars such as HE~1327$-$2326, the large observed carbon overabundances in these most metal-poor stars have been attributed to significant C-N-O rich yields of the supernova explosions of the massive Population III first stars. Specifically, faint first supernovae are assumed to eject very little iron due to a fallback mechanism \citep{UmedaNomoto:2002} which results in a large [C/Fe] ratio of the surrounding gas from which the most iron-poor stars then formed \cite{cooke14}. As such, carbon abundances of individual stars, as well as larger samples, provide important clues about the properties of their progenitor stars, such as potential mass loss during stellar evolution, stellar mass, supernova explosion energy and explosion mechanism. 

When it comes to determining final stellar carbon abundances ready for interpretation, it is important to note that carbon (along with nitrogen but which is very difficult to measure in stars with [Fe/H] $<-$2) is the only element whose surface abundance changes as the star evolves. As a star ascends the red giant branch, its surface carbon abundance reduces as a result of conversion of carbon into nitrogen via the CN cycle operating in deeper layers, and carbon-poor gas being flushed to the surface \citep{gratton00}. However, by using stellar evolutionary models, the observed (depleted) carbon abundances can be corrected for this effect \citep{placco14}. The observed (corrected) carbon abundance then reflects the natal carbon abundance of the birth gas clouds. Then the observed abundance can be correctly interpreted with regards to how any nucleosynthesis processes of the first stars and/or supernovae may have provided the carbon. 

Many CEMP stars also show other elemental abundance signatures. Hence, there exist several  subgroups (discussed further below). The stars discussed thus far directly trace the carbon abundance of their (Population III) progenitors and are termed CEMP-no stars, with “no" referring to “normal neutron-capture element abundances", i.e. they show no unusual or specific heavy element pattern. Some CEMP-no stars are also enriched in $\alpha$-elements Mg and Si \citep{aoki_mg}. Examples of these groups are listed in Table~\ref{top50}.

The CEMP-r stars depict a relatively smaller group of CEMP stars that are enhanced in heavy elements formed in the rapid (r-) neutron-capture process, such as during a neutron star merger discussed in more detail in Section~\ref{rproc}. The presence of r-process elements in these CEMP stars indicates that they formed from gas previously enriched in material from (at least) two nucleosynthesis events, i.e. a supernovae and a neutron star merger. Any of these events may have different enrichment timelines which might explain why no r-process stars have yet been found with [Fe/H] $<-$3.5, in stark contrast to the CEMP-no stars that are found down to [Fe/H] $<-$7 \citep{keller14, Nordlander17}. 

An entirely different group of CEMP stars are the CEMP-s stars. They are easily identified by their rather large carbon overabundances in combination with large enhancements in heavy elements made in the slow (s-) neutron-capture process. However, the carbon abundance does not stem from the star's natal gas cloud but instead from a mass transfer event from a former binary companion that underwent its asymptotic giant branch (AGB) phase which is characterized by significant production of carbon as well as s-process elements. Through stellar winds or Roche lobe overflow, this carbon-rich material is transferred across the orbit and onto the slightly less massive secondary star. As a result, a CEMP-s star is created whose abundances reflect the stellar evolution and nucleosynthesis processes occuring in the former (evolved) primary star of the system rather than any events that enriched the birth gas cloud. More details on s-process stars are provided in Section~\ref{sproc}.

A related group are the CEMP-r+s stars. These objects show enhancements in carbon and both the r- and s-process elements. They basically are CEMP-r stars that are part of a binary systems that has undergone a mass transfer event. As such, their abundance signature is the product of three different nucleosynthesis events and sites. Thus far, only one such star has been found that shows this specific combination of the two neutron-capture element processes, likely because r-process stars are extremely rare \citep{gull18}.

Finally, CEMP-i stars show neutron-capture element patterns attributed to the intermediate i-process (referring to the neutron flux necessary for neutron-capture, and it being in between that of the s- and the r-process). This process yields an element pattern different from that of the r- and the s-process and does vary from star to star but has been successfully modeled \citep{hampel16,clarkson18}.

\subsection{$r$-process enhanced metal-poor stars}\label{rproc}

\begin{figure}[t]
\centering
\includegraphics[width=.99\textwidth]{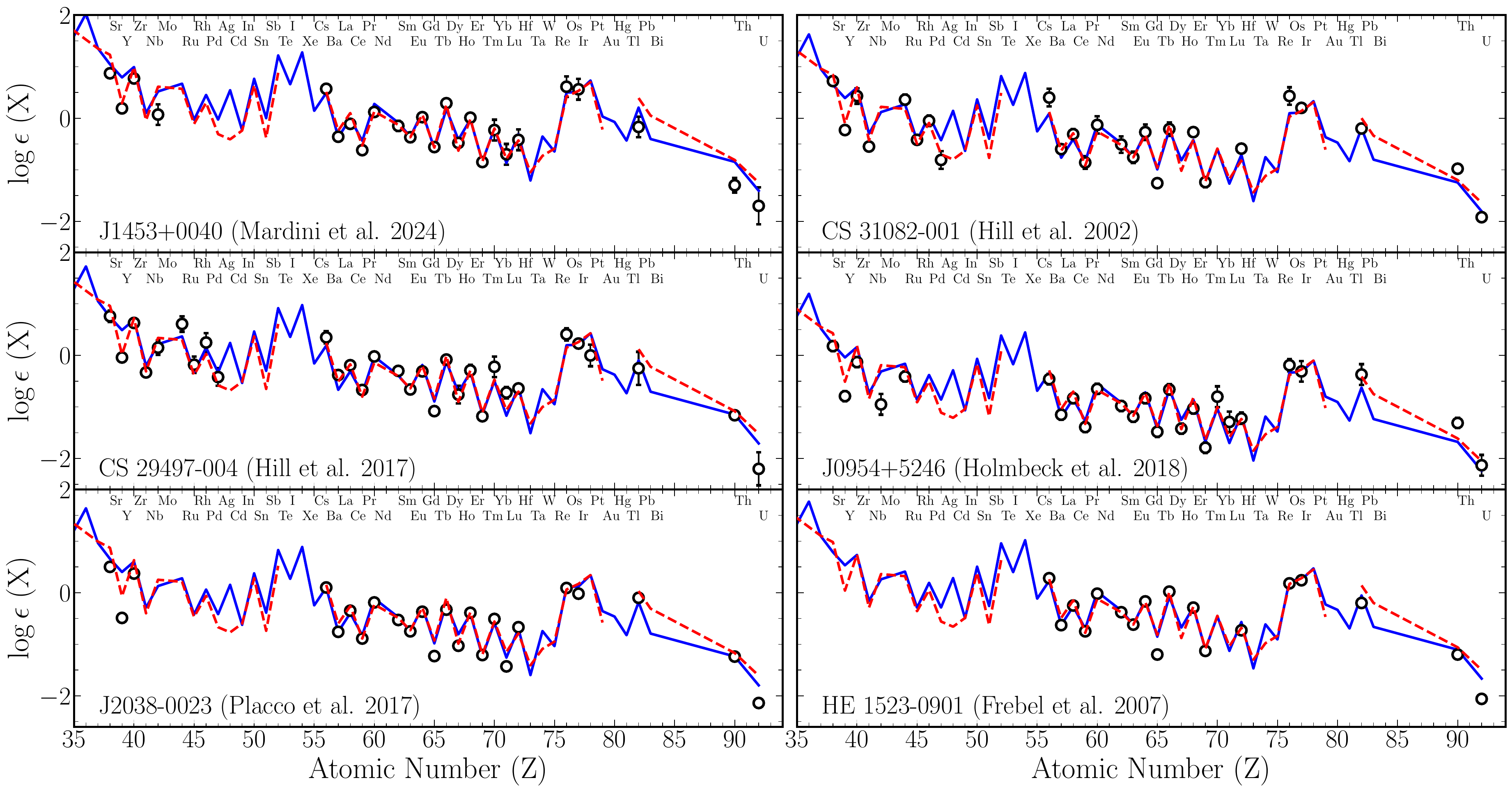}
\caption{R-process abundance patterns of six metal-poor r-process enhanced stars with detected actinide elements thorium (Th) and uranium (U). The stellar elemental abundances are compared with the scaled solar r-process pattern (blue solid line) and the abundances of HD222925 \citep{roederer22}, all scaled to europium. There is remarkable agreement for elements heavier than Ba (Z $>$56). Actinide elements show deviations because of their decay over cosmic time but additionally, the stars do show some scatter that has been termed “actinide boost". Deviations also often occur among light neutron-capture elements strontium (Sr), yttrium (Y) and zirconium (Zr), possibly due to other nucleosynthesis process contributing such as the limited r-process. Based on \citet{mardini_ustar}.}
\label{rproc_fig}
\end{figure}

A subgroup of metal-poor stars show significant overabundances in heavy elements associated with the rapid neutron-capture (r-) process. The main r-process occurs within 1-2 seconds in extremely neutron-rich environments where seed nuclei rapidly grow into unstable neutron-rich nuclei from the neutron bombardment. They $\beta$-decay into heavy stable isotopes after the neutron flux ceases. These r-process stars are then thought to have formed from gas previously enriched by an r-process event. They typically have low metallicities of $-3<$ [Fe/H] $<-$1.5 and exhibit prominent absorption lines of most heavy r-process elements in their high-resolution spectra, such as samarium (Sm), europium (Eu), gadolinium (Gd), and dysprosium (Dy). Altogether, the heavy element abundances follow a characteristic, universal pattern that extremely well aligns with the solar r-process pattern\footnote{The solar r-process pattern can be obtained from subtracting a calculated s-process component from the measured total abundances.} when scaled to the stellar abundances, such as Eu. The r-process pattern shows three prominent peaks as a result of the neutron-capture onto seed nuclei during the r-process depending on the properties of the involved isotopes, including those with closed shells. Figure~\ref{rproc_fig} shows examples of six r-process stars and their observed stellar abundance patterns.

The overall level of the observed r-process enhancement is usually quantified by means of the [Eu/Fe] abundance. Stars with [Eu/Fe] $>$ 0.3 are considered as mildly enhanced and called r-I stars. They are found across a broad metallicity range with [Fe/H] $\lesssim$ $-$1.5. About a quarter of all metal-poor stars show this signature \citep{hansen18}. Stars with [Eu/Fe] $>$ 0.7 are termed strongly enhanced r-II stars \citep{araa, frebel18}, and r-III stars have an extreme [Eu/Fe] $>$ 2.0 \citep{cain20, ezzeddine20}. r-II and r-III stars typically have [Fe/H] $\sim-$3.0, and they occur at a rate of about 5\% among metal-poor stars \citep{holmbeck20}. Some r-process stars also display large carbon abundances, making them CEMP-r stars, such as CS22892-052 \citep{sneden03}. Others appear to be in binary systems \citep{hansen16}. 

The vast majority of r-process stars is found in the Galactic halo but they have also been discovered in dwarf galaxies, globular clusters and some stellar streams. An extreme example is the ultra-faint dwarf galaxy Reticulum\,II \citep{ji16b, Roederer16b} that contains almost exclusively r-II stars. Only the two most metal-poor stars are not r-process enhanced. Given that this small dwarf galaxy has survived to the present, limits on the available gas mass can be placed which constrains the origin of the observed abundances. It thus appears likely that a neutron star merger enriched Reticulum\,II before the long-lived low-mass stars now observed were formed at very early times. In the halo, only very few r-process stars have metallicities of [Fe/H] $<-$3.0 and none have been found below [Fe/H] $\sim -$3.5. The existence of Reticulum II at least qualitatively supports this finding, as it suggests some (albeit short) time delay in the production of r-process elements relative to that of iron which is ejected in every core-collapse supernova explosions.

Other dwarf spheroidal galaxies contain r-I stars \citep{lemasle14, hill19}, although their origins are less clear. The globular clusters M92 \citep{yong14, kirby23} hosts a population of r-process stars, enriched by events occurring prior to their formation. Finding r-process stars in all these different locations and populations suggests that early r-process events occurred regularly across the early universe, although perhaps not too frequently. In general, for the r-process to operate, astrophysical environments with a high neutron flux are required, such as during core-collapse supernovae or neutron star mergers. Following the discovery of Reticulum\,II and the gravitational wave detection event GW1710817 of a merging pair of neutron stars \citep{LIGOGW170817a, LIGOGW170817b}, neutron star mergers are considered a major r-process site (e.g. \citealt{drout17, coulter17, kilpatrick17, shappee17}). Yet the body of [Eu/Fe] abundances in metal-poor stars suggests that multiple r-process sites are required to explain the data \citep{cote19}, such as contributions from core-collapse supernovae, neutron star mergers, and potentially also other, more exotic supernova events such as magneto-rotationally driven supernovae \citep{moesta18,yong21}. 

Enrichment in r-process elements from neutron star mergers are principally assumed to operate on a relatively longer timescales of up to a billion years, although mergers as fast as several tens of Myr are predicted also \citep{maoz24}. Supernovae, on the other hand, provide fast enrichment, given the short lifetimes of massive stars of only millions to tens of millions of years. Understanding the environmental conditions as well as the associated enrichment timescales is vital for interpreting observed heavy element abundances of r-process stars given that these events must have, by design, occurred prior to the stars’ formation. As such, r-process stars serve as tracers of the astrophysical sites responsible for early r-process element production. 

Utilizing astrometric data, e.g. from Gaia, the spatial distribution and orbital histories of r-process stars in the halo are currently being mapped. As a result, some r-process stars are found to be part of kinematically linked stellar groups \citep{roederer18_kin,gudin21}, suggesting a common origin, perhaps within a former dwarf galaxy that was accreted long ago. Using the existence of an r-process signature as a new chemical tagging approach \citep{brauer19} offers a promising way of tracing r-process stars back to their birth sites. This provides further constraints on the nature and environment needed to support r-process nucleosynthesis in the early universe, and further reveals how the Galactic halo may have been assembled from accreted dwarf galaxies. 

The actinide elements thorium (Th) and uranium (U) can be detected in many of the r-process stars. These elements are radioactive with half-lives of 14\,Gyr (Th) and 4.6\,Gyr (U). Measuring their current abundances makes it principally possible to age date the event that produced these nuclei, provided there is information on how much was produced. This would require knowledge of the astrophysical site and environmental details that supported the r-process, but which remain highly uncertain. In addition, those metal-poor stars with available thorium abundances are showing a large relative scatter relative to stable r-process element abundances of 0.9\,dex scatter in $\log\epsilon(\mathrm{Th/Eu})$ \citep{mardini_ustar}. This behavior has been termed “actinide boost" and might be a result of environmental or site differences. Hence, precision age dating of r-process stars remains challenging, if not impossible, for the time being. Alternatively, assuming an old age for these stars allows for an empirical prediction of different production ratios. Either way, as more details become more constrained, cosmo-chronometry of ancient r-process stars has the potential to confirm the old ages of these metal-poor stars, to date potentially long-accreted stars and even globular clusters, and to more fully map the history of the Milky Way and its early chemical evolution and assembly process.  

Finally, during the direct decay of trans-uranian nuclei produced during an r-process event,  
fission fragments are produced, i.e. elements around the second r-process peak. Small overabundances of elements ruthenium (Ru), rhodium (Rh), palladium (Pb) and silver (Ag) \citep{roederer23_fission} have been detected as an excess to the individual r-process patterns in a sample of stars, confirming fission as part of r-process nucleosynthesis. Paired with the suggestion that the light neutron-capture elements found in all these r-process stars might arise from a limited r-process (see below), it remains to be seen whether the r-process, as it is observed in stars, is truly has a universal elemental abundance pattern after all.

\subsection{Limited r-process metal-poor stars}

A group of metal-poor stars also displays overabundances in heavy elements but primarily among light neutron-capture elements, such as strontium (Sr), yttrium (Y) and zirconium (Zr). Heavier elements progressively become less abundant. This behavior has been described as the weak or limited r-process \citep{wanajo01, Travaglio04, frebel18} and thought of as an r-process with a lower neutron flux that does not produce the full elemental distribution of the main r-process. Core-collapse supernovae are a likely production site as they have been shown to not produce the main r-process abundance pattern \citep{Arcones11}. 

\begin{figure}[t]
\centering
\includegraphics[width=.96\textwidth]{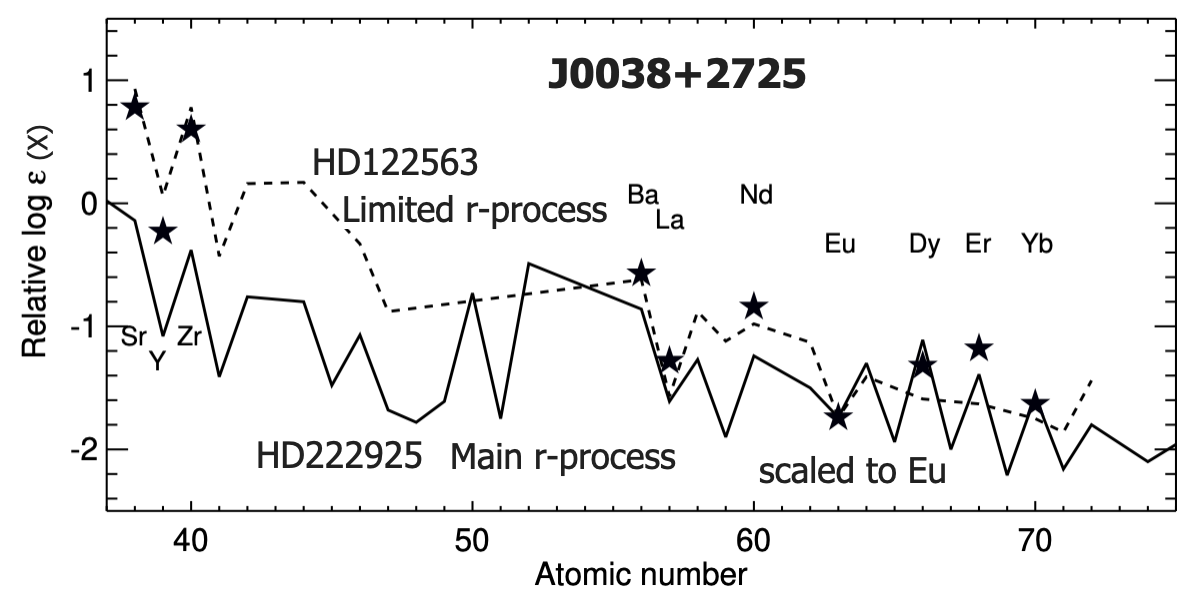}
\caption{Elemental abundances of the limited r-process star J0038+2725 (black star symbols; \citealt{xylakis24}) in comparison with the abundance patterns of the classic limited r-process star HD122562 (dashed line; \citealt{honda06}) and the main r-process star HD222925 (solid line; \citealt{roederer22}). Based on \citealt{xylakis24}.}
\label{limited}
\end{figure}

The well-studied bright metal-poor giant HD122563 is the best observed example of this nucleosynthesis signature \citep{honda06}. Figure~\ref{limited} shows its abundance pattern, along with an r-process pattern (scaled to Eu), and an r-I star \citep{xylakis24} for comparison. Based on HD122563's abundance pattern, limited r-process stars are selected to have [Eu/Fe] $<$ 0.3, [Sr/Ba] $>$ 0.5, and [Sr/Eu] $>$ 0.0. Hundreds of them have been identified but only relatively few have been studied in detail \citep{xylakis24}. 

Interestingly, the role of limited r-process stars may have yet to become fully utilized. There is mounting evidence that in the abundance patterns observed in the main r-process stars may be a combination of a baseline limited r-process pattern \citep{roederer23_fission}, likely produced by supernovae, plus a heavier-element-only main pattern. Studying large samples of limited r-process stars will be needed to better establish the observed signature and how variable it might be amongst stars.

\subsection{$s$-process enhanced metal-poor stars}\label{sproc}

Another subgroup of metal-poor stars displaying large enhancements in heavy elements made in the slow neutron-capture (s-) process which produces heavy nuclei when seed nuclei capture a neutron only after the previously unstable nucleus has $\beta$-decayed. High-resolution spectra of these stars display strong lines of s-process elements, such as strontium (Sr), yttrium (Y), zirconium (Zr), barium (Ba), and lanthanum (La). The heavy element abundances form an abundance pattern characterized by high Ba and low Eu and heavier element abundances, also with three peaks although they are shifted with respect to the r-process peaks \citep{sneden08, frebel18}. 

Accordingly, s-process metal-poor stars are classified by the combination of [Ba/Fe] $>$ 1.0 and [Ba/Eu] $>$ 0.5, in combination with a significant lead (Pb) abundance at the level of [Pb/Ba] $>$1.5 \citep{araa}. Compared to the r-process, the s-process abundance pattern is not universal because the availability of seed nuclei, e.g. Fe, required for the s-process to occur depends on metallicity. At low [Fe/H], large numbers of neutrons are available per seed nucleus to allow the s-process to run to completion and producing large amounts of Pb. This is frequently observed in s-process stars whereas the more metal-rich counterparts do not show such enormous Pb overabundances \citep{2001vaneck, aoki02}. This renders the s-process pattern metallicity dependent, especially among the heaviest elements \citep{gallino1998}. It should be noted, thought, that the s-process stars are not among the most metal-poor ones; they start to appear around [Fe/H] $\sim-$2.8, with the majority having $-$2.5 $<$ [Fe/H] $<-$2.0.

The s-process stars are members of binary systems and as such have a very specific origin. When repeat radial velocity measurements of a star show variations over time, their binary nature is revealed. The other star in the system is thought to be a slightly more massive companion that already went through its asymptotic giant branch (AGB) phase and has since evolved into a white dwarf. Given that the white dwarf is very faint in comparison, the absorption lines in the spectrum can be attributed to just the observed star. 

In general, during their AGB phase, low and intermediate mass stars undergo various nucleosynthesis processes that produce neutrons in the core region above the helium burning shell. These neutrons serve as a sustained neutron source for the s-process to operate on timescales of 10,000 years. Convective processes then continuously dredge up the freshly synthesized s-process elements to the surface, along with copious amounts of carbon. For example, the accompanying carbon overabundances typically exceed [C/Fe] = 2 but they are not a signature of any previous Population III star, as it is the case of the CEMP stars.

Mass transfer across binary systems is common and can occur through a strong stellar wind across wider binary systems or through direct Roche lobe overflow across close pairs. The observed large amounts of s-process elements in the metal-poor star, as well as the carbon, are thus the result of such a mass transfer event after the  companion star produced these elements. Hence, the observed heavy elements and carbon are not reflective of the star's birth gas cloud, unlike all other lighter elements. 

In addition to the fact that the s-process signature can be cleanly observed as produced by AGB stars in these s-process metal-poor binary stars, the s-process elements in AGB stars (in binary systems and not) are constantly dispersed into the surrounding gas via their stellar winds. This has resulted in an ever increasing (low-level) background enrichment of the Galaxy in s-process elements. Studies of ordinary metal-poor stars at different metallicities have been used to map out the slow rise of s-process material in the Galaxy, from the low-metallicity environment starting at about [Fe/H] $\sim-$2.8 to solar metallicity \citep{simmerer04}. Given that the s-process occurs only during the end phase of stellar evolution of low or intermediate mass stars,  s-process elements are formed with a delay and not found among the most metal-poor stars, and can only be measured in ordinary metal-poor stars from that somewhat higher [Fe/H] abundance. In those cases, the s-process element do reflect the conditions of the birth gas cloud; however, it is essentially impossible to trace them back to just one nucleosynthesis event (one AGB star).

\subsection{Higher metallicity CH stars and Ba stars} \label{chba} 

The more metal-rich counterparts of CEMP stars are the CH stars that have been known as a distinct group of stars for many decades. Characterized by a significant overabundance of carbon of [C/Fe] $>$ 1, they show very strong molecular absorption line bands of CH, e.g. at $\sim4300$\,${\AA}$, in their spectra. Originally, it was these strong observed features that led to their identification in low-resolution spectra \citep{keenan42}. With their sub-solar metallicities ([Fe/H] $\lesssim-$1) these stars are on the metal-rich end of Population II. They often have high velocities, and as such, are typically found in the halo. Possibly all of them are in binary systems since they also show enrichment in heavy elements, just like the CEMP-s stars. Their large carbon abundances are a strong indicator of a previous mass transfer event and that they were not born that way.

Barium (Ba) stars are similar to CH stars and another group of chemically distinct (peculiar) stars. Part of binary systems also, they are largely found in the Galactic disk, as opposed to the CH stars in the halo. They display large overabundances of barium, [Ba/Fe] $>$ 1, and other s-process elements that arise from mass transfer across their binary system. As a result, absorption lines of Ba, in particular the strongest Ba II line at 4554\,{\AA} can be easily identified even in low-resolution spectra which is not possible in the case of metal-poor stars. 

Ba stars tend to be red giants or subgiants. At the associated cooler surface temperatures, any absorption feature is more pronounced (as abundances are temperature dependent) which makes it possible for the Ba lines to become so strong in the spectra. The pronounced Ba features are the name sake of these stars and make Ba stars easily distinguishable from other types of stars. From high-resolution spectra, s-process element abundances can be measured to obtain their overall heavy element abundance pattern.

\subsection{Chemically peculiar stars}

Chemically peculiar stars mark another related group of stars that is interesting to mention in this context as they show heavy elements in their outer atmosphere but their origins are unlike those of typical ancient metal-poor stars. Some of these stars also show heavy neutron-capture elements in their spectra but not because the star was born from gas enriched by a specific neutron-capture nucleosynthesis event. Instead, their outermost layers have been internally enriched in a variety of heavy elements after the star formed.

Hot main-sequence stars do not have an extended convection zone like red giant stars that mixes material throughout the star. However, some of stars have unusually large magnetic fields (hundreds of gauss to several kilogauss) that appear to stabilize the outer atmospheric layers against mixing. This allows for processes such as gravitational settling to occur which enables some elements, particularly helium (He), nitrogen (N) and oxygen (O), to sink down to lower layers in the star. In turn, other elements, such as manganese (Mn), strontium (Sr), yttrium (Y) and zirconium (Zr), neodymium (Nd), praeseodymium (Pr) float up through radiative levitation and accumulate in the surface layer. Some even show lines of uranium (U) in their spectra \citep{shulyak10}. Since only the surface is ever observed, these processes result in an unusual --chemically peculiar-- elemental abundance signature that remains to be fully understood. This sets them apart from other, ordinary main sequence stars.

There are four main classes of chemically peculiar stars with different spectral characteristics. Ap and Am stars show strong and weaker magnetic fields, respectively, are slow rotators, display large overabundances of various heavy elements, and have high surface temperatures from 7,000 to 15,000\,K. The magnetic fields are thought to be left over from the stars' formation processes. A well-studied Ap star, HD 101065 (“Przybylski's Star") is listed in Table~\ref{top50}. The other two classes include the HgMn star that are similar to Ap stars but do not have the strong magnetic field and they do show large mercury (Hg) surface abundances. He-weak stars are even hotter but show less pronounced He lines than normal stars. 

Chemically peculiar stars are common, about 15\% of hot main sequence stars show these chemical peculiarities \citep{romanyuk07}. Stars cooler than these are also regarded chemically peculiar. Some are the CH and Ba stars described in Section~\ref{chba}. Their surface abundances are the result of mixing of nuclear burning products from the interior to the surface, such is the case for metal-poor CH stars, and Ba stars when the star is in a binary system.

\section{Metal-poor stars in globular clusters} \label{gc}

Globular clusters are very dense stellar systems with up to a million member stars, all concentrated within a small volume of tens to a hundred parsec in diameter. They are also some of the oldest known objects in the universe, with ages exceeding 10 billion years. Contrary to individual stars, the age of an entire stellar population can be derived from its color-magnitude-diagram by matching the turnoff region to an isochrone of known age and metallicity \citep{bastian18}. This allows for clusters stars to have both age and metallicity measurements which is unattainable for field stars, or for any single stars. 

There are at least 150 globular clusters in Galaxy, including some that are thought to have once been associated with the now disrupting Sagittarius dwarf galaxy and the progenitors of other stellar streams. Yet, it remains unclear how globular cluster formation fits into the picture of hierarchical structure formation and the assembly of the Milky Way. Most other large galaxies host significant numbers of globular clusters, including Andromeda which boasts nearly 500 of them. Some of the most interesting globular clusters are listed in Table~\ref{top50}.

For some elements, globular cluster members display chemical abundance ratios different from those observed in halo stars. The most striking feature are the Na-O and Mg-Al anti-correlations that occur as a result of internal enrichment processes \citep{gratton01} and vary for the different stellar populations present within a cluster. This behavior has also been linked to the corresponding photometric sequences \citep{piotto07,milone12}. However, there is no significant spread among member star metallicities of a given cluster. Neutron-capture elements are also frequently observed in globular cluster stars, showing a rich and complex internal and external enrichment history that remains to be fully understood. Signatures from the s- and r-process and combinations thereof have been observed (e.g., \citealt{yong14}). 

Globular clusters have metallicities in the range of $-$2.5 $\lesssim$ [Fe/H] $\lesssim-$1.0. There is generally a very small spread ($\sim$0.05\,dex) in [Fe/H] among members as expected for a single age population, although larger spreads are increasingly found for some, suggesting that clusters contain multiple stellar populations of slightly different ages, and hence, metallicities. This has been confirmed with their color-magnitude diagrams where the sequences spread out and even separate in some cases (e.g. Omega Centauri; \citealt{milone17}). While low, the cluster [Fe/H] values collectively do not reach the low levels found in numerous halo and dwarf galaxy extremely metal-poor stars. This “metallicity floor" has robustly appeared among all clusters, implying that these objects did form from gas that was already significantly enriched at the earliest times. 

However, some recent discoveries have challenged this picture, or at least offer additional insights into how star clusters may have formed in the early universe. The stellar stream C-19 is the tidal remnant of an erstwhile extremely metal-poor globular cluster that was found to have an average [Fe/H] =$-$3.4 \citep{martin22}. This implies that star clusters can form from extremely metal-poor gas but perhaps most, if not all of them, got rapidly accreted by the proto-Milky Way. Separately, a single metal-poor star with [Fe/H] =$-$3 was found that shows all the chemical abundance characteristics of a globular cluster \citep{bandyopadhyay24}, adding to the increasing inventory of halo stars that appear to have originated in early extremely metal-poor globular clusters (e.g. \citealt{roederer19_sylger}). Finally, a star with [Fe/H] $\sim-$2.9 was discovered in one of Andromeda's globular clusters pointing to a more universal scenario of low-metallicity cluster formation \citep{larsen20}.


\section{Metal-poor stars in dwarf galaxies} \label{dwgal}

The Milky Way is orbited by dozens of ancient dwarf satellite galaxies that span a range of properties \citep{tolstoy04, simon19, frebel_ji23}. The ultra-faint dwarf galaxies are the smallest and least luminous galaxies in the universe. 
They are gas-poor and consist of only a few thousand stars (stellar masses of 10$^{3-5}$\,M$_{\odot}$, \citealt{kirby13}). Their stars are all metal-poor, with [Fe/H] $\sim-$4 to $-1$, as a result of only one star formation episode occurring before losing all the gas due to blowout and/or reionization. They are about 12\,billion years old \citep{brown12} and their stars preserve the signatures of early chemical enrichment that occurred in their local region. Furthermore, these systems are highly dark matter dominated objects (with total dynamical masses of 10$^{6-7}$\,M$_{\odot}$, \citealt{strigari08}). Some of the most important dwarf galaxies are listed in Table~\ref{top50}. 

The classical dwarf spheroidal galaxies are more massive (stellar masses of 10$^{5-7}$\,M$_{\odot}$; \citealt{kirby13}) and luminous than the ultra-faint dwarfs, and also contain a significant dark matter halo 10-100 times more massive than their stellar mass (total dynamical masses of 10$^{7-9}$\,M$_{\odot}$). These systems have experienced multiple star burst phases, contain stars with [Fe/H] $\sim-$4 to 0, similar to the range found in the halo, and are gas-poor as a result of gas blowout by supernovae. Since they were large enough to sustain their mass during reionization, and are similarly ancient, and their stars reflect an extended phase of chemical evolution. 

Ultra-faint dwarf galaxies are some of the most primitive systems and can be regarded as the surviving analogs of the earliest galaxies, such as Segue 1, which may well be a surviving first galaxy \citep{frebel14}, and Tucana II which appears to be the merger product of two first galaxies \citep{chiti21_tuc, tarumi21}. All these ultra-faint dwarfs contain exclusively metal-poor stars which makes them interesting targets for stellar archaeology. More so, that concept can be extended to “dwarf galaxy archaeology". Any of these entire dwarf galaxy becomes a tool for reconstructing the conditions of star formation and nucleosynthesis in the early universe, given the constraints provided by the fact that the galaxy is still intact at present. This has led to the conclusion that one of the faintest galaxies, the r-process enriched ultra-faint dwarf Reticulum II, was likely enriched by a neutron star merger. Estimates on the stellar mass of the system enabled an additional constraint that cannot be invoked in the case of a single r-process star located in e.g. the halo.

Dwarf galaxy member stars can be confirmed through various techniques including narrow-band photometry and radial velocity measurements. Given their distances of dozens to hundreds of kpc into the outer halo of the Galaxy, dwarf galaxy stars are usually very faint and challenging to observe with high spectral resolution even with the largest telescopes. Hence, only relatively few intrinsically luminous stars (all located on the giant branch of their system) have been studied for their detailed chemical abundances. Overall, dwarf galaxy stars exhibit abundance ratios aligned with what is found for metal-poor halo stars. This is observed across many different systems, suggesting that supernova enrichment of the lighter elements (Z$<$30) operates in a universal way, at least in the early phases of any given galaxy's history \citep{araa, psss, fn15}. Differences only exist among neutron-capture elements whose abundances are typically lower in the dwarf galaxy stars compared to average metal-poor halo stars \citep{frebel_ji23}. This likely reflects the simple enrichment history of dwarf galaxies but it also simultaneously showcases the utility of these systems for constraining nucleosynthesis and its sites, such as that related to the r-process. 

Stars in the classical dwarfs have metallicities up to the solar value which establishes their extended chemical evolution. Again, the abundance ratios are comparable to those of halo stars with one exception. The [$\alpha$/Fe] ratio as a function of [Fe/H] turns down at a lower metallicity ([Fe/H] $\sim-2.5$ to $-2.0$) than what is observed for the halo (at [Fe/H] $\sim-1.5$ to $-1.0$) because in these smaller systems, chemical evolution proceeded slower. As a result, the dwarf galaxy could only enrich to the lower [Fe/H] in the time it took for the first supernovae type Ia to explode and produce iron. 

Recently, metal-poor stars have also been discovered in the Large Magellanic Cloud, the largest satellite of the Milky Way, with stars reaching down to [Fe/H] $\sim-$4 \citep{reggiani21, chiti24}. Interestingly, this gas poor, more massive dwarf galaxy appears to be on its first infall \citep{besla07} into the Milky Way, along with its companion, the Small Magellanic Cloud. This implies that especially its ancient metal-poor population likely formed in a cosmic region that was separate from what led to the early formation of the Milky Way. Stellar archaeology can thus be extended to “extragalactic stellar archaeology" to test the universality of (early) chemical enrichment and evolution across a much larger volume of space. 

Another noteworthy case of a metal-poor dwarf galaxy is the Atari disk, which is likely an early disrupted dwarf galaxy that is now located within and spread throughout the thick disk of the Milky Way \citep{mardini22_atari}. Its kinematic signature is similar to that of the thick disk; hence, it has long been thought to me the metal-weak tail of the thick disk \citep{norris85, carollo10}. Indeed, Atari boast an impressive number of metal-poor stars down to [Fe/H] = $-$5 (e.g. SDSSJ102915+172927, \citealt{caffau11}) which strongly supports its ancient nature and accretion origin. A bigger, somewhat more metal-rich analog that built the halo is the Gaia-Sausage-Enceladus systems \citep{Belokurov18}.
These kinds of long dissolved systems thus provide important information to further understand the formation of the early Galactic disk and halo, and the onset of the hierarchical assembly of large galaxies.



\section{Summary and outlook}

Ancient, long-lived metal-poor stars are present in all components of our home galaxy, the Milky Way, including the halo, disk, bulge, and different kinds of dwarf galaxies, globular clusters and stellar streams. Born a few hundred million years after the Big Bang and during a time that marked the very beginning of the chemical evolution, these stars display very low abundances of elements heavier and hydrogen and helium, making them “metal-poor”.  Studying the chemical composition of these stars with high-resolution spectroscopy reveals direct information about the conditions of the early universe. Each of these stars has long preserved the local chemical signature of their individual birth gas clouds in their stellar atmosphere which is now being observed. This approach forms the basis of the field of stellar archaeology. 

There are many different subgroups of metal-poor stars in the halo, dwarf galaxies, and globular clusters, including ordinary metal-poor stars and the most iron-poor stars, carbon-enhanced stars, r-, limited r- and s-process stars, CH and Ba stars and peculiar stars. Table~\ref{top50} lists numerous of astrophysically interesting metal-poor stars in different locations as well as example stars for all these subgroups. Each of the subgroups provides information on a different element production history an astrophysical sites that occurred prior to the stars' birth. Mapping the orbital histories of these stars further helps to discern their origins, e.g. in small early galaxies. Individual as well as larger samples of these metal-poor stars thus enable the reconstruction of nucleosynthesis processes, the chemical enrichment history of the Galaxy, early star formation processes, and various aspects of the assembly and evolution of the Milky. 

While much progress has been made in stellar archaeology over the past 25 years \citep{araa, fn15, frebel_ji23}, much more is yet to come. Following the observations of one billion stars with Gaia to map the Milky Way, massive multiplexing wide-field spectroscopic surveys are underway or will start soon to provide spectra for chemical abundance studies, such as SDSS-V \citep{Kollmeier17}, WEAVE \citep{Dalton14} and 4MOST \citep{deJong19}. Along with the next-generation of large 30-meter-class telescopes, these new datasets promise a new level of understanding of our Milky Way, its history, and its place in the universe.

\begin{ack}[Acknowledgments] 
~A.F. warmly thanks her colleagues for many years of fruitful and inspiring collaborations, including 
T.C. Beers, 
K. Brauer,
A. Chiti,
R. Ezzeddine,
T. Hansen,
E. Holmbeck,
A.P. Ji, 
M. Mardini, 
X. Ou,
V. Placco,
I.U. Roederer,
and
C. Sakari.
She thanks A. Chiti and M. Mardini for providing figures for this entry, and acknowledges support from NSF-AAG grants AST-1716251 and AST-2307436. 
\end{ack}



\seealso{
A list of websites related to the results described in this article: \\
Literature compilation of metal-poor stars: JINAbase: https://jinabase.pythonanywhere.com \\
Literature compilation of metal-poor stars: SAGA database: https://sagadatabase.jp \\ 
Atomic line data: VALD http://vald.astro.univie.ac.at  \\
Atomic line data: NIST http://www.physics.nist.gov/PhysRefData/ASD/lines\_form.html \\
Linelist generator: linemake https://github.com/vmplacco/linemake  \\
Model atmosphere: Kurucs models: http://kurucz.harvard.edu \\
Model atmosphere: MARCS models: https://marcs.astro.uu.se, \\
LTE line analysis: MOOG http://www.as.utexas.edu/$\Tilde{~}$chris/moog.html\\ 
LTE line analysis: TURBOspectrum https://github.com/bertrandplez/Turbospectrum2019\\
Chemical abundance analysis: Spectroscopy Made Hard  https://github.com/andycasey/smhr  \\
Stellar evolutionary effects: Carbon correction https://vplacco.pythonanywhere.com \\
NLTE corrections: Inspect-stars https://www.inspect-stars.com \\
Orbital history analysis: ORIENT https://github.com/Mohammad-Mardini/The-ORIENT \\
Orbital history analysis: galpy https://github.com/jobovy/galpy \\
Supernova fitting: Starfit https://starfit.org \\
}

\begin{thebibliography*}{161}
\providecommand{\bibtype}[1]{}
\providecommand{\natexlab}[1]{#1}
{\catcode`\|=0\catcode`\#=12\catcode`\@=11\catcode`\\=12
|immediate|write|@auxout{\expandafter\ifx\csname
  natexlab\endcsname\relax\gdef\natexlab#1{#1}\fi}}
\renewcommand{\url}[1]{{\tt #1}}
\providecommand{\urlprefix}{URL }
\expandafter\ifx\csname urlstyle\endcsname\relax
  \providecommand{\doi}[1]{doi:\discretionary{}{}{}#1}\else
  \providecommand{\doi}{doi:\discretionary{}{}{}\begingroup
  \urlstyle{rm}\Url}\fi
\providecommand{\bibinfo}[2]{#2}
\providecommand{\eprint}[2][]{\url{#2}}

\bibtype{Article}%
\bibitem[{Abbott} et al.(2017{\natexlab{a}})]{LIGOGW170817a}
\bibinfo{author}{{Abbott} BP}, \bibinfo{author}{{Abbott} R},
  \bibinfo{author}{{Abbott} TD}, \bibinfo{author}{{Acernese} F},
  \bibinfo{author}{{Ackley} K}, \bibinfo{author}{{Adams} C},
  \bibinfo{author}{{Adams} T}, \bibinfo{author}{{Addesso} P},
  \bibinfo{author}{{Adhikari} RX}, \bibinfo{author}{{Adya} VB} and
  \bibinfo{author}{et~al.} (\bibinfo{year}{2017}{\natexlab{a}}),
  \bibinfo{month}{Oct.}
\bibinfo{title}{{GW170817: Observation of Gravitational Waves from a Binary
  Neutron Star Inspiral}}.
\bibinfo{journal}{{\em Physical Review Letters}} \bibinfo{volume}{119}
  (\bibinfo{number}{16}), \bibinfo{eid}{161101}.
  \bibinfo{doi}{\doi{10.1103/PhysRevLett.119.161101}}.
\eprint{1710.05832}.

\bibtype{Article}%
\bibitem[{Abbott} et al.(2017{\natexlab{b}})]{LIGOGW170817b}
\bibinfo{author}{{Abbott} BP}, \bibinfo{author}{{Abbott} R},
  \bibinfo{author}{{Abbott} TD}, \bibinfo{author}{{Acernese} F},
  \bibinfo{author}{{Ackley} K}, \bibinfo{author}{{Adams} C},
  \bibinfo{author}{{Adams} T}, \bibinfo{author}{{Addesso} P},
  \bibinfo{author}{{Adhikari} RX}, \bibinfo{author}{{Adya} VB} and
  \bibinfo{author}{et~al.} (\bibinfo{year}{2017}{\natexlab{b}}),
  \bibinfo{month}{Oct.}
\bibinfo{title}{{Multi-messenger Observations of a Binary Neutron Star
  Merger}}.
\bibinfo{journal}{{\em ApJl}} \bibinfo{volume}{848}, \bibinfo{eid}{L12}.
  \bibinfo{doi}{\doi{10.3847/2041-8213/aa91c9}}.
\eprint{1710.05833}.

\bibtype{Article}%
\bibitem[{Abohalima} and {Frebel}(2017)]{jinabase}
\bibinfo{author}{{Abohalima} A} and  \bibinfo{author}{{Frebel} A}
  (\bibinfo{year}{2017}), \bibinfo{month}{Nov.}
\bibinfo{title}{{JINAbase: A database for chemical abundances of metal-poor
  stars}}.
\bibinfo{journal}{{\em ArXiv e-prints}} \eprint{1711.04410}.

\bibtype{Article}%
\bibitem[{Aoki} et al.(2000)]{aoki_lead2000}
\bibinfo{author}{{Aoki} W}, \bibinfo{author}{{Norris} JE},
  \bibinfo{author}{{Ryan} SG}, \bibinfo{author}{{Beers} TC} and
  \bibinfo{author}{{Ando} H} (\bibinfo{year}{2000}), \bibinfo{month}{Jun.}
\bibinfo{title}{{Detection of Lead in the Carbon-rich, Very Metal-poor Star LP
  625-44: A Strong Constrai nt on S-Process Nucleosynthesis at Low
  Metallicity}}.
\bibinfo{journal}{{\em ApJ}} \bibinfo{volume}{536}: \bibinfo{pages}{L97--L100}.
  \bibinfo{doi}{\doi{10.1086/312740}}.

\bibtype{Article}%
\bibitem[{Aoki} et al.(2002{\natexlab{a}})]{aoki_mg}
\bibinfo{author}{{Aoki} W}, \bibinfo{author}{{Norris} JE},
  \bibinfo{author}{{Ryan} SG}, \bibinfo{author}{{Beers} TC} and
  \bibinfo{author}{{Ando} H} (\bibinfo{year}{2002}{\natexlab{a}}),
  \bibinfo{month}{Sep.}
\bibinfo{title}{{Chemical Composition of the Carbon-rich, Extremely Metal Poor
  Star CS 29498-043: A New Class of Extremely Metal Poor Stars with Excesses of
  Magnesium and Silicon}}.
\bibinfo{journal}{{\em ApJ}} \bibinfo{volume}{576}:
  \bibinfo{pages}{L141--L144}. \bibinfo{doi}{\doi{10.1086/343761}}.

\bibtype{Article}%
\bibitem[{Aoki} et al.(2002{\natexlab{b}})]{aoki02}
\bibinfo{author}{{Aoki} W}, \bibinfo{author}{{Norris} JE},
  \bibinfo{author}{{Ryan} SG}, \bibinfo{author}{{Beers} TC} and
  \bibinfo{author}{{Ando} H} (\bibinfo{year}{2002}{\natexlab{b}}),
  \bibinfo{month}{Sep.}
\bibinfo{title}{{Chemical Composition of the Carbon-rich, Extremely Metal Poor
  Star CS 29498-043: A New Class of Extremely Metal Poor Stars with Excesses of
  Magnesium and Silicon}}.
\bibinfo{journal}{{\em ApJl}} \bibinfo{volume}{576}:
  \bibinfo{pages}{L141--L144}. \bibinfo{doi}{\doi{10.1086/343761}}.
\eprint{astro-ph/0208019}.

\bibtype{Article}%
\bibitem[{Aoki} et al.(2006)]{Aokihe1327}
\bibinfo{author}{{Aoki} W}, \bibinfo{author}{{Frebel} A},
  \bibinfo{author}{{Christlieb} N}, \bibinfo{author}{{Norris} JE},
  \bibinfo{author}{{Beers} TC}, \bibinfo{author}{{Minezaki} T},
  \bibinfo{author}{{Barklem} PS}, \bibinfo{author}{{Honda} S},
  \bibinfo{author}{{Takada-Hidai} M}, \bibinfo{author}{{Asplund} M},
  \bibinfo{author}{{Ryan} SG}, \bibinfo{author}{{Tsangarides} S},
  \bibinfo{author}{{Eriksson} K}, \bibinfo{author}{{Steinhauer} A},
  \bibinfo{author}{{Deliyannis} CP}, \bibinfo{author}{{Nomoto} K},
  \bibinfo{author}{{Fujimoto} MY}, \bibinfo{author}{{Ando} H},
  \bibinfo{author}{{Yoshii} Y} and  \bibinfo{author}{{Kajino} T}
  (\bibinfo{year}{2006}), \bibinfo{month}{Mar.}
\bibinfo{title}{{HE 1327-2326, an Unevolved Star with [Fe/H] < -5.0. I. A
  Comprehensive Abundance Anal ysis}}.
\bibinfo{journal}{{\em ApJ}} \bibinfo{volume}{639}: \bibinfo{pages}{897--917}.
  \bibinfo{doi}{\doi{10.1086/497906}}.

\bibtype{Article}%
\bibitem[{Aoki} et al.(2007{\natexlab{a}})]{aoki_cemp_2007}
\bibinfo{author}{{Aoki} W}, \bibinfo{author}{{Beers} TC},
  \bibinfo{author}{{Christlieb} N}, \bibinfo{author}{{Norris} JE},
  \bibinfo{author}{{Ryan} SG} and  \bibinfo{author}{{Tsangarides} S}
  (\bibinfo{year}{2007}{\natexlab{a}}), \bibinfo{month}{Jan.}
\bibinfo{title}{{Carbon-enhanced Metal-poor Stars. I. Chemical Compositions of
  26 Stars}}.
\bibinfo{journal}{{\em ApJ}} \bibinfo{volume}{655}: \bibinfo{pages}{492--521}.
  \bibinfo{doi}{\doi{10.1086/509817}}.
\eprint{arXiv:astro-ph/0609702}.

\bibtype{Article}%
\bibitem[{Aoki} et al.(2007{\natexlab{b}})]{aoki07}
\bibinfo{author}{{Aoki} W}, \bibinfo{author}{{Beers} TC},
  \bibinfo{author}{{Christlieb} N}, \bibinfo{author}{{Norris} JE},
  \bibinfo{author}{{Ryan} SG} and  \bibinfo{author}{{Tsangarides} S}
  (\bibinfo{year}{2007}{\natexlab{b}}), \bibinfo{month}{Jan.}
\bibinfo{title}{{Carbon-enhanced Metal-poor Stars. I. Chemical Compositions of
  26 Stars}}.
\bibinfo{journal}{{\em ApJ}} \bibinfo{volume}{655}: \bibinfo{pages}{492--521}.
\eprint{arXiv:astro-ph/0609702}.

\bibtype{Article}%
\bibitem[{Arcones} and {Montes}(2011)]{Arcones11}
\bibinfo{author}{{Arcones} A} and  \bibinfo{author}{{Montes} F}
  (\bibinfo{year}{2011}), \bibinfo{month}{Apr.}
\bibinfo{title}{{Production of Light-element Primary Process Nuclei in
  Neutrino-driven Winds}}.
\bibinfo{journal}{{\em ApJ}} \bibinfo{volume}{731}, \bibinfo{eid}{5}.
  \bibinfo{doi}{\doi{10.1088/0004-637X/731/1/5}}.
\eprint{1007.1275}.

\bibtype{Article}%
\bibitem[{Baade}(1944)]{baade44}
\bibinfo{author}{{Baade} W} (\bibinfo{year}{1944}), \bibinfo{month}{Sep.}
\bibinfo{title}{{The Resolution of Messier 32, NGC 205, and the Central Region
  of the Andromeda Nebula.}}
\bibinfo{journal}{{\em ApJ}} \bibinfo{volume}{100}: \bibinfo{pages}{137--146}.
  \bibinfo{doi}{\doi{10.1086/144650}}.

\bibtype{Article}%
\bibitem[{Bandyopadhyay} et al.(2024)]{bandyopadhyay24}
\bibinfo{author}{{Bandyopadhyay} A}, \bibinfo{author}{{Ezzeddine} R},
  \bibinfo{author}{{Allende Prieto} C}, \bibinfo{author}{{Aria} N},
  \bibinfo{author}{{Shah} SP}, \bibinfo{author}{{Beers} TC},
  \bibinfo{author}{{Frebel} A}, \bibinfo{author}{{Hansen} TT},
  \bibinfo{author}{{Holmbeck} EM}, \bibinfo{author}{{Placco} VM},
  \bibinfo{author}{{Roederer} IU} and  \bibinfo{author}{{Sakari} CM}
  (\bibinfo{year}{2024}), \bibinfo{month}{Aug.}
\bibinfo{title}{{The $R$-Process Alliance: Fifth Data Release from the Search
  for $R$-Process-Enhanced Metal-poor Stars in the Galactic Halo with the
  GTC}}.
\bibinfo{journal}{{\em arXiv e-prints}} ,
  \bibinfo{eid}{arXiv:2408.03731}\bibinfo{doi}{\doi{10.48550/arXiv.2408.03731}}.
\eprint{2408.03731}.

\bibtype{Article}%
\bibitem[{Bastian} and {Lardo}(2018)]{bastian18}
\bibinfo{author}{{Bastian} N} and  \bibinfo{author}{{Lardo} C}
  (\bibinfo{year}{2018}), \bibinfo{month}{Sep.}
\bibinfo{title}{{Multiple Stellar Populations in Globular Clusters}}.
\bibinfo{journal}{{\em AR\&A}} \bibinfo{volume}{56}: \bibinfo{pages}{83--136}.
  \bibinfo{doi}{\doi{10.1146/annurev-astro-081817-051839}}.
\eprint{1712.01286}.

\bibtype{Article}%
\bibitem[{Beers} and {Christlieb}(2005)]{araa}
\bibinfo{author}{{Beers} TC} and  \bibinfo{author}{{Christlieb} N}
  (\bibinfo{year}{2005}).
\bibinfo{title}{The discovery and analysis of very metal-poor stars in the
  galaxy}.
\bibinfo{journal}{{\em ARA\&A}} \bibinfo{volume}{43}:
  \bibinfo{pages}{531--580}.

\bibtype{Article}%
\bibitem[{Beers} et al.(1999)]{BeersCaKII}
\bibinfo{author}{{Beers} TC}, \bibinfo{author}{{Rossi} S},
  \bibinfo{author}{{Norris} JE}, \bibinfo{author}{{Ryan} SG} and
  \bibinfo{author}{{Shefler} T} (\bibinfo{year}{1999}), \bibinfo{month}{Feb.}
\bibinfo{title}{{Estimation of Stellar Metal Abundance. II. A Recalibration of
  the Ca II K Technique, an d the Autocorrelation Function Method}}.
\bibinfo{journal}{{\em AJ}} \bibinfo{volume}{117}: \bibinfo{pages}{981--1009}.
  \bibinfo{doi}{\doi{10.1086/300727}}.

\bibtype{Article}%
\bibitem[{Behara} et al.(2010)]{behara10}
\bibinfo{author}{{Behara} NT}, \bibinfo{author}{{Bonifacio} P},
  \bibinfo{author}{{Ludwig} HG}, \bibinfo{author}{{Sbordone} L},
  \bibinfo{author}{{Gonz{\'a}lez Hern{\'a}ndez} JI} and
  \bibinfo{author}{{Caffau} E} (\bibinfo{year}{2010}), \bibinfo{month}{Apr.}
\bibinfo{title}{{Three carbon-enhanced metal-poor dwarf stars from the SDSS.
  Chemical abundances from CO$^{5}$BOLD 3D hydrodynamical model atmospheres}}.
\bibinfo{journal}{{\em A\&A}} \bibinfo{volume}{513}, \bibinfo{eid}{A72}.
  \bibinfo{doi}{\doi{10.1051/0004-6361/200913213}}.
\eprint{1002.1670}.

\bibtype{Article}%
\bibitem[{Belokurov} et al.(2018)]{Belokurov18}
\bibinfo{author}{{Belokurov} V}, \bibinfo{author}{{Erkal} D},
  \bibinfo{author}{{Evans} NW}, \bibinfo{author}{{Koposov} SE} and
  \bibinfo{author}{{Deason} AJ} (\bibinfo{year}{2018}), \bibinfo{month}{Jul.}
\bibinfo{title}{{Co-formation of the disc and the stellar halo}}.
\bibinfo{journal}{{\em MNRAS}} \bibinfo{volume}{478}:
  \bibinfo{pages}{611--619}. \bibinfo{doi}{\doi{10.1093/mnras/sty982}}.
\eprint{1802.03414}.

\bibtype{Article}%
\bibitem[{Bensby} et al.(2005)]{bensby05}
\bibinfo{author}{{Bensby} T}, \bibinfo{author}{{Feltzing} S},
  \bibinfo{author}{{Lundstr{\"o}m} I} and  \bibinfo{author}{{Ilyin} I}
  (\bibinfo{year}{2005}), \bibinfo{month}{Apr.}
\bibinfo{title}{{{\ensuremath{\alpha}}-, r-, and s-process element trends in
  the Galactic thin and thick disks}}.
\bibinfo{journal}{{\em A\&A}} \bibinfo{volume}{433} (\bibinfo{number}{1}):
  \bibinfo{pages}{185--203}. \bibinfo{doi}{\doi{10.1051/0004-6361:20040332}}.
\eprint{astro-ph/0412132}.

\bibtype{Article}%
\bibitem[{Bergemann} et al.(2012)]{bergemann12}
\bibinfo{author}{{Bergemann} M}, \bibinfo{author}{{Lind} K},
  \bibinfo{author}{{Collet} R}, \bibinfo{author}{{Magic} Z} and
  \bibinfo{author}{{Asplund} M} (\bibinfo{year}{2012}), \bibinfo{month}{Nov.}
\bibinfo{title}{{Non-LTE line formation of Fe in late-type stars - I. Standard
  stars with 1D and 3D model atmospheres}}.
\bibinfo{journal}{{\em MNRAS}} \bibinfo{volume}{427}: \bibinfo{pages}{27--49}.
  \bibinfo{doi}{\doi{10.1111/j.1365-2966.2012.21687.x}}.
\eprint{1207.2455}.

\bibtype{Article}%
\bibitem[{Besla} et al.(2007)]{besla07}
\bibinfo{author}{{Besla} G}, \bibinfo{author}{{Kallivayalil} N},
  \bibinfo{author}{{Hernquist} L}, \bibinfo{author}{{Robertson} B},
  \bibinfo{author}{{Cox} TJ}, \bibinfo{author}{{van der Marel} RP} and
  \bibinfo{author}{{Alcock} C} (\bibinfo{year}{2007}), \bibinfo{month}{Oct.}
\bibinfo{title}{{Are the Magellanic Clouds on Their First Passage about the
  Milky Way?}}
\bibinfo{journal}{{\em ApJ}} \bibinfo{volume}{668}: \bibinfo{pages}{949--967}.
  \bibinfo{doi}{\doi{10.1086/521385}}.
\eprint{arXiv:astro-ph/0703196}.

\bibtype{Article}%
\bibitem[{Brauer} et al.(2019)]{brauer19}
\bibinfo{author}{{Brauer} K}, \bibinfo{author}{{Ji} AP},
  \bibinfo{author}{{Frebel} A}, \bibinfo{author}{{Dooley} GA},
  \bibinfo{author}{{G{\'o}mez} FA} and  \bibinfo{author}{{O'Shea} BW}
  (\bibinfo{year}{2019}), \bibinfo{month}{Feb.}
\bibinfo{title}{{The Origin of r-process Enhanced Metal-poor Halo Stars In
  Now-destroyed Ultra-faint Dwarf Galaxies}}.
\bibinfo{journal}{{\em ApJ}} \bibinfo{volume}{871} (\bibinfo{number}{2}),
  \bibinfo{eid}{247}. \bibinfo{doi}{\doi{10.3847/1538-4357/aafafb}}.

\bibtype{Article}%
\bibitem[{Bromm} and {Larson}(2004)]{brommARAA}
\bibinfo{author}{{Bromm} V} and  \bibinfo{author}{{Larson} RB}
  (\bibinfo{year}{2004}), \bibinfo{month}{Sep.}
\bibinfo{title}{{The First Stars}}.
\bibinfo{journal}{{\em ARA\&A}} \bibinfo{volume}{42}: \bibinfo{pages}{79--118}.

\bibtype{Article}%
\bibitem[{Bromm} and {Loeb}(2003)]{brommloeb03}
\bibinfo{author}{{Bromm} V} and  \bibinfo{author}{{Loeb} A}
  (\bibinfo{year}{2003}), \bibinfo{month}{Oct.}
\bibinfo{title}{{The formation of the first low-mass stars from gas with low
  carbon and oxygen abundance s}}.
\bibinfo{journal}{{\em Nature}} \bibinfo{volume}{425}:
  \bibinfo{pages}{812--814}.

\bibtype{Article}%
\bibitem[{Bromm} et al.(2002)]{bromm02}
\bibinfo{author}{{Bromm} V}, \bibinfo{author}{{Coppi} PS} and
  \bibinfo{author}{{Larson} RB} (\bibinfo{year}{2002}), \bibinfo{month}{Jan.}
\bibinfo{title}{{The Formation of the First Stars. I. The Primordial
  Star-forming Cloud}}.
\bibinfo{journal}{{\em ApJ}} \bibinfo{volume}{564}: \bibinfo{pages}{23--51}.
  \bibinfo{doi}{\doi{10.1086/323947}}.

\bibtype{Article}%
\bibitem[{Brown} et al.(2012)]{brown12}
\bibinfo{author}{{Brown} TM}, \bibinfo{author}{{Tumlinson} J},
  \bibinfo{author}{{Geha} M}, \bibinfo{author}{{Kirby} EN},
  \bibinfo{author}{{VandenBerg} DA}, \bibinfo{author}{{Mu{\~n}oz} RR},
  \bibinfo{author}{{Kalirai} JS}, \bibinfo{author}{{Simon} JD},
  \bibinfo{author}{{Avila} RJ}, \bibinfo{author}{{Guhathakurta} P},
  \bibinfo{author}{{Renzini} A} and  \bibinfo{author}{{Ferguson} HC}
  (\bibinfo{year}{2012}), \bibinfo{month}{Jul.}
\bibinfo{title}{{The Primeval Populations of the Ultra-faint Dwarf Galaxies}}.
\bibinfo{journal}{{\em ApJL}} \bibinfo{volume}{753}, \bibinfo{eid}{L21}.
  \bibinfo{doi}{\doi{10.1088/2041-8205/753/1/L21}}.
\eprint{1206.0941}.

\bibtype{Article}%
\bibitem[{Brown} et al.(2014)]{brown14}
\bibinfo{author}{{Brown} TM}, \bibinfo{author}{{Tumlinson} J},
  \bibinfo{author}{{Geha} M}, \bibinfo{author}{{Simon} JD},
  \bibinfo{author}{{Vargas} LC}, \bibinfo{author}{{VandenBerg} DA},
  \bibinfo{author}{{Kirby} EN}, \bibinfo{author}{{Kalirai} JS},
  \bibinfo{author}{{Avila} RJ}, \bibinfo{author}{{Gennaro} M},
  \bibinfo{author}{{Ferguson} HC}, \bibinfo{author}{{Mu{\~n}oz} RR},
  \bibinfo{author}{{Guhathakurta} P} and  \bibinfo{author}{{Renzini} A}
  (\bibinfo{year}{2014}), \bibinfo{month}{Dec.}
\bibinfo{title}{{The Quenching of the Ultra-faint Dwarf Galaxies in the
  Reionization Era}}.
\bibinfo{journal}{{\em ApJ}} \bibinfo{volume}{796}, \bibinfo{eid}{91}.
  \bibinfo{doi}{\doi{10.1088/0004-637X/796/2/91}}.
\eprint{1410.0681}.

\bibtype{Article}%
\bibitem[{Caffau} et al.(2011)]{caffau11}
\bibinfo{author}{{Caffau} E}, \bibinfo{author}{{Bonifacio} P},
  \bibinfo{author}{{Fran{\c c}ois} P}, \bibinfo{author}{{Sbordone} L},
  \bibinfo{author}{{Monaco} L}, \bibinfo{author}{{Spite} M},
  \bibinfo{author}{{Spite} F}, \bibinfo{author}{{Ludwig} HG},
  \bibinfo{author}{{Cayrel} R}, \bibinfo{author}{{Zaggia} S},
  \bibinfo{author}{{Hammer} F}, \bibinfo{author}{{Randich} S},
  \bibinfo{author}{{Molaro} P} and  \bibinfo{author}{{Hill} V}
  (\bibinfo{year}{2011}), \bibinfo{month}{Sep.}
\bibinfo{title}{{An extremely primitive star in the Galactic halo}}.
\bibinfo{journal}{{\em Nature}} \bibinfo{volume}{477}: \bibinfo{pages}{67--69}.
  \bibinfo{doi}{\doi{10.1038/nature10377}}.

\bibtype{Article}%
\bibitem[{Cain} et al.(2020)]{cain20}
\bibinfo{author}{{Cain} M}, \bibinfo{author}{{Frebel} A}, \bibinfo{author}{{Ji}
  AP}, \bibinfo{author}{{Placco} VM}, \bibinfo{author}{{Ezzeddine} R},
  \bibinfo{author}{{Roederer} IU}, \bibinfo{author}{{Hattori} K},
  \bibinfo{author}{{Beers} TC}, \bibinfo{author}{{Mel{\'e}ndez} J},
  \bibinfo{author}{{Hansen} TT} and  \bibinfo{author}{{Sakari} CM}
  (\bibinfo{year}{2020}), \bibinfo{month}{Jul.}
\bibinfo{title}{{The R-Process Alliance: A Very Metal-poor, Extremely
  r-process-enhanced Star with [Eu/Fe] = + 2.2, and the Class of r-III Stars}}.
\bibinfo{journal}{{\em ApJ}} \bibinfo{volume}{898} (\bibinfo{number}{1}),
  \bibinfo{eid}{40}. \bibinfo{doi}{\doi{10.3847/1538-4357/ab97ba}}.

\bibtype{Article}%
\bibitem[{Carney} et al.(1996)]{carney96}
\bibinfo{author}{{Carney} BW}, \bibinfo{author}{{Laird} JB},
  \bibinfo{author}{{Latham} DW} and  \bibinfo{author}{{Aguilar} LA}
  (\bibinfo{year}{1996}), \bibinfo{month}{Aug.}
\bibinfo{title}{{A Survey of Proper Motion Stars. XIII. The Halo Population}}.
\bibinfo{journal}{{\em AJ}} \bibinfo{volume}{112}: \bibinfo{pages}{668}.
  \bibinfo{doi}{\doi{10.1086/118042}}.

\bibtype{Article}%
\bibitem[{Carney} et al.(1997)]{carney_alphapoor}
\bibinfo{author}{{Carney} BW}, \bibinfo{author}{{Wright} JS},
  \bibinfo{author}{{Sneden} C}, \bibinfo{author}{{Laird} JB},
  \bibinfo{author}{{Aguilar} LA} and  \bibinfo{author}{{Latham} DW}
  (\bibinfo{year}{1997}), \bibinfo{month}{Jul.}
\bibinfo{title}{{Discovery of an ''alpha'' Element-Poor Halo Star in a Search
  for Very Low- Metallicity Disk Stars}}.
\bibinfo{journal}{{\em AJ}} \bibinfo{volume}{114}: \bibinfo{pages}{363--375}.
  \bibinfo{doi}{\doi{10.1086/118480}}.

\bibtype{Article}%
\bibitem[{Carollo} et al.(2007)]{carollo}
\bibinfo{author}{{Carollo} D}, \bibinfo{author}{{Beers} TC},
  \bibinfo{author}{{Lee} YS}, \bibinfo{author}{{Chiba} M},
  \bibinfo{author}{{Norris} JE}, \bibinfo{author}{{Wilhelm} R},
  \bibinfo{author}{{Sivarani} T}, \bibinfo{author}{{Marsteller} B},
  \bibinfo{author}{{Munn} JA}, \bibinfo{author}{{Bailer-Jones} CAL},
  \bibinfo{author}{{Fiorentin} PR} and  \bibinfo{author}{{York} DG}
  (\bibinfo{year}{2007}), \bibinfo{month}{Dec.}
\bibinfo{title}{{Two stellar components in the halo of the Milky Way}}.
\bibinfo{journal}{{\em Nature}} \bibinfo{volume}{450}:
  \bibinfo{pages}{1020--1025}. \bibinfo{doi}{\doi{10.1038/nature06460}}.
\eprint{arXiv:0706.3005}.

\bibtype{Article}%
\bibitem[{Carollo} et al.(2010)]{carollo10}
\bibinfo{author}{{Carollo} D}, \bibinfo{author}{{Beers} TC},
  \bibinfo{author}{{Chiba} M}, \bibinfo{author}{{Norris} JE},
  \bibinfo{author}{{Freeman} KC}, \bibinfo{author}{{Lee} YS},
  \bibinfo{author}{{Ivezi{\'c}} {\v Z}}, \bibinfo{author}{{Rockosi} CM} and
  \bibinfo{author}{{Yanny} B} (\bibinfo{year}{2010}), \bibinfo{month}{Mar.}
\bibinfo{title}{{Structure and Kinematics of the Stellar Halos and Thick Disks
  of the Milky Way Based on Calibration Stars from Sloan Digital Sky Survey
  DR7}}.
\bibinfo{journal}{{\em ApJ}} \bibinfo{volume}{712}: \bibinfo{pages}{692--727}.
  \bibinfo{doi}{\doi{10.1088/0004-637X/712/1/692}}.
\eprint{0909.3019}.

\bibtype{Article}%
\bibitem[{Casagrande} and {VandenBerg}(2014)]{Casagrande14}
\bibinfo{author}{{Casagrande} L} and  \bibinfo{author}{{VandenBerg} DA}
  (\bibinfo{year}{2014}), \bibinfo{month}{Oct.}
\bibinfo{title}{{Synthetic stellar photometry - I. General considerations and
  new transformations for broad-band systems}}.
\bibinfo{journal}{{\em MNRAS}} \bibinfo{volume}{444}:
  \bibinfo{pages}{392--419}. \bibinfo{doi}{\doi{10.1093/mnras/stu1476}}.
\eprint{1407.6095}.

\bibtype{Article}%
\bibitem[{Cayrel} et al.(2004)]{cayrel04}
\bibinfo{author}{{Cayrel} R}, \bibinfo{author}{{Depagne} E},
  \bibinfo{author}{{Spite} M}, \bibinfo{author}{{Hill} V},
  \bibinfo{author}{{Spite} F}, \bibinfo{author}{{Fran{\c c}ois} P},
  \bibinfo{author}{{Plez} B}, \bibinfo{author}{{Beers} T},
  \bibinfo{author}{{Primas} F}, \bibinfo{author}{{Andersen} J},
  \bibinfo{author}{{Barbuy} B}, \bibinfo{author}{{Bonifacio} P},
  \bibinfo{author}{{Molaro} P} and  \bibinfo{author}{{Nordstr{\"o}m} B}
  (\bibinfo{year}{2004}), \bibinfo{month}{Mar.}
\bibinfo{title}{{First stars V - Abundance patterns from C to Zn and supernova
  yields in the early Galaxy}}.
\bibinfo{journal}{{\em A\&A}} \bibinfo{volume}{416}:
  \bibinfo{pages}{1117--1138}. \bibinfo{doi}{\doi{10.1051/0004-6361:20034074}}.
\eprint{astro-ph/0311082}.

\bibtype{Article}%
\bibitem[{Chamberlain} and {Aller}(1951)]{chamberlain51}
\bibinfo{author}{{Chamberlain} JW} and  \bibinfo{author}{{Aller} LH}
  (\bibinfo{year}{1951}), \bibinfo{month}{Jul.}
\bibinfo{title}{{The Atmospheres of A-Type Subdwarfs and 95 Leonis.}}
\bibinfo{journal}{{\em ApJ}} \bibinfo{volume}{114}: \bibinfo{pages}{52--72}.
  \bibinfo{doi}{\doi{10.1086/145451}}.

\bibtype{Article}%
\bibitem[{Chiti} et al.(2018)]{chiti18}
\bibinfo{author}{{Chiti} A}, \bibinfo{author}{{Simon} JD},
  \bibinfo{author}{{Frebel} A}, \bibinfo{author}{{Thompson} IB},
  \bibinfo{author}{{Shectman} SA}, \bibinfo{author}{{Mateo} M},
  \bibinfo{author}{{Bailey} John~I. I}, \bibinfo{author}{{Crane} JD} and
  \bibinfo{author}{{Walker} M} (\bibinfo{year}{2018}), \bibinfo{month}{Apr.}
\bibinfo{title}{{Detection of a Population of Carbon-enhanced Metal-poor Stars
  in the Sculptor Dwarf Spheroidal Galaxy}}.
\bibinfo{journal}{{\em ApJ}} \bibinfo{volume}{856} (\bibinfo{number}{2}),
  \bibinfo{eid}{142}. \bibinfo{doi}{\doi{10.3847/1538-4357/aab663}}.

\bibtype{Article}%
\bibitem[{Chiti} et al.(2020)]{chiti20}
\bibinfo{author}{{Chiti} A}, \bibinfo{author}{{Frebel} A},
  \bibinfo{author}{{Jerjen} H}, \bibinfo{author}{{Kim} D} and
  \bibinfo{author}{{Norris} JE} (\bibinfo{year}{2020}), \bibinfo{month}{Mar.}
\bibinfo{title}{{Stellar Metallicities from SkyMapper Photometry I: A Study of
  the Tucana II Ultra-faint Dwarf Galaxy}}.
\bibinfo{journal}{{\em ApJ}} \bibinfo{volume}{891} (\bibinfo{number}{1}),
  \bibinfo{eid}{8}. \bibinfo{doi}{\doi{10.3847/1538-4357/ab6d72}}.

\bibtype{Article}%
\bibitem[{Chiti} et al.(2021)]{chiti21_tuc}
\bibinfo{author}{{Chiti} A}, \bibinfo{author}{{Frebel} A},
  \bibinfo{author}{{Simon} JD}, \bibinfo{author}{{Erkal} D},
  \bibinfo{author}{{Chang} LJ}, \bibinfo{author}{{Necib} L},
  \bibinfo{author}{{Ji} AP}, \bibinfo{author}{{Jerjen} H},
  \bibinfo{author}{{Kim} D} and  \bibinfo{author}{{Norris} JE}
  (\bibinfo{year}{2021}), \bibinfo{month}{Apr.}
\bibinfo{title}{{An extended halo around an ancient dwarf galaxy}}.
\bibinfo{journal}{{\em Nature Astronomy}} \bibinfo{volume}{5}:
  \bibinfo{pages}{392--400}. \bibinfo{doi}{\doi{10.1038/s41550-020-01285-w}}.

\bibtype{Article}%
\bibitem[{Chiti} et al.(2024)]{chiti24}
\bibinfo{author}{{Chiti} A}, \bibinfo{author}{{Mardini} M},
  \bibinfo{author}{{Limberg} G}, \bibinfo{author}{{Frebel} A},
  \bibinfo{author}{{Ji} AP}, \bibinfo{author}{{Reggiani} H},
  \bibinfo{author}{{Ferguson} P}, \bibinfo{author}{{Andales} HD},
  \bibinfo{author}{{Brauer} K}, \bibinfo{author}{{Li} TS} and
  \bibinfo{author}{{Simon} JD} (\bibinfo{year}{2024}), \bibinfo{month}{May}.
\bibinfo{title}{{Enrichment by extragalactic first stars in the Large
  Magellanic Cloud}}.
\bibinfo{journal}{{\em Nature Astronomy}} \bibinfo{volume}{8}:
  \bibinfo{pages}{637--647}. \bibinfo{doi}{\doi{10.1038/s41550-024-02223-w}}.

\bibtype{Article}%
\bibitem[Christlieb(2003)]{Christlieb:2003}
\bibinfo{author}{Christlieb N} (\bibinfo{year}{2003}).
\bibinfo{title}{Finding the most metal-poor stars of the galactic halo with the
  hamburg/eso survey}.
\bibinfo{journal}{{\em Rev. Mod. Astron.}} \bibinfo{volume}{16}:
  \bibinfo{pages}{191--206}.

\bibtype{Article}%
\bibitem[{Christlieb} et al.(2002)]{christliebetal02}
\bibinfo{author}{{Christlieb} N}, \bibinfo{author}{{Bessell} MS},
  \bibinfo{author}{{Beers} TC}, \bibinfo{author}{{Gustafsson} B},
  \bibinfo{author}{{Korn} A}, \bibinfo{author}{{Barklem} PS},
  \bibinfo{author}{{Karlsson} T}, \bibinfo{author}{{Mizuno-Wiedner} M} and
  \bibinfo{author}{{Rossi} S} (\bibinfo{year}{2002}), \bibinfo{month}{Oct.}
\bibinfo{title}{{A stellar relic from the early Milky Way}}.
\bibinfo{journal}{{\em Nature}} \bibinfo{volume}{419}:
  \bibinfo{pages}{904--906}. \bibinfo{doi}{\doi{10.1038/nature01142}}.
\eprint{arXiv:astro-ph/0211274}.

\bibtype{Article}%
\bibitem[{Clarkson} et al.(2018)]{clarkson18}
\bibinfo{author}{{Clarkson} O}, \bibinfo{author}{{Herwig} F} and
  \bibinfo{author}{{Pignatari} M} (\bibinfo{year}{2018}), \bibinfo{month}{Feb.}
\bibinfo{title}{{Pop III i-process nucleosynthesis and the elemental abundances
  of SMSS J0313-6708 and the most iron-poor stars}}.
\bibinfo{journal}{{\em MNRAS}} \bibinfo{volume}{474}:
  \bibinfo{pages}{L37--L41}. \bibinfo{doi}{\doi{10.1093/mnrasl/slx190}}.
\eprint{1710.01763}.

\bibtype{Article}%
\bibitem[{Cooke} and {Madau}(2014)]{cooke14}
\bibinfo{author}{{Cooke} RJ} and  \bibinfo{author}{{Madau} P}
  (\bibinfo{year}{2014}), \bibinfo{month}{Aug.}
\bibinfo{title}{{Carbon-enhanced Metal-poor Stars: Relics from the Dark Ages}}.
\bibinfo{journal}{{\em ApJ}} \bibinfo{volume}{791}, \bibinfo{eid}{116}.
  \bibinfo{doi}{\doi{10.1088/0004-637X/791/2/116}}.
\eprint{1405.7369}.

\bibtype{Article}%
\bibitem[{Cordoni} et al.(2021)]{cordoni21}
\bibinfo{author}{{Cordoni} G}, \bibinfo{author}{{Da Costa} GS},
  \bibinfo{author}{{Yong} D}, \bibinfo{author}{{Mackey} AD},
  \bibinfo{author}{{Marino} AF}, \bibinfo{author}{{Monty} S},
  \bibinfo{author}{{Nordlander} T}, \bibinfo{author}{{Norris} JE},
  \bibinfo{author}{{Asplund} M}, \bibinfo{author}{{Bessell} MS},
  \bibinfo{author}{{Casey} AR}, \bibinfo{author}{{Frebel} A},
  \bibinfo{author}{{Lind} K}, \bibinfo{author}{{Murphy} SJ},
  \bibinfo{author}{{Schmidt} BP}, \bibinfo{author}{{Gao} XD},
  \bibinfo{author}{{Xylakis-Dornbusch} T}, \bibinfo{author}{{Amarsi} AM} and
  \bibinfo{author}{{Milone} AP} (\bibinfo{year}{2021}), \bibinfo{month}{May}.
\bibinfo{title}{{Exploring the Galaxy's halo and very metal-weak thick disc
  with SkyMapper and Gaia DR2}}.
\bibinfo{journal}{{\em MNRAS}} \bibinfo{volume}{503} (\bibinfo{number}{2}):
  \bibinfo{pages}{2539--2561}. \bibinfo{doi}{\doi{10.1093/mnras/staa3417}}.

\bibtype{Article}%
\bibitem[{C{\^o}t{\'e}} et al.(2019)]{cote19}
\bibinfo{author}{{C{\^o}t{\'e}} B}, \bibinfo{author}{{Eichler} M},
  \bibinfo{author}{{Arcones} A}, \bibinfo{author}{{Hansen} CJ},
  \bibinfo{author}{{Simonetti} P}, \bibinfo{author}{{Frebel} A},
  \bibinfo{author}{{Fryer} CL}, \bibinfo{author}{{Pignatari} M},
  \bibinfo{author}{{Reichert} M}, \bibinfo{author}{{Belczynski} K} and
  \bibinfo{author}{{Matteucci} F} (\bibinfo{year}{2019}), \bibinfo{month}{Apr.}
\bibinfo{title}{{Neutron Star Mergers Might Not Be the Only Source of r-process
  Elements in the Milky Way}}.
\bibinfo{journal}{{\em ApJ}} \bibinfo{volume}{875} (\bibinfo{number}{2}),
  \bibinfo{eid}{106}. \bibinfo{doi}{\doi{10.3847/1538-4357/ab10db}}.

\bibtype{Article}%
\bibitem[{Coulter} et al.(2017)]{coulter17}
\bibinfo{author}{{Coulter} DA}, \bibinfo{author}{{Foley} RJ},
  \bibinfo{author}{{Kilpatrick} CD}, \bibinfo{author}{{Drout} MR},
  \bibinfo{author}{{Piro} AL}, \bibinfo{author}{{Shappee} BJ},
  \bibinfo{author}{{Siebert} MR}, \bibinfo{author}{{Simon} JD},
  \bibinfo{author}{{Ulloa} N}, \bibinfo{author}{{Kasen} D},
  \bibinfo{author}{{Madore} BF}, \bibinfo{author}{{Murguia-Berthier} A},
  \bibinfo{author}{{Pan} YC}, \bibinfo{author}{{Prochaska} JX},
  \bibinfo{author}{{Ramirez-Ruiz} E}, \bibinfo{author}{{Rest} A} and
  \bibinfo{author}{{Rojas-Bravo} C} (\bibinfo{year}{2017}),
  \bibinfo{month}{Dec.}
\bibinfo{title}{{Swope Supernova Survey 2017a (SSS17a), the optical counterpart
  to a gravitational wave source}}.
\bibinfo{journal}{{\em Science}} \bibinfo{volume}{358}:
  \bibinfo{pages}{1556--1558}. \bibinfo{doi}{\doi{10.1126/science.aap9811}}.
\eprint{1710.05452}.

\bibtype{Inproceedings}%
\bibitem[{Dalton} et al.(2014)]{Dalton14}
\bibinfo{author}{{Dalton} G}, \bibinfo{author}{{Trager} S},
  \bibinfo{author}{{Abrams} DC}, \bibinfo{author}{{Bonifacio} P},
  \bibinfo{author}{{L{\'o}pez Aguerri} JA}, \bibinfo{author}{{Middleton} K},
  \bibinfo{author}{{Benn} C}, \bibinfo{author}{{Dee} K},
  \bibinfo{author}{{Say{\`e}de} F}, \bibinfo{author}{{Lewis} I},
  \bibinfo{author}{{Pragt} J}, \bibinfo{author}{{Pico} S},
  \bibinfo{author}{{Walton} N}, \bibinfo{author}{{Rey} J},
  \bibinfo{author}{{Allende Prieto} C}, \bibinfo{author}{{Pe{\~n}ate} J},
  \bibinfo{author}{{Lhome} E}, \bibinfo{author}{{Ag{\'o}cs} T},
  \bibinfo{author}{{Alonso} J}, \bibinfo{author}{{Terrett} D},
  \bibinfo{author}{{Brock} M}, \bibinfo{author}{{Gilbert} J},
  \bibinfo{author}{{Ridings} A}, \bibinfo{author}{{Guinouard} I},
  \bibinfo{author}{{Verheijen} M}, \bibinfo{author}{{Tosh} I},
  \bibinfo{author}{{Rogers} K}, \bibinfo{author}{{Steele} I},
  \bibinfo{author}{{Stuik} R}, \bibinfo{author}{{Tromp} N},
  \bibinfo{author}{{Jasko} A}, \bibinfo{author}{{Kragt} J},
  \bibinfo{author}{{Lesman} D}, \bibinfo{author}{{Mottram} C},
  \bibinfo{author}{{Bates} S}, \bibinfo{author}{{Gribbin} F},
  \bibinfo{author}{{Fernando Rodriguez} L}, \bibinfo{author}{{Delgado} JM},
  \bibinfo{author}{{Martin} C}, \bibinfo{author}{{Cano} D},
  \bibinfo{author}{{Navarro} R}, \bibinfo{author}{{Irwin} M},
  \bibinfo{author}{{Lewis} J}, \bibinfo{author}{{Gonzalez Solares} E},
  \bibinfo{author}{{O'Mahony} N}, \bibinfo{author}{{Bianco} A},
  \bibinfo{author}{{Zurita} C}, \bibinfo{author}{{ter Horst} R},
  \bibinfo{author}{{Molinari} E}, \bibinfo{author}{{Lodi} M},
  \bibinfo{author}{{Guerra} J}, \bibinfo{author}{{Vallenari} A} and
  \bibinfo{author}{{Baruffolo} A} (\bibinfo{year}{2014}),
  \bibinfo{month}{Jul.}, \bibinfo{title}{{Project overview and update on WEAVE:
  the next generation wide-field spectroscopy facility for the William Herschel
  Telescope}}, \bibinfo{editor}{{Ramsay} SK}, \bibinfo{editor}{{McLean} IS} and
   \bibinfo{editor}{{Takami} H}, (Eds.), \bibinfo{booktitle}{Ground-based and
  Airborne Instrumentation for Astronomy V}, \bibinfo{series}{Society of
  Photo-Optical Instrumentation Engineers (SPIE) Conference Series},
  \bibinfo{volume}{9147}, pp. \bibinfo{pages}{91470L}, \eprint{1412.0843}.

\bibtype{Article}%
\bibitem[{de Jong} et al.(2019)]{deJong19}
\bibinfo{author}{{de Jong} RS}, \bibinfo{author}{{Agertz} O},
  \bibinfo{author}{{Berbel} AA}, \bibinfo{author}{{Aird} J},
  \bibinfo{author}{{Alexander} DA}, \bibinfo{author}{{Amarsi} A},
  \bibinfo{author}{{Anders} F}, \bibinfo{author}{{Andrae} R},
  \bibinfo{author}{{Ansarinejad} B}, \bibinfo{author}{{Ansorge} W},
  \bibinfo{author}{{Antilogus} P}, \bibinfo{author}{{Anwand-Heerwart} H},
  \bibinfo{author}{{Arentsen} A}, \bibinfo{author}{{Arnadottir} A},
  \bibinfo{author}{{Asplund} M}, \bibinfo{author}{{Auger} M},
  \bibinfo{author}{{Azais} N}, \bibinfo{author}{{Baade} D},
  \bibinfo{author}{{Baker} G}, \bibinfo{author}{{Baker} S},
  \bibinfo{author}{{Balbinot} E}, \bibinfo{author}{{Baldry} IK},
  \bibinfo{author}{{Banerji} M}, \bibinfo{author}{{Barden} S},
  \bibinfo{author}{{Barklem} P}, \bibinfo{author}{{Barth{\'e}l{\'e}my-Mazot}
  E}, \bibinfo{author}{{Battistini} C}, \bibinfo{author}{{Bauer} S},
  \bibinfo{author}{{Bell} CPM}, \bibinfo{author}{{Bellido-Tirado} O},
  \bibinfo{author}{{Bellstedt} S}, \bibinfo{author}{{Belokurov} V},
  \bibinfo{author}{{Bensby} T}, \bibinfo{author}{{Bergemann} M},
  \bibinfo{author}{{Bestenlehner} JM}, \bibinfo{author}{{Bielby} R},
  \bibinfo{author}{{Bilicki} M}, \bibinfo{author}{{Blake} C},
  \bibinfo{author}{{Bland-Hawthorn} J}, \bibinfo{author}{{Boeche} C},
  \bibinfo{author}{{Boland} W}, \bibinfo{author}{{Boller} T},
  \bibinfo{author}{{Bongard} S}, \bibinfo{author}{{Bongiorno} A},
  \bibinfo{author}{{Bonifacio} P}, \bibinfo{author}{{Boudon} D},
  \bibinfo{author}{{Brooks} D}, \bibinfo{author}{{Brown} MJI},
  \bibinfo{author}{{Brown} R}, \bibinfo{author}{{Br{\"u}ggen} M},
  \bibinfo{author}{{Brynnel} J}, \bibinfo{author}{{Brzeski} J},
  \bibinfo{author}{{Buchert} T}, \bibinfo{author}{{Buschkamp} P},
  \bibinfo{author}{{Caffau} E}, \bibinfo{author}{{Caillier} P},
  \bibinfo{author}{{Carrick} J}, \bibinfo{author}{{Casagrande} L},
  \bibinfo{author}{{Case} S}, \bibinfo{author}{{Casey} A},
  \bibinfo{author}{{Cesarini} I}, \bibinfo{author}{{Cescutti} G},
  \bibinfo{author}{{Chapuis} D}, \bibinfo{author}{{Chiappini} C},
  \bibinfo{author}{{Childress} M}, \bibinfo{author}{{Christlieb} N},
  \bibinfo{author}{{Church} R}, \bibinfo{author}{{Cioni} MRL},
  \bibinfo{author}{{Cluver} M}, \bibinfo{author}{{Colless} M},
  \bibinfo{author}{{Collett} T}, \bibinfo{author}{{Comparat} J},
  \bibinfo{author}{{Cooper} A}, \bibinfo{author}{{Couch} W},
  \bibinfo{author}{{Courbin} F}, \bibinfo{author}{{Croom} S},
  \bibinfo{author}{{Croton} D}, \bibinfo{author}{{Daguis{\'e}} E},
  \bibinfo{author}{{Dalton} G}, \bibinfo{author}{{Davies} LJM},
  \bibinfo{author}{{Davis} T}, \bibinfo{author}{{de Laverny} P},
  \bibinfo{author}{{Deason} A}, \bibinfo{author}{{Dionies} F},
  \bibinfo{author}{{Disseau} K}, \bibinfo{author}{{Doel} P},
  \bibinfo{author}{{D{\"o}scher} D}, \bibinfo{author}{{Driver} SP},
  \bibinfo{author}{{Dwelly} T}, \bibinfo{author}{{Eckert} D},
  \bibinfo{author}{{Edge} A}, \bibinfo{author}{{Edvardsson} B},
  \bibinfo{author}{{Youssoufi} DE}, \bibinfo{author}{{Elhaddad} A},
  \bibinfo{author}{{Enke} H}, \bibinfo{author}{{Erfanianfar} G},
  \bibinfo{author}{{Farrell} T}, \bibinfo{author}{{Fechner} T},
  \bibinfo{author}{{Feiz} C}, \bibinfo{author}{{Feltzing} S},
  \bibinfo{author}{{Ferreras} I}, \bibinfo{author}{{Feuerstein} D},
  \bibinfo{author}{{Feuillet} D}, \bibinfo{author}{{Finoguenov} A},
  \bibinfo{author}{{Ford} D}, \bibinfo{author}{{Fotopoulou} S},
  \bibinfo{author}{{Fouesneau} M}, \bibinfo{author}{{Frenk} C},
  \bibinfo{author}{{Frey} S}, \bibinfo{author}{{Gaessler} W},
  \bibinfo{author}{{Geier} S}, \bibinfo{author}{{Gentile Fusillo} N},
  \bibinfo{author}{{Gerhard} O}, \bibinfo{author}{{Giannantonio} T},
  \bibinfo{author}{{Giannone} D}, \bibinfo{author}{{Gibson} B},
  \bibinfo{author}{{Gillingham} P},
  \bibinfo{author}{{Gonz{\'a}lez-Fern{\'a}ndez} C},
  \bibinfo{author}{{Gonzalez-Solares} E}, \bibinfo{author}{{Gottloeber} S},
  \bibinfo{author}{{Gould} A}, \bibinfo{author}{{Grebel} EK},
  \bibinfo{author}{{Gueguen} A}, \bibinfo{author}{{Guiglion} G},
  \bibinfo{author}{{Haehnelt} M}, \bibinfo{author}{{Hahn} T},
  \bibinfo{author}{{Hansen} CJ}, \bibinfo{author}{{Hartman} H},
  \bibinfo{author}{{Hauptner} K}, \bibinfo{author}{{Hawkins} K},
  \bibinfo{author}{{Haynes} D}, \bibinfo{author}{{Haynes} R},
  \bibinfo{author}{{Heiter} U}, \bibinfo{author}{{Helmi} A},
  \bibinfo{author}{{Aguayo} CH}, \bibinfo{author}{{Hewett} P},
  \bibinfo{author}{{Hinton} S}, \bibinfo{author}{{Hobbs} D},
  \bibinfo{author}{{Hoenig} S}, \bibinfo{author}{{Hofman} D},
  \bibinfo{author}{{Hook} I}, \bibinfo{author}{{Hopgood} J},
  \bibinfo{author}{{Hopkins} A}, \bibinfo{author}{{Hourihane} A},
  \bibinfo{author}{{Howes} L}, \bibinfo{author}{{Howlett} C},
  \bibinfo{author}{{Huet} T}, \bibinfo{author}{{Irwin} M},
  \bibinfo{author}{{Iwert} O}, \bibinfo{author}{{Jablonka} P},
  \bibinfo{author}{{Jahn} T}, \bibinfo{author}{{Jahnke} K},
  \bibinfo{author}{{Jarno} A}, \bibinfo{author}{{Jin} S},
  \bibinfo{author}{{Jofre} P}, \bibinfo{author}{{Johl} D},
  \bibinfo{author}{{Jones} D}, \bibinfo{author}{{J{\"o}nsson} H},
  \bibinfo{author}{{Jordan} C}, \bibinfo{author}{{Karovicova} I},
  \bibinfo{author}{{Khalatyan} A}, \bibinfo{author}{{Kelz} A},
  \bibinfo{author}{{Kennicutt} R}, \bibinfo{author}{{King} D},
  \bibinfo{author}{{Kitaura} F}, \bibinfo{author}{{Klar} J},
  \bibinfo{author}{{Klauser} U}, \bibinfo{author}{{Kneib} JP},
  \bibinfo{author}{{Koch} A}, \bibinfo{author}{{Koposov} S},
  \bibinfo{author}{{Kordopatis} G}, \bibinfo{author}{{Korn} A},
  \bibinfo{author}{{Kosmalski} J}, \bibinfo{author}{{Kotak} R},
  \bibinfo{author}{{Kovalev} M}, \bibinfo{author}{{Kreckel} K},
  \bibinfo{author}{{Kripak} Y}, \bibinfo{author}{{Krumpe} M},
  \bibinfo{author}{{Kuijken} K}, \bibinfo{author}{{Kunder} A},
  \bibinfo{author}{{Kushniruk} I}, \bibinfo{author}{{Lam} MI},
  \bibinfo{author}{{Lamer} G}, \bibinfo{author}{{Laurent} F},
  \bibinfo{author}{{Lawrence} J}, \bibinfo{author}{{Lehmitz} M},
  \bibinfo{author}{{Lemasle} B}, \bibinfo{author}{{Lewis} J},
  \bibinfo{author}{{Li} B}, \bibinfo{author}{{Lidman} C},
  \bibinfo{author}{{Lind} K}, \bibinfo{author}{{Liske} J},
  \bibinfo{author}{{Lizon} JL}, \bibinfo{author}{{Loveday} J},
  \bibinfo{author}{{Ludwig} HG}, \bibinfo{author}{{McDermid} RM},
  \bibinfo{author}{{Maguire} K}, \bibinfo{author}{{Mainieri} V},
  \bibinfo{author}{{Mali} S}, \bibinfo{author}{{Mandel} H},
  \bibinfo{author}{{Mandel} K}, \bibinfo{author}{{Mannering} L},
  \bibinfo{author}{{Martell} S}, \bibinfo{author}{{Martinez Delgado} D},
  \bibinfo{author}{{Matijevic} G}, \bibinfo{author}{{McGregor} H},
  \bibinfo{author}{{McMahon} R}, \bibinfo{author}{{McMillan} P},
  \bibinfo{author}{{Mena} O}, \bibinfo{author}{{Merloni} A},
  \bibinfo{author}{{Meyer} MJ}, \bibinfo{author}{{Michel} C},
  \bibinfo{author}{{Micheva} G}, \bibinfo{author}{{Migniau} JE},
  \bibinfo{author}{{Minchev} I}, \bibinfo{author}{{Monari} G},
  \bibinfo{author}{{Muller} R}, \bibinfo{author}{{Murphy} D},
  \bibinfo{author}{{Muthukrishna} D}, \bibinfo{author}{{Nandra} K},
  \bibinfo{author}{{Navarro} R}, \bibinfo{author}{{Ness} M},
  \bibinfo{author}{{Nichani} V}, \bibinfo{author}{{Nichol} R},
  \bibinfo{author}{{Nicklas} H}, \bibinfo{author}{{Niederhofer} F},
  \bibinfo{author}{{Norberg} P}, \bibinfo{author}{{Obreschkow} D},
  \bibinfo{author}{{Oliver} S}, \bibinfo{author}{{Owers} M},
  \bibinfo{author}{{Pai} N}, \bibinfo{author}{{Pankratow} S},
  \bibinfo{author}{{Parkinson} D}, \bibinfo{author}{{Paschke} J},
  \bibinfo{author}{{Paterson} R}, \bibinfo{author}{{Pecontal} A},
  \bibinfo{author}{{Parry} I}, \bibinfo{author}{{Phillips} D},
  \bibinfo{author}{{Pillepich} A}, \bibinfo{author}{{Pinard} L},
  \bibinfo{author}{{Pirard} J}, \bibinfo{author}{{Piskunov} N},
  \bibinfo{author}{{Plank} V}, \bibinfo{author}{{Pl{\"u}schke} D},
  \bibinfo{author}{{Pons} E}, \bibinfo{author}{{Popesso} P},
  \bibinfo{author}{{Power} C}, \bibinfo{author}{{Pragt} J},
  \bibinfo{author}{{Pramskiy} A}, \bibinfo{author}{{Pryer} D},
  \bibinfo{author}{{Quattri} M}, \bibinfo{author}{{Queiroz} ABdA},
  \bibinfo{author}{{Quirrenbach} A}, \bibinfo{author}{{Rahurkar} S},
  \bibinfo{author}{{Raichoor} A}, \bibinfo{author}{{Ramstedt} S},
  \bibinfo{author}{{Rau} A}, \bibinfo{author}{{Recio-Blanco} A},
  \bibinfo{author}{{Reiss} R}, \bibinfo{author}{{Renaud} F},
  \bibinfo{author}{{Revaz} Y}, \bibinfo{author}{{Rhode} P},
  \bibinfo{author}{{Richard} J}, \bibinfo{author}{{Richter} AD},
  \bibinfo{author}{{Rix} HW}, \bibinfo{author}{{Robotham} ASG},
  \bibinfo{author}{{Roelfsema} R}, \bibinfo{author}{{Romaniello} M},
  \bibinfo{author}{{Rosario} D}, \bibinfo{author}{{Rothmaier} F},
  \bibinfo{author}{{Roukema} B}, \bibinfo{author}{{Ruchti} G},
  \bibinfo{author}{{Rupprecht} G}, \bibinfo{author}{{Rybizki} J},
  \bibinfo{author}{{Ryde} N}, \bibinfo{author}{{Saar} A},
  \bibinfo{author}{{Sadler} E}, \bibinfo{author}{{Sahl{\'e}n} M},
  \bibinfo{author}{{Salvato} M}, \bibinfo{author}{{Sassolas} B},
  \bibinfo{author}{{Saunders} W}, \bibinfo{author}{{Saviauk} A},
  \bibinfo{author}{{Sbordone} L}, \bibinfo{author}{{Schmidt} T},
  \bibinfo{author}{{Schnurr} O}, \bibinfo{author}{{Scholz} RD},
  \bibinfo{author}{{Schwope} A}, \bibinfo{author}{{Seifert} W},
  \bibinfo{author}{{Shanks} T}, \bibinfo{author}{{Sheinis} A},
  \bibinfo{author}{{Sivov} T}, \bibinfo{author}{{Sk{\'u}lad{\'o}ttir} {\'A}},
  \bibinfo{author}{{Smartt} S}, \bibinfo{author}{{Smedley} S},
  \bibinfo{author}{{Smith} G}, \bibinfo{author}{{Smith} R},
  \bibinfo{author}{{Sorce} J}, \bibinfo{author}{{Spitler} L},
  \bibinfo{author}{{Starkenburg} E}, \bibinfo{author}{{Steinmetz} M},
  \bibinfo{author}{{Stilz} I}, \bibinfo{author}{{Storm} J},
  \bibinfo{author}{{Sullivan} M}, \bibinfo{author}{{Sutherland} W},
  \bibinfo{author}{{Swann} E}, \bibinfo{author}{{Tamone} A},
  \bibinfo{author}{{Taylor} EN}, \bibinfo{author}{{Teillon} J},
  \bibinfo{author}{{Tempel} E}, \bibinfo{author}{{ter Horst} R},
  \bibinfo{author}{{Thi} WF}, \bibinfo{author}{{Tolstoy} E},
  \bibinfo{author}{{Trager} S}, \bibinfo{author}{{Traven} G},
  \bibinfo{author}{{Tremblay} PE}, \bibinfo{author}{{Tresse} L},
  \bibinfo{author}{{Valentini} M}, \bibinfo{author}{{van de Weygaert} R},
  \bibinfo{author}{{van den Ancker} M}, \bibinfo{author}{{Veljanoski} J},
  \bibinfo{author}{{Venkatesan} S}, \bibinfo{author}{{Wagner} L},
  \bibinfo{author}{{Wagner} K}, \bibinfo{author}{{Walcher} CJ},
  \bibinfo{author}{{Waller} L}, \bibinfo{author}{{Walton} N},
  \bibinfo{author}{{Wang} L}, \bibinfo{author}{{Winkler} R},
  \bibinfo{author}{{Wisotzki} L}, \bibinfo{author}{{Worley} CC},
  \bibinfo{author}{{Worseck} G}, \bibinfo{author}{{Xiang} M},
  \bibinfo{author}{{Xu} W}, \bibinfo{author}{{Yong} D}, \bibinfo{author}{{Zhao}
  C}, \bibinfo{author}{{Zheng} J}, \bibinfo{author}{{Zscheyge} F} and
  \bibinfo{author}{{Zucker} D} (\bibinfo{year}{2019}), \bibinfo{month}{Mar.}
\bibinfo{title}{{4MOST: Project overview and information for the First Call for
  Proposals}}.
\bibinfo{journal}{{\em The Messenger}} \bibinfo{volume}{175}:
  \bibinfo{pages}{3--11}. \bibinfo{doi}{\doi{10.18727/0722-6691/5117}}.
\eprint{1903.02464}.

\bibtype{Article}%
\bibitem[{De Silva} et al.(2015)]{DeSilva15}
\bibinfo{author}{{De Silva} GM}, \bibinfo{author}{{Freeman} KC},
  \bibinfo{author}{{Bland-Hawthorn} J}, \bibinfo{author}{{Martell} S},
  \bibinfo{author}{{de Boer} EW}, \bibinfo{author}{{Asplund} M},
  \bibinfo{author}{{Keller} S}, \bibinfo{author}{{Sharma} S},
  \bibinfo{author}{{Zucker} DB}, \bibinfo{author}{{Zwitter} T},
  \bibinfo{author}{{Anguiano} B}, \bibinfo{author}{{Bacigalupo} C},
  \bibinfo{author}{{Bayliss} D}, \bibinfo{author}{{Beavis} MA},
  \bibinfo{author}{{Bergemann} M}, \bibinfo{author}{{Campbell} S},
  \bibinfo{author}{{Cannon} R}, \bibinfo{author}{{Carollo} D},
  \bibinfo{author}{{Casagrande} L}, \bibinfo{author}{{Casey} AR},
  \bibinfo{author}{{Da Costa} G}, \bibinfo{author}{{D'Orazi} V},
  \bibinfo{author}{{Dotter} A}, \bibinfo{author}{{Duong} L},
  \bibinfo{author}{{Heger} A}, \bibinfo{author}{{Ireland} MJ},
  \bibinfo{author}{{Kafle} PR}, \bibinfo{author}{{Kos} J},
  \bibinfo{author}{{Lattanzio} J}, \bibinfo{author}{{Lewis} GF},
  \bibinfo{author}{{Lin} J}, \bibinfo{author}{{Lind} K},
  \bibinfo{author}{{Munari} U}, \bibinfo{author}{{Nataf} DM},
  \bibinfo{author}{{O'Toole} S}, \bibinfo{author}{{Parker} Q},
  \bibinfo{author}{{Reid} W}, \bibinfo{author}{{Schlesinger} KJ},
  \bibinfo{author}{{Sheinis} A}, \bibinfo{author}{{Simpson} JD},
  \bibinfo{author}{{Stello} D}, \bibinfo{author}{{Ting} YS},
  \bibinfo{author}{{Traven} G}, \bibinfo{author}{{Watson} F},
  \bibinfo{author}{{Wittenmyer} R}, \bibinfo{author}{{Yong} D} and
  \bibinfo{author}{{{\v{Z}}erjal} M} (\bibinfo{year}{2015}),
  \bibinfo{month}{May}.
\bibinfo{title}{{The GALAH survey: scientific motivation}}.
\bibinfo{journal}{{\em MNRAS}} \bibinfo{volume}{449} (\bibinfo{number}{3}):
  \bibinfo{pages}{2604--2617}. \bibinfo{doi}{\doi{10.1093/mnras/stv327}}.
\eprint{1502.04767}.

\bibtype{Article}%
\bibitem[Depagne et al.(2002)]{depagne02}
\bibinfo{author}{Depagne E}, \bibinfo{author}{Hill V}, \bibinfo{author}{Spite
  M}, \bibinfo{author}{Spite F}, \bibinfo{author}{Plez B},
  \bibinfo{author}{Beers T}, \bibinfo{author}{Barbuy B},
  \bibinfo{author}{Cayrel R}, \bibinfo{author}{Andersen J},
  \bibinfo{author}{Bonifacio P}, \bibinfo{author}{Francois P},
  \bibinfo{author}{Nordstr{\" o}m B} and  \bibinfo{author}{Primas F}
  (\bibinfo{year}{2002}).
\bibinfo{title}{First stars. ii. elemental abundances in the extremely
  metal-poor star cs~22949--037. a diagnostic of early massive supernovae}.
\bibinfo{journal}{{\em A\&A}} \bibinfo{volume}{390}: \bibinfo{pages}{187--198}.

\bibtype{Article}%
\bibitem[{Dotter} et al.(2008)]{dcj08}
\bibinfo{author}{{Dotter} A}, \bibinfo{author}{{Chaboyer} B},
  \bibinfo{author}{{Jevremovi{\'c}} D}, \bibinfo{author}{{Kostov} V},
  \bibinfo{author}{{Baron} E} and  \bibinfo{author}{{Ferguson} JW}
  (\bibinfo{year}{2008}), \bibinfo{month}{Sep.}
\bibinfo{title}{{The Dartmouth Stellar Evolution Database}}.
\bibinfo{journal}{{\em ApJS}} \bibinfo{volume}{178}: \bibinfo{pages}{89--101}.
  \bibinfo{doi}{\doi{10.1086/589654}}.
\eprint{0804.4473}.

\bibtype{Article}%
\bibitem[{Drout} et al.(2017)]{drout17}
\bibinfo{author}{{Drout} MR}, \bibinfo{author}{{Piro} AL},
  \bibinfo{author}{{Shappee} BJ}, \bibinfo{author}{{Kilpatrick} CD},
  \bibinfo{author}{{Simon} JD}, \bibinfo{author}{{Contreras} C},
  \bibinfo{author}{{Coulter} DA}, \bibinfo{author}{{Foley} RJ},
  \bibinfo{author}{{Siebert} MR}, \bibinfo{author}{{Morrell} N},
  \bibinfo{author}{{Boutsia} K}, \bibinfo{author}{{Di Mille} F},
  \bibinfo{author}{{Holoien} TWS}, \bibinfo{author}{{Kasen} D},
  \bibinfo{author}{{Kollmeier} JA}, \bibinfo{author}{{Madore} BF},
  \bibinfo{author}{{Monson} AJ}, \bibinfo{author}{{Murguia-Berthier} A},
  \bibinfo{author}{{Pan} YC}, \bibinfo{author}{{Prochaska} JX},
  \bibinfo{author}{{Ramirez-Ruiz} E}, \bibinfo{author}{{Rest} A},
  \bibinfo{author}{{Adams} C}, \bibinfo{author}{{Alatalo} K},
  \bibinfo{author}{{Ba{\~n}ados} E}, \bibinfo{author}{{Baughman} J},
  \bibinfo{author}{{Beers} TC}, \bibinfo{author}{{Bernstein} RA},
  \bibinfo{author}{{Bitsakis} T}, \bibinfo{author}{{Campillay} A},
  \bibinfo{author}{{Hansen} TT}, \bibinfo{author}{{Higgs} CR},
  \bibinfo{author}{{Ji} AP}, \bibinfo{author}{{Maravelias} G},
  \bibinfo{author}{{Marshall} JL}, \bibinfo{author}{{Bidin} CM},
  \bibinfo{author}{{Prieto} JL}, \bibinfo{author}{{Rasmussen} KC},
  \bibinfo{author}{{Rojas-Bravo} C}, \bibinfo{author}{{Strom} AL},
  \bibinfo{author}{{Ulloa} N}, \bibinfo{author}{{Vargas-Gonz{\'a}lez} J},
  \bibinfo{author}{{Wan} Z} and  \bibinfo{author}{{Whitten} DD}
  (\bibinfo{year}{2017}), \bibinfo{month}{Dec.}
\bibinfo{title}{{Light curves of the neutron star merger GW170817/SSS17a:
  Implications for r-process nucleosynthesis}}.
\bibinfo{journal}{{\em Science}} \bibinfo{volume}{358}:
  \bibinfo{pages}{1570--1574}. \bibinfo{doi}{\doi{10.1126/science.aaq0049}}.
\eprint{1710.05443}.

\bibtype{Article}%
\bibitem[{Ezzeddine} et al.(2018)]{ezzeddine18_H}
\bibinfo{author}{{Ezzeddine} R}, \bibinfo{author}{{Merle} T},
  \bibinfo{author}{{Plez} B}, \bibinfo{author}{{Gebran} M},
  \bibinfo{author}{{Th{\'e}venin} F} and  \bibinfo{author}{{Van der Swaelmen}
  M} (\bibinfo{year}{2018}), \bibinfo{month}{Oct.}
\bibinfo{title}{{An empirical recipe for inelastic hydrogen-atom collisions in
  non-LTE calculations}}.
\bibinfo{journal}{{\em A\&A}} \bibinfo{volume}{618}, \bibinfo{eid}{A141}.
  \bibinfo{doi}{\doi{10.1051/0004-6361/201630352}}.

\bibtype{Article}%
\bibitem[{Ezzeddine} et al.(2020)]{ezzeddine20}
\bibinfo{author}{{Ezzeddine} R}, \bibinfo{author}{{Rasmussen} K},
  \bibinfo{author}{{Frebel} A}, \bibinfo{author}{{Chiti} A},
  \bibinfo{author}{{Hinojisa} K}, \bibinfo{author}{{Placco} VM},
  \bibinfo{author}{{Ji} AP}, \bibinfo{author}{{Beers} TC},
  \bibinfo{author}{{Hansen} TT}, \bibinfo{author}{{Roederer} IU},
  \bibinfo{author}{{Sakari} CM} and  \bibinfo{author}{{Melendez} J}
  (\bibinfo{year}{2020}), \bibinfo{month}{Aug.}
\bibinfo{title}{{The R-Process Alliance: First Magellan/MIKE Release from the
  Southern Search for R-process-enhanced Stars}}.
\bibinfo{journal}{{\em ApJ}} \bibinfo{volume}{898} (\bibinfo{number}{2}),
  \bibinfo{eid}{150}. \bibinfo{doi}{\doi{10.3847/1538-4357/ab9d1a}}.

\bibtype{Article}%
\bibitem[{Fran{\c{c}}ois} et al.(2020)]{francois20}
\bibinfo{author}{{Fran{\c{c}}ois} P}, \bibinfo{author}{{Wanajo} S},
  \bibinfo{author}{{Caffau} E}, \bibinfo{author}{{Prantzos} N},
  \bibinfo{author}{{Aoki} W}, \bibinfo{author}{{Aoki} M},
  \bibinfo{author}{{Bonifacio} P}, \bibinfo{author}{{Spite} M} and
  \bibinfo{author}{{Spite} F} (\bibinfo{year}{2020}), \bibinfo{month}{Oct.}
\bibinfo{title}{{Detailed abundances in a sample of very metal-poor stars}}.
\bibinfo{journal}{{\em A\&A}} \bibinfo{volume}{642}, \bibinfo{eid}{A25}.
  \bibinfo{doi}{\doi{10.1051/0004-6361/202038028}}.

\bibtype{Article}%
\bibitem[{Frebel}(2018)]{frebel18}
\bibinfo{author}{{Frebel} A} (\bibinfo{year}{2018}), \bibinfo{month}{Jun.}
\bibinfo{title}{{From Nuclei to the Cosmos: Tracing Heavy-Element Production
  with the Oldest Stars}}.
\bibinfo{journal}{{\em Annual Reviews of Nuclear and Particle Science}}
  \eprint{1806.08955}.

\bibtype{Article}%
\bibitem[{Frebel} and {Ji}(2023)]{frebel_ji23}
\bibinfo{author}{{Frebel} A} and  \bibinfo{author}{{Ji} AP}
  (\bibinfo{year}{2023}), \bibinfo{month}{Feb.}
\bibinfo{title}{{Observations of R-Process Stars in the Milky Way and Dwarf
  Galaxies}}.
\bibinfo{journal}{{\em arXiv e-prints}} ,
  \bibinfo{eid}{arXiv:2302.09188}\bibinfo{doi}{\doi{10.48550/arXiv.2302.09188}}.
\eprint{2302.09188}.

\bibtype{inbook}%
\bibitem[{Frebel} and {Norris}(2013)]{psss}
\bibinfo{author}{{Frebel} A} and  \bibinfo{author}{{Norris} JE}
  (\bibinfo{year}{2013}).
\bibinfo{title}{{Metal-Poor Stars and the Chemical Enrichment of the
  Universe}}.
 \bibinfo{pages}{55--114}.
\bibinfo{doi}{\doi{10.1007/978-94-007-5612-0_3}}.

\bibtype{Article}%
\bibitem[{Frebel} and {Norris}(2015{\natexlab{a}})]{fn15}
\bibinfo{author}{{Frebel} A} and  \bibinfo{author}{{Norris} JE}
  (\bibinfo{year}{2015}{\natexlab{a}}), \bibinfo{month}{Aug.}
\bibinfo{title}{{Near-Field Cosmology with Extremely Metal-Poor Stars}}.
\bibinfo{journal}{{\em ARA\&A}} \bibinfo{volume}{53}:
  \bibinfo{pages}{631--688}.
  \bibinfo{doi}{\doi{10.1146/annurev-astro-082214-122423}}.
\eprint{1501.06921}.

\bibtype{Article}%
\bibitem[{Frebel} and {Norris}(2015{\natexlab{b}})]{frebel15}
\bibinfo{author}{{Frebel} A} and  \bibinfo{author}{{Norris} JE}
  (\bibinfo{year}{2015}{\natexlab{b}}), \bibinfo{month}{Aug.}
\bibinfo{title}{{Near-Field Cosmology with Metal-Poor Stars}}.
\bibinfo{journal}{{\em ARA\&A}} \bibinfo{volume}{53}:
  \bibinfo{pages}{631--688}.
\eprint{1501.06921}.

\bibtype{Article}%
\bibitem[{Frebel} et al.(2005{\natexlab{a}})]{frebel05}
\bibinfo{author}{{Frebel} A}, \bibinfo{author}{{Aoki} W},
  \bibinfo{author}{{Christlieb} N}, \bibinfo{author}{{Ando} H},
  \bibinfo{author}{{Asplund} M}, \bibinfo{author}{{Barklem} PS},
  \bibinfo{author}{{Beers} TC}, \bibinfo{author}{{Eriksson} K},
  \bibinfo{author}{{Fechner} C}, \bibinfo{author}{{Fujimoto} MY},
  \bibinfo{author}{{Honda} S}, \bibinfo{author}{{Kajino} T},
  \bibinfo{author}{{Minezaki} T}, \bibinfo{author}{{Nomoto} K},
  \bibinfo{author}{{Norris} JE}, \bibinfo{author}{{Ryan} SG},
  \bibinfo{author}{{Takada-Hidai} M}, \bibinfo{author}{{Tsangarides} S} and
  \bibinfo{author}{{Yoshii} Y} (\bibinfo{year}{2005}{\natexlab{a}}),
  \bibinfo{month}{Apr.}
\bibinfo{title}{{Nucleosynthetic signatures of the first stars}}.
\bibinfo{journal}{{\em Nature}} \bibinfo{volume}{434}:
  \bibinfo{pages}{871--873}. \bibinfo{doi}{\doi{10.1038/nature03455}}.
\eprint{arXiv:astro-ph/0503021}.

\bibtype{Article}%
\bibitem[{Frebel} et al.(2005{\natexlab{b}})]{HE1327_Nature}
\bibinfo{author}{{Frebel} A}, \bibinfo{author}{{Aoki} W},
  \bibinfo{author}{{Christlieb} N}, \bibinfo{author}{{Ando} H},
  \bibinfo{author}{{Asplund} M}, \bibinfo{author}{{Barklem} PS},
  \bibinfo{author}{{Beers} TC}, \bibinfo{author}{{Eriksson} K},
  \bibinfo{author}{{Fechner} C}, \bibinfo{author}{{Fujimoto} MY},
  \bibinfo{author}{{Honda} S}, \bibinfo{author}{{Kajino} T},
  \bibinfo{author}{{Minezaki} T}, \bibinfo{author}{{Nomoto} K},
  \bibinfo{author}{{Norris} JE}, \bibinfo{author}{{Ryan} SG},
  \bibinfo{author}{{Takada-Hidai} M}, \bibinfo{author}{{Tsangarides} S} and
  \bibinfo{author}{{Yoshii} Y} (\bibinfo{year}{2005}{\natexlab{b}}),
  \bibinfo{month}{Apr.}
\bibinfo{title}{{Nucleosynthetic signatures of the first stars}}.
\bibinfo{journal}{{\em Nature}} \bibinfo{volume}{434}:
  \bibinfo{pages}{871--873}. \bibinfo{doi}{\doi{10.1038/nature03455}}.

\bibtype{Article}%
\bibitem[{Frebel} et al.(2006)]{frebel_bmps}
\bibinfo{author}{{Frebel} A}, \bibinfo{author}{{Christlieb} N},
  \bibinfo{author}{{Norris} JE}, \bibinfo{author}{{Beers} TC},
  \bibinfo{author}{{Bessell} MS}, \bibinfo{author}{{Rhee} J},
  \bibinfo{author}{{Fechner} C}, \bibinfo{author}{{Marsteller} B},
  \bibinfo{author}{{Rossi} S}, \bibinfo{author}{{Thom} C},
  \bibinfo{author}{{Wisotzki} L} and  \bibinfo{author}{{Reimers} D}
  (\bibinfo{year}{2006}), \bibinfo{month}{Dec.}
\bibinfo{title}{{Bright Metal-poor Stars from the Hamburg/ESO Survey. I.
  Selection and Follow-up Observations from 329 Fields}}.
\bibinfo{journal}{{\em ApJ}} \bibinfo{volume}{652}:
  \bibinfo{pages}{1585--1603}. \bibinfo{doi}{\doi{10.1086/508506}}.

\bibtype{Article}%
\bibitem[{Frebel} et al.(2007)]{dtrans}
\bibinfo{author}{{Frebel} A}, \bibinfo{author}{{Johnson} JL} and
  \bibinfo{author}{{Bromm} V} (\bibinfo{year}{2007}), \bibinfo{month}{Sep.}
\bibinfo{title}{{Probing the formation of the first low-mass stars with stellar
  archaeology}}.
\bibinfo{journal}{{\em MNRAS}} \bibinfo{volume}{380}:
  \bibinfo{pages}{L40--L44}.
  \bibinfo{doi}{\doi{10.1111/j.1745-3933.2007.00344.x}}.
\eprint{arXiv:astro-ph/0701395}.

\bibtype{Article}%
\bibitem[{Frebel} et al.(2013)]{frebel13}
\bibinfo{author}{{Frebel} A}, \bibinfo{author}{{Casey} AR},
  \bibinfo{author}{{Jacobson} HR} and  \bibinfo{author}{{Yu} Q}
  (\bibinfo{year}{2013}), \bibinfo{month}{May}.
\bibinfo{title}{{Deriving Stellar Effective Temperatures of Metal-poor Stars
  with the Excitation Potential Method}}.
\bibinfo{journal}{{\em ApJ}} \bibinfo{volume}{769}, \bibinfo{eid}{57}.
  \bibinfo{doi}{\doi{10.1088/0004-637X/769/1/57}}.
\eprint{1304.2396}.

\bibtype{Article}%
\bibitem[{Frebel} et al.(2014)]{frebel14}
\bibinfo{author}{{Frebel} A}, \bibinfo{author}{{Simon} JD} and
  \bibinfo{author}{{Kirby} EN} (\bibinfo{year}{2014}), \bibinfo{month}{May}.
\bibinfo{title}{{Segue 1: An Unevolved Fossil Galaxy from the Early Universe}}.
\bibinfo{journal}{{\em ApJ}} \bibinfo{volume}{786}, \bibinfo{eid}{74}.
  \bibinfo{doi}{\doi{10.1088/0004-637X/786/1/74}}.
\eprint{1403.6116}.

\bibtype{Article}%
\bibitem[{Gaia Collaboration} et al.(2021)]{gaia21}
\bibinfo{author}{{Gaia Collaboration}}, \bibinfo{author}{{Brown} AGA},
  \bibinfo{author}{{Vallenari} A}, \bibinfo{author}{{Prusti} T},
  \bibinfo{author}{{de Bruijne} JHJ}, \bibinfo{author}{{Babusiaux} C},
  \bibinfo{author}{{Biermann} M}, \bibinfo{author}{{Creevey} OL},
  \bibinfo{author}{{Evans} DW}, \bibinfo{author}{{Eyer} L},
  \bibinfo{author}{{Hutton} A}, \bibinfo{author}{{Jansen} F},
  \bibinfo{author}{{Jordi} C}, \bibinfo{author}{{Klioner} SA},
  \bibinfo{author}{{Lammers} U}, \bibinfo{author}{{Lindegren} L},
  \bibinfo{author}{{Luri} X}, \bibinfo{author}{{Mignard} F},
  \bibinfo{author}{{Panem} C}, \bibinfo{author}{{Pourbaix} D},
  \bibinfo{author}{{Randich} S}, \bibinfo{author}{{Sartoretti} P},
  \bibinfo{author}{{Soubiran} C}, \bibinfo{author}{{Walton} NA},
  \bibinfo{author}{{Arenou} F}, \bibinfo{author}{{Bailer-Jones} CAL},
  \bibinfo{author}{{Bastian} U}, \bibinfo{author}{{Cropper} M},
  \bibinfo{author}{{Drimmel} R}, \bibinfo{author}{{Katz} D},
  \bibinfo{author}{{Lattanzi} MG}, \bibinfo{author}{{van Leeuwen} F},
  \bibinfo{author}{{Bakker} J}, \bibinfo{author}{{Cacciari} C},
  \bibinfo{author}{{Casta{\~n}eda} J}, \bibinfo{author}{{De Angeli} F},
  \bibinfo{author}{{Ducourant} C}, \bibinfo{author}{{Fabricius} C},
  \bibinfo{author}{{Fouesneau} M}, \bibinfo{author}{{Fr{\'e}mat} Y},
  \bibinfo{author}{{Guerra} R}, \bibinfo{author}{{Guerrier} A},
  \bibinfo{author}{{Guiraud} J}, \bibinfo{author}{{Jean-Antoine Piccolo} A},
  \bibinfo{author}{{Masana} E}, \bibinfo{author}{{Messineo} R},
  \bibinfo{author}{{Mowlavi} N}, \bibinfo{author}{{Nicolas} C},
  \bibinfo{author}{{Nienartowicz} K}, \bibinfo{author}{{Pailler} F},
  \bibinfo{author}{{Panuzzo} P}, \bibinfo{author}{{Riclet} F},
  \bibinfo{author}{{Roux} W}, \bibinfo{author}{{Seabroke} GM},
  \bibinfo{author}{{Sordo} R}, \bibinfo{author}{{Tanga} P},
  \bibinfo{author}{{Th{\'e}venin} F}, \bibinfo{author}{{Gracia-Abril} G},
  \bibinfo{author}{{Portell} J}, \bibinfo{author}{{Teyssier} D},
  \bibinfo{author}{{Altmann} M}, \bibinfo{author}{{Andrae} R},
  \bibinfo{author}{{Bellas-Velidis} I}, \bibinfo{author}{{Benson} K},
  \bibinfo{author}{{Berthier} J}, \bibinfo{author}{{Blomme} R},
  \bibinfo{author}{{Brugaletta} E}, \bibinfo{author}{{Burgess} PW},
  \bibinfo{author}{{Busso} G}, \bibinfo{author}{{Carry} B},
  \bibinfo{author}{{Cellino} A}, \bibinfo{author}{{Cheek} N},
  \bibinfo{author}{{Clementini} G}, \bibinfo{author}{{Damerdji} Y},
  \bibinfo{author}{{Davidson} M}, \bibinfo{author}{{Delchambre} L},
  \bibinfo{author}{{Dell'Oro} A},
  \bibinfo{author}{{Fern{\'a}ndez-Hern{\'a}ndez} J},
  \bibinfo{author}{{Galluccio} L}, \bibinfo{author}{{Garc{\'\i}a-Lario} P},
  \bibinfo{author}{{Garcia-Reinaldos} M},
  \bibinfo{author}{{Gonz{\'a}lez-N{\'u}{\~n}ez} J}, \bibinfo{author}{{Gosset}
  E}, \bibinfo{author}{{Haigron} R}, \bibinfo{author}{{Halbwachs} JL},
  \bibinfo{author}{{Hambly} NC}, \bibinfo{author}{{Harrison} DL},
  \bibinfo{author}{{Hatzidimitriou} D}, \bibinfo{author}{{Heiter} U},
  \bibinfo{author}{{Hern{\'a}ndez} J}, \bibinfo{author}{{Hestroffer} D},
  \bibinfo{author}{{Hodgkin} ST}, \bibinfo{author}{{Holl} B},
  \bibinfo{author}{{Jan{\ss}en} K}, \bibinfo{author}{{Jevardat de Fombelle} G},
  \bibinfo{author}{{Jordan} S}, \bibinfo{author}{{Krone-Martins} A},
  \bibinfo{author}{{Lanzafame} AC}, \bibinfo{author}{{L{\"o}ffler} W},
  \bibinfo{author}{{Lorca} A}, \bibinfo{author}{{Manteiga} M},
  \bibinfo{author}{{Marchal} O}, \bibinfo{author}{{Marrese} PM},
  \bibinfo{author}{{Moitinho} A}, \bibinfo{author}{{Mora} A},
  \bibinfo{author}{{Muinonen} K}, \bibinfo{author}{{Osborne} P},
  \bibinfo{author}{{Pancino} E}, \bibinfo{author}{{Pauwels} T},
  \bibinfo{author}{{Petit} JM}, \bibinfo{author}{{Recio-Blanco} A},
  \bibinfo{author}{{Richards} PJ}, \bibinfo{author}{{Riello} M},
  \bibinfo{author}{{Rimoldini} L}, \bibinfo{author}{{Robin} AC},
  \bibinfo{author}{{Roegiers} T}, \bibinfo{author}{{Rybizki} J},
  \bibinfo{author}{{Sarro} LM}, \bibinfo{author}{{Siopis} C},
  \bibinfo{author}{{Smith} M}, \bibinfo{author}{{Sozzetti} A},
  \bibinfo{author}{{Ulla} A}, \bibinfo{author}{{Utrilla} E},
  \bibinfo{author}{{van Leeuwen} M}, \bibinfo{author}{{van Reeven} W},
  \bibinfo{author}{{Abbas} U}, \bibinfo{author}{{Abreu Aramburu} A},
  \bibinfo{author}{{Accart} S}, \bibinfo{author}{{Aerts} C},
  \bibinfo{author}{{Aguado} JJ}, \bibinfo{author}{{Ajaj} M},
  \bibinfo{author}{{Altavilla} G}, \bibinfo{author}{{{\'A}lvarez} MA},
  \bibinfo{author}{{{\'A}lvarez Cid-Fuentes} J}, \bibinfo{author}{{Alves} J},
  \bibinfo{author}{{Anderson} RI}, \bibinfo{author}{{Anglada Varela} E},
  \bibinfo{author}{{Antoja} T}, \bibinfo{author}{{Audard} M},
  \bibinfo{author}{{Baines} D}, \bibinfo{author}{{Baker} SG},
  \bibinfo{author}{{Balaguer-N{\'u}{\~n}ez} L}, \bibinfo{author}{{Balbinot} E},
  \bibinfo{author}{{Balog} Z}, \bibinfo{author}{{Barache} C},
  \bibinfo{author}{{Barbato} D}, \bibinfo{author}{{Barros} M},
  \bibinfo{author}{{Barstow} MA}, \bibinfo{author}{{Bartolom{\'e}} S},
  \bibinfo{author}{{Bassilana} JL}, \bibinfo{author}{{Bauchet} N},
  \bibinfo{author}{{Baudesson-Stella} A}, \bibinfo{author}{{Becciani} U},
  \bibinfo{author}{{Bellazzini} M}, \bibinfo{author}{{Bernet} M},
  \bibinfo{author}{{Bertone} S}, \bibinfo{author}{{Bianchi} L},
  \bibinfo{author}{{Blanco-Cuaresma} S}, \bibinfo{author}{{Boch} T},
  \bibinfo{author}{{Bombrun} A}, \bibinfo{author}{{Bossini} D},
  \bibinfo{author}{{Bouquillon} S}, \bibinfo{author}{{Bragaglia} A},
  \bibinfo{author}{{Bramante} L}, \bibinfo{author}{{Breedt} E},
  \bibinfo{author}{{Bressan} A}, \bibinfo{author}{{Brouillet} N},
  \bibinfo{author}{{Bucciarelli} B}, \bibinfo{author}{{Burlacu} A},
  \bibinfo{author}{{Busonero} D}, \bibinfo{author}{{Butkevich} AG},
  \bibinfo{author}{{Buzzi} R}, \bibinfo{author}{{Caffau} E},
  \bibinfo{author}{{Cancelliere} R}, \bibinfo{author}{{C{\'a}novas} H},
  \bibinfo{author}{{Cantat-Gaudin} T}, \bibinfo{author}{{Carballo} R},
  \bibinfo{author}{{Carlucci} T}, \bibinfo{author}{{Carnerero} MI},
  \bibinfo{author}{{Carrasco} JM}, \bibinfo{author}{{Casamiquela} L},
  \bibinfo{author}{{Castellani} M}, \bibinfo{author}{{Castro-Ginard} A},
  \bibinfo{author}{{Castro Sampol} P}, \bibinfo{author}{{Chaoul} L},
  \bibinfo{author}{{Charlot} P}, \bibinfo{author}{{Chemin} L},
  \bibinfo{author}{{Chiavassa} A}, \bibinfo{author}{{Cioni} MRL},
  \bibinfo{author}{{Comoretto} G}, \bibinfo{author}{{Cooper} WJ},
  \bibinfo{author}{{Cornez} T}, \bibinfo{author}{{Cowell} S},
  \bibinfo{author}{{Crifo} F}, \bibinfo{author}{{Crosta} M},
  \bibinfo{author}{{Crowley} C}, \bibinfo{author}{{Dafonte} C},
  \bibinfo{author}{{Dapergolas} A}, \bibinfo{author}{{David} M},
  \bibinfo{author}{{David} P}, \bibinfo{author}{{de Laverny} P},
  \bibinfo{author}{{De Luise} F}, \bibinfo{author}{{De March} R},
  \bibinfo{author}{{De Ridder} J}, \bibinfo{author}{{de Souza} R},
  \bibinfo{author}{{de Teodoro} P}, \bibinfo{author}{{de Torres} A},
  \bibinfo{author}{{del Peloso} EF}, \bibinfo{author}{{del Pozo} E},
  \bibinfo{author}{{Delbo} M}, \bibinfo{author}{{Delgado} A},
  \bibinfo{author}{{Delgado} HE}, \bibinfo{author}{{Delisle} JB},
  \bibinfo{author}{{Di Matteo} P}, \bibinfo{author}{{Diakite} S},
  \bibinfo{author}{{Diener} C}, \bibinfo{author}{{Distefano} E},
  \bibinfo{author}{{Dolding} C}, \bibinfo{author}{{Eappachen} D},
  \bibinfo{author}{{Edvardsson} B}, \bibinfo{author}{{Enke} H},
  \bibinfo{author}{{Esquej} P}, \bibinfo{author}{{Fabre} C},
  \bibinfo{author}{{Fabrizio} M}, \bibinfo{author}{{Faigler} S},
  \bibinfo{author}{{Fedorets} G}, \bibinfo{author}{{Fernique} P},
  \bibinfo{author}{{Fienga} A}, \bibinfo{author}{{Figueras} F},
  \bibinfo{author}{{Fouron} C}, \bibinfo{author}{{Fragkoudi} F},
  \bibinfo{author}{{Fraile} E}, \bibinfo{author}{{Franke} F},
  \bibinfo{author}{{Gai} M}, \bibinfo{author}{{Garabato} D},
  \bibinfo{author}{{Garcia-Gutierrez} A}, \bibinfo{author}{{Garc{\'\i}a-Torres}
  M}, \bibinfo{author}{{Garofalo} A}, \bibinfo{author}{{Gavras} P},
  \bibinfo{author}{{Gerlach} E}, \bibinfo{author}{{Geyer} R},
  \bibinfo{author}{{Giacobbe} P}, \bibinfo{author}{{Gilmore} G},
  \bibinfo{author}{{Girona} S}, \bibinfo{author}{{Giuffrida} G},
  \bibinfo{author}{{Gomel} R}, \bibinfo{author}{{Gomez} A},
  \bibinfo{author}{{Gonzalez-Santamaria} I},
  \bibinfo{author}{{Gonz{\'a}lez-Vidal} JJ}, \bibinfo{author}{{Granvik} M},
  \bibinfo{author}{{Guti{\'e}rrez-S{\'a}nchez} R}, \bibinfo{author}{{Guy} LP},
  \bibinfo{author}{{Hauser} M}, \bibinfo{author}{{Haywood} M},
  \bibinfo{author}{{Helmi} A}, \bibinfo{author}{{Hidalgo} SL},
  \bibinfo{author}{{Hilger} T}, \bibinfo{author}{{H{\l}adczuk} N},
  \bibinfo{author}{{Hobbs} D}, \bibinfo{author}{{Holland} G},
  \bibinfo{author}{{Huckle} HE}, \bibinfo{author}{{Jasniewicz} G},
  \bibinfo{author}{{Jonker} PG}, \bibinfo{author}{{Juaristi Campillo} J},
  \bibinfo{author}{{Julbe} F}, \bibinfo{author}{{Karbevska} L},
  \bibinfo{author}{{Kervella} P}, \bibinfo{author}{{Khanna} S},
  \bibinfo{author}{{Kochoska} A}, \bibinfo{author}{{Kontizas} M},
  \bibinfo{author}{{Kordopatis} G}, \bibinfo{author}{{Korn} AJ},
  \bibinfo{author}{{Kostrzewa-Rutkowska} Z}, \bibinfo{author}{{Kruszy{\'n}ska}
  K}, \bibinfo{author}{{Lambert} S}, \bibinfo{author}{{Lanza} AF},
  \bibinfo{author}{{Lasne} Y}, \bibinfo{author}{{Le Campion} JF},
  \bibinfo{author}{{Le Fustec} Y}, \bibinfo{author}{{Lebreton} Y},
  \bibinfo{author}{{Lebzelter} T}, \bibinfo{author}{{Leccia} S},
  \bibinfo{author}{{Leclerc} N}, \bibinfo{author}{{Lecoeur-Taibi} I},
  \bibinfo{author}{{Liao} S}, \bibinfo{author}{{Licata} E},
  \bibinfo{author}{{Lindstr{\o}m} EP}, \bibinfo{author}{{Lister} TA},
  \bibinfo{author}{{Livanou} E}, \bibinfo{author}{{Lobel} A},
  \bibinfo{author}{{Madrero Pardo} P}, \bibinfo{author}{{Managau} S},
  \bibinfo{author}{{Mann} RG}, \bibinfo{author}{{Marchant} JM},
  \bibinfo{author}{{Marconi} M}, \bibinfo{author}{{Marcos Santos} MMS},
  \bibinfo{author}{{Marinoni} S}, \bibinfo{author}{{Marocco} F},
  \bibinfo{author}{{Marshall} DJ}, \bibinfo{author}{{Martin Polo} L},
  \bibinfo{author}{{Mart{\'\i}n-Fleitas} JM}, \bibinfo{author}{{Masip} A},
  \bibinfo{author}{{Massari} D}, \bibinfo{author}{{Mastrobuono-Battisti} A},
  \bibinfo{author}{{Mazeh} T}, \bibinfo{author}{{McMillan} PJ},
  \bibinfo{author}{{Messina} S}, \bibinfo{author}{{Michalik} D},
  \bibinfo{author}{{Millar} NR}, \bibinfo{author}{{Mints} A},
  \bibinfo{author}{{Molina} D}, \bibinfo{author}{{Molinaro} R},
  \bibinfo{author}{{Moln{\'a}r} L}, \bibinfo{author}{{Montegriffo} P},
  \bibinfo{author}{{Mor} R}, \bibinfo{author}{{Morbidelli} R},
  \bibinfo{author}{{Morel} T}, \bibinfo{author}{{Morris} D},
  \bibinfo{author}{{Mulone} AF}, \bibinfo{author}{{Munoz} D},
  \bibinfo{author}{{Muraveva} T}, \bibinfo{author}{{Murphy} CP},
  \bibinfo{author}{{Musella} I}, \bibinfo{author}{{Noval} L},
  \bibinfo{author}{{Ord{\'e}novic} C}, \bibinfo{author}{{Orr{\`u}} G},
  \bibinfo{author}{{Osinde} J}, \bibinfo{author}{{Pagani} C},
  \bibinfo{author}{{Pagano} I}, \bibinfo{author}{{Palaversa} L},
  \bibinfo{author}{{Palicio} PA}, \bibinfo{author}{{Panahi} A},
  \bibinfo{author}{{Pawlak} M}, \bibinfo{author}{{Pe{\~n}alosa Esteller} X},
  \bibinfo{author}{{Penttil{\"a}} A}, \bibinfo{author}{{Piersimoni} AM},
  \bibinfo{author}{{Pineau} FX}, \bibinfo{author}{{Plachy} E},
  \bibinfo{author}{{Plum} G}, \bibinfo{author}{{Poggio} E},
  \bibinfo{author}{{Poretti} E}, \bibinfo{author}{{Poujoulet} E},
  \bibinfo{author}{{Pr{\v{s}}a} A}, \bibinfo{author}{{Pulone} L},
  \bibinfo{author}{{Racero} E}, \bibinfo{author}{{Ragaini} S},
  \bibinfo{author}{{Rainer} M}, \bibinfo{author}{{Raiteri} CM},
  \bibinfo{author}{{Rambaux} N}, \bibinfo{author}{{Ramos} P},
  \bibinfo{author}{{Ramos-Lerate} M}, \bibinfo{author}{{Re Fiorentin} P},
  \bibinfo{author}{{Regibo} S}, \bibinfo{author}{{Reyl{\'e}} C},
  \bibinfo{author}{{Ripepi} V}, \bibinfo{author}{{Riva} A},
  \bibinfo{author}{{Rixon} G}, \bibinfo{author}{{Robichon} N},
  \bibinfo{author}{{Robin} C}, \bibinfo{author}{{Roelens} M},
  \bibinfo{author}{{Rohrbasser} L}, \bibinfo{author}{{Romero-G{\'o}mez} M},
  \bibinfo{author}{{Rowell} N}, \bibinfo{author}{{Royer} F},
  \bibinfo{author}{{Rybicki} KA}, \bibinfo{author}{{Sadowski} G},
  \bibinfo{author}{{Sagrist{\`a} Sell{\'e}s} A}, \bibinfo{author}{{Sahlmann}
  J}, \bibinfo{author}{{Salgado} J}, \bibinfo{author}{{Salguero} E},
  \bibinfo{author}{{Samaras} N}, \bibinfo{author}{{Sanchez Gimenez} V},
  \bibinfo{author}{{Sanna} N}, \bibinfo{author}{{Santove{\~n}a} R},
  \bibinfo{author}{{Sarasso} M}, \bibinfo{author}{{Schultheis} M},
  \bibinfo{author}{{Sciacca} E}, \bibinfo{author}{{Segol} M},
  \bibinfo{author}{{Segovia} JC}, \bibinfo{author}{{S{\'e}gransan} D},
  \bibinfo{author}{{Semeux} D}, \bibinfo{author}{{Shahaf} S},
  \bibinfo{author}{{Siddiqui} HI}, \bibinfo{author}{{Siebert} A},
  \bibinfo{author}{{Siltala} L}, \bibinfo{author}{{Slezak} E},
  \bibinfo{author}{{Smart} RL}, \bibinfo{author}{{Solano} E},
  \bibinfo{author}{{Solitro} F}, \bibinfo{author}{{Souami} D},
  \bibinfo{author}{{Souchay} J}, \bibinfo{author}{{Spagna} A},
  \bibinfo{author}{{Spoto} F}, \bibinfo{author}{{Steele} IA},
  \bibinfo{author}{{Steidelm{\"u}ller} H}, \bibinfo{author}{{Stephenson} CA},
  \bibinfo{author}{{S{\"u}veges} M}, \bibinfo{author}{{Szabados} L},
  \bibinfo{author}{{Szegedi-Elek} E}, \bibinfo{author}{{Taris} F},
  \bibinfo{author}{{Tauran} G}, \bibinfo{author}{{Taylor} MB},
  \bibinfo{author}{{Teixeira} R}, \bibinfo{author}{{Thuillot} W},
  \bibinfo{author}{{Tonello} N}, \bibinfo{author}{{Torra} F},
  \bibinfo{author}{{Torra} J}, \bibinfo{author}{{Turon} C},
  \bibinfo{author}{{Unger} N}, \bibinfo{author}{{Vaillant} M},
  \bibinfo{author}{{van Dillen} E}, \bibinfo{author}{{Vanel} O},
  \bibinfo{author}{{Vecchiato} A}, \bibinfo{author}{{Viala} Y},
  \bibinfo{author}{{Vicente} D}, \bibinfo{author}{{Voutsinas} S},
  \bibinfo{author}{{Weiler} M}, \bibinfo{author}{{Wevers} T},
  \bibinfo{author}{{Wyrzykowski} {\L}}, \bibinfo{author}{{Yoldas} A},
  \bibinfo{author}{{Yvard} P}, \bibinfo{author}{{Zhao} H},
  \bibinfo{author}{{Zorec} J}, \bibinfo{author}{{Zucker} S},
  \bibinfo{author}{{Zurbach} C} and  \bibinfo{author}{{Zwitter} T}
  (\bibinfo{year}{2021}), \bibinfo{month}{May}.
\bibinfo{title}{{Gaia Early Data Release 3. Summary of the contents and survey
  properties}}.
\bibinfo{journal}{{\em A\&A}} \bibinfo{volume}{649}, \bibinfo{eid}{A1}.
  \bibinfo{doi}{\doi{10.1051/0004-6361/202039657}}.
\eprint{2012.01533}.

\bibtype{Article}%
\bibitem[{Gallagher} et al.(2017)]{gallagher17}
\bibinfo{author}{{Gallagher} AJ}, \bibinfo{author}{{Caffau} E},
  \bibinfo{author}{{Bonifacio} P}, \bibinfo{author}{{Ludwig} HG},
  \bibinfo{author}{{Steffen} M}, \bibinfo{author}{{Homeier} D} and
  \bibinfo{author}{{Plez} B} (\bibinfo{year}{2017}), \bibinfo{month}{Feb.}
\bibinfo{title}{{An in-depth spectroscopic examination of molecular bands from
  3D hydrodynamical model atmospheres. II. Carbon-enhanced metal-poor 3D model
  atmospheres}}.
\bibinfo{journal}{{\em A\&A}} \bibinfo{volume}{598}, \bibinfo{eid}{L10}.
  \bibinfo{doi}{\doi{10.1051/0004-6361/201630272}}.
\eprint{1701.09102}.

\bibtype{Article}%
\bibitem[{Gallino} et al.(1998)]{gallino1998}
\bibinfo{author}{{Gallino} R}, \bibinfo{author}{{Arlandini} C},
  \bibinfo{author}{{Busso} M}, \bibinfo{author}{{Lugaro} M},
  \bibinfo{author}{{Travaglio} C}, \bibinfo{author}{{Straniero} O},
  \bibinfo{author}{{Chieffi} A} and  \bibinfo{author}{{Limongi} M}
  (\bibinfo{year}{1998}), \bibinfo{month}{Apr.}
\bibinfo{title}{{Evolution and Nucleosynthesis in Low-Mass Asymptotic Giant
  Branch Stars. II. Neutron Capture and the s-Process}}.
\bibinfo{journal}{{\em ApJ}} \bibinfo{volume}{497}: \bibinfo{pages}{388}.
  \bibinfo{doi}{\doi{10.1086/305437}}.

\bibtype{Article}%
\bibitem[{Gratton} et al.(2000)]{gratton00}
\bibinfo{author}{{Gratton} RG}, \bibinfo{author}{{Sneden} C},
  \bibinfo{author}{{Carretta} E} and  \bibinfo{author}{{Bragaglia} A}
  (\bibinfo{year}{2000}), \bibinfo{month}{Feb.}
\bibinfo{title}{{Mixing along the red giant branch in metal-poor field stars}}.
\bibinfo{journal}{{\em A\&A}} \bibinfo{volume}{354}: \bibinfo{pages}{169--187}.

\bibtype{Article}%
\bibitem[{Gratton} et al.(2001)]{gratton01}
\bibinfo{author}{{Gratton} RG}, \bibinfo{author}{{Bonifacio} P},
  \bibinfo{author}{{Bragaglia} A}, \bibinfo{author}{{Carretta} E},
  \bibinfo{author}{{Castellani} V}, \bibinfo{author}{{Centurion} M},
  \bibinfo{author}{{Chieffi} A}, \bibinfo{author}{{Claudi} R},
  \bibinfo{author}{{Clementini} G}, \bibinfo{author}{{D'Antona} F},
  \bibinfo{author}{{Desidera} S}, \bibinfo{author}{{Fran{\c{c}}ois} P},
  \bibinfo{author}{{Grundahl} F}, \bibinfo{author}{{Lucatello} S},
  \bibinfo{author}{{Molaro} P}, \bibinfo{author}{{Pasquini} L},
  \bibinfo{author}{{Sneden} C}, \bibinfo{author}{{Spite} F} and
  \bibinfo{author}{{Straniero} O} (\bibinfo{year}{2001}), \bibinfo{month}{Apr.}
\bibinfo{title}{{The O-Na and Mg-Al anticorrelations in turn-off and early
  subgiants in globular clusters}}.
\bibinfo{journal}{{\em A\&A}} \bibinfo{volume}{369}: \bibinfo{pages}{87--98}.
  \bibinfo{doi}{\doi{10.1051/0004-6361:20010144}}.
\eprint{astro-ph/0012457}.

\bibtype{Book}%
\bibitem[{Gray}(2005)]{Gray05}
\bibinfo{author}{{Gray} DF} (\bibinfo{year}{2005}), \bibinfo{month}{Sep.}
\bibinfo{title}{{The Observation and Analysis of Stellar Photospheres, 3rd
  Edition}}, \bibinfo{publisher}{Cambridge: Cambridge University Press}.

\bibtype{Article}%
\bibitem[{Gudin} et al.(2021)]{gudin21}
\bibinfo{author}{{Gudin} D}, \bibinfo{author}{{Shank} D},
  \bibinfo{author}{{Beers} TC}, \bibinfo{author}{{Yuan} Z},
  \bibinfo{author}{{Limberg} G}, \bibinfo{author}{{Roederer} IU},
  \bibinfo{author}{{Placco} V}, \bibinfo{author}{{Holmbeck} EM},
  \bibinfo{author}{{Dietz} S}, \bibinfo{author}{{Rasmussen} KC},
  \bibinfo{author}{{Hansen} TT}, \bibinfo{author}{{Sakari} CM},
  \bibinfo{author}{{Ezzeddine} R} and  \bibinfo{author}{{Frebel} A}
  (\bibinfo{year}{2021}), \bibinfo{month}{Feb.}
\bibinfo{title}{{The R-Process Alliance: Chemodynamically Tagged Groups of Halo
  r-process-enhanced Stars Reveal a Shared Chemical-evolution History}}.
\bibinfo{journal}{{\em ApJ}} \bibinfo{volume}{908} (\bibinfo{number}{1}),
  \bibinfo{eid}{79}. \bibinfo{doi}{\doi{10.3847/1538-4357/abd7ed}}.

\bibtype{Article}%
\bibitem[{Gull} et al.(2018)]{gull18}
\bibinfo{author}{{Gull} M}, \bibinfo{author}{{Frebel} A},
  \bibinfo{author}{{Cain} MG}, \bibinfo{author}{{Placco} VM},
  \bibinfo{author}{{Ji} AP}, \bibinfo{author}{{Abate} C},
  \bibinfo{author}{{Ezzeddine} R}, \bibinfo{author}{{Karakas} AI},
  \bibinfo{author}{{Hansen} TT}, \bibinfo{author}{{Sakari} C},
  \bibinfo{author}{{Holmbeck} EM}, \bibinfo{author}{{Santucci} RM},
  \bibinfo{author}{{Casey} AR} and  \bibinfo{author}{{Beers} TC}
  (\bibinfo{year}{2018}), \bibinfo{month}{Aug.}
\bibinfo{title}{{The R-Process Alliance: Discovery of the First Metal-poor Star
  with a Combined r- and s-process Element Signature}}.
\bibinfo{journal}{{\em ApJ}} \bibinfo{volume}{862} (\bibinfo{number}{2}),
  \bibinfo{eid}{174}. \bibinfo{doi}{\doi{10.3847/1538-4357/aacbc3}}.

\bibtype{Article}%
\bibitem[{Gustafsson} et al.(2008)]{Gustafsson08}
\bibinfo{author}{{Gustafsson} B}, \bibinfo{author}{{Edvardsson} B},
  \bibinfo{author}{{Eriksson} K}, \bibinfo{author}{{J{\o}rgensen} UG},
  \bibinfo{author}{{Nordlund} {\AA}} and  \bibinfo{author}{{Plez} B}
  (\bibinfo{year}{2008}), \bibinfo{month}{Aug.}
\bibinfo{title}{{A grid of MARCS model atmospheres for late-type stars. I.
  Methods and general properties}}.
\bibinfo{journal}{{\em A\&A}} \bibinfo{volume}{486}: \bibinfo{pages}{951--970}.
  \bibinfo{doi}{\doi{10.1051/0004-6361:200809724}}.
\eprint{0805.0554}.

\bibtype{Article}%
\bibitem[{Hampel} et al.(2016)]{hampel16}
\bibinfo{author}{{Hampel} M}, \bibinfo{author}{{Stancliffe} RJ},
  \bibinfo{author}{{Lugaro} M} and  \bibinfo{author}{{Meyer} BS}
  (\bibinfo{year}{2016}), \bibinfo{month}{Nov.}
\bibinfo{title}{{The Intermediate Neutron-capture Process and Carbon-enhanced
  Metal-poor Stars}}.
\bibinfo{journal}{{\em ApJ}} \bibinfo{volume}{831}, \bibinfo{eid}{171}.
  \bibinfo{doi}{\doi{10.3847/0004-637X/831/2/171}}.
\eprint{1608.08634}.

\bibtype{Article}%
\bibitem[{Hansen} et al.(2016)]{hansen16}
\bibinfo{author}{{Hansen} TT}, \bibinfo{author}{{Andersen} J},
  \bibinfo{author}{{Nordstr{\"o}m} B}, \bibinfo{author}{{Beers} TC},
  \bibinfo{author}{{Placco} VM}, \bibinfo{author}{{Yoon} J} and
  \bibinfo{author}{{Buchhave} LA} (\bibinfo{year}{2016}), \bibinfo{month}{Feb.}
\bibinfo{title}{{The role of binaries in the enrichment of the early Galactic
  halo. II. Carbon-enhanced metal-poor stars: CEMP-no stars}}.
\bibinfo{journal}{{\em A\&A}} \bibinfo{volume}{586}, \bibinfo{eid}{A160}.
  \bibinfo{doi}{\doi{10.1051/0004-6361/201527235}}.
\eprint{1511.08197}.

\bibtype{Article}%
\bibitem[{Hansen} et al.(2018)]{hansen18}
\bibinfo{author}{{Hansen} TT}, \bibinfo{author}{{Holmbeck} EM},
  \bibinfo{author}{{Beers} TC}, \bibinfo{author}{{Placco} VM},
  \bibinfo{author}{{Roederer} IU}, \bibinfo{author}{{Frebel} A},
  \bibinfo{author}{{Sakari} CM}, \bibinfo{author}{{Simon} JD} and
  \bibinfo{author}{{Thompson} IB} (\bibinfo{year}{2018}), \bibinfo{month}{Apr.}
\bibinfo{title}{{The $R$-Process Alliance: First Release from the Southern
  Search for $r$-Process-Enhanced Stars in the Galactic Halo}}.
\bibinfo{journal}{{\em ArXiv e-prints}} \eprint{1804.03114}.

\bibtype{Article}%
\bibitem[{Hill} et al.(2019)]{hill19}
\bibinfo{author}{{Hill} V}, \bibinfo{author}{{Sk{\'u}lad{\'o}ttir} {\'A}},
  \bibinfo{author}{{Tolstoy} E}, \bibinfo{author}{{Venn} KA},
  \bibinfo{author}{{Shetrone} MD}, \bibinfo{author}{{Jablonka} P},
  \bibinfo{author}{{Primas} F}, \bibinfo{author}{{Battaglia} G},
  \bibinfo{author}{{de Boer} TJL}, \bibinfo{author}{{Fran{\c{c}}ois} P},
  \bibinfo{author}{{Helmi} A}, \bibinfo{author}{{Kaufer} A},
  \bibinfo{author}{{Letarte} B}, \bibinfo{author}{{Starkenburg} E} and
  \bibinfo{author}{{Spite} M} (\bibinfo{year}{2019}), \bibinfo{month}{Jun.}
\bibinfo{title}{{VLT/FLAMES high-resolution chemical abundances in Sculptor: a
  textbook dwarf spheroidal galaxy}}.
\bibinfo{journal}{{\em A\&A}} \bibinfo{volume}{626}, \bibinfo{eid}{A15}.
  \bibinfo{doi}{\doi{10.1051/0004-6361/201833950}}.

\bibtype{Article}%
\bibitem[{Holmbeck} et al.(2020)]{holmbeck20}
\bibinfo{author}{{Holmbeck} EM}, \bibinfo{author}{{Hansen} TT},
  \bibinfo{author}{{Beers} TC}, \bibinfo{author}{{Placco} VM},
  \bibinfo{author}{{Whitten} DD}, \bibinfo{author}{{Rasmussen} KC},
  \bibinfo{author}{{Roederer} IU}, \bibinfo{author}{{Ezzeddine} R},
  \bibinfo{author}{{Sakari} CM}, \bibinfo{author}{{Frebel} A},
  \bibinfo{author}{{Drout} MR}, \bibinfo{author}{{Simon} JD},
  \bibinfo{author}{{Thompson} IB}, \bibinfo{author}{{Bland-Hawthorn} J},
  \bibinfo{author}{{Gibson} BK}, \bibinfo{author}{{Grebel} EK},
  \bibinfo{author}{{Kordopatis} G}, \bibinfo{author}{{Kunder} A},
  \bibinfo{author}{{Mel{\'e}ndez} J}, \bibinfo{author}{{Navarro} JF},
  \bibinfo{author}{{Reid} WA}, \bibinfo{author}{{Seabroke} G},
  \bibinfo{author}{{Steinmetz} M}, \bibinfo{author}{{Watson} F} and
  \bibinfo{author}{{Wyse} RF{\.{G}}} (\bibinfo{year}{2020}),
  \bibinfo{month}{Aug.}
\bibinfo{title}{{The R-Process Alliance: Fourth Data Release from the Search
  for R-process-enhanced Stars in the Galactic Halo}}.
\bibinfo{journal}{{\em ApJs}} \bibinfo{volume}{249} (\bibinfo{number}{2}),
  \bibinfo{eid}{30}. \bibinfo{doi}{\doi{10.3847/1538-4365/ab9c19}}.

\bibtype{Article}%
\bibitem[{Honda} et al.(2006)]{honda06}
\bibinfo{author}{{Honda} S}, \bibinfo{author}{{Aoki} W},
  \bibinfo{author}{{Ishimaru} Y}, \bibinfo{author}{{Wanajo} S} and
  \bibinfo{author}{{Ryan} SG} (\bibinfo{year}{2006}), \bibinfo{month}{Jun.}
\bibinfo{title}{{Neutron-Capture Elements in the Very Metal Poor Star HD
  122563}}.
\bibinfo{journal}{{\em ApJ}} \bibinfo{volume}{643}:
  \bibinfo{pages}{1180--1189}. \bibinfo{doi}{\doi{10.1086/503195}}.
\eprint{arXiv:astro-ph/0602107}.

\bibtype{Article}%
\bibitem[{Howes} et al.(2014)]{howes14}
\bibinfo{author}{{Howes} L}, \bibinfo{author}{{Asplund} M},
  \bibinfo{author}{{Casey} AR}, \bibinfo{author}{{Keller} SC},
  \bibinfo{author}{{Yong} D}, \bibinfo{author}{{Gilmore} G},
  \bibinfo{author}{{Lind} K}, \bibinfo{author}{{Worley} C},
  \bibinfo{author}{{Bessell} MS}, \bibinfo{author}{{Casagrande} L},
  \bibinfo{author}{{Marino} AF}, \bibinfo{author}{{Nataf} DM},
  \bibinfo{author}{{Owen} CI}, \bibinfo{author}{{Da Costa} GS},
  \bibinfo{author}{{Schmidt} BP}, \bibinfo{author}{{Tisserand} P},
  \bibinfo{author}{{Randich} S}, \bibinfo{author}{{Feltzing} S},
  \bibinfo{author}{{Vallenari} A}, \bibinfo{author}{{Allende Prieto} C},
  \bibinfo{author}{{Bensby} T}, \bibinfo{author}{{Flaccomio} E},
  \bibinfo{author}{{Korn} AJ}, \bibinfo{author}{{Pancino} E},
  \bibinfo{author}{{Recio-Blanco} A}, \bibinfo{author}{{Smiljanic} R},
  \bibinfo{author}{{Bergemann} M}, \bibinfo{author}{{Costado} MT},
  \bibinfo{author}{{Damiani} F}, \bibinfo{author}{{Heiter} U},
  \bibinfo{author}{{Hill} V}, \bibinfo{author}{{Hourihane} A},
  \bibinfo{author}{{Jofr{\'e}} P}, \bibinfo{author}{{Lardo} C},
  \bibinfo{author}{{de Laverny} P}, \bibinfo{author}{{Magrini} L},
  \bibinfo{author}{{Maiorca} E}, \bibinfo{author}{{Masseron} T},
  \bibinfo{author}{{Morbidelli} L}, \bibinfo{author}{{Sacco} GG},
  \bibinfo{author}{{Minniti} D} and  \bibinfo{author}{{Zoccali} M}
  (\bibinfo{year}{2014}), \bibinfo{month}{Sep.}
\bibinfo{title}{{The Gaia-ESO Survey: the most metal-poor stars in the Galactic
  bulge}}.
\bibinfo{journal}{{\em ArXiv:}} \bibinfo{volume}{1409.7952}.
\eprint{1409.7952}.

\bibtype{Article}%
\bibitem[{Jacobson} et al.(2015)]{jacobson15}
\bibinfo{author}{{Jacobson} HR}, \bibinfo{author}{{Keller} S},
  \bibinfo{author}{{Frebel} A}, \bibinfo{author}{{Casey} AR},
  \bibinfo{author}{{Asplund} M}, \bibinfo{author}{{Bessell} MS},
  \bibinfo{author}{{Da Costa} GS}, \bibinfo{author}{{Lind} K},
  \bibinfo{author}{{Marino} AF}, \bibinfo{author}{{Norris} JE},
  \bibinfo{author}{{Pena} JM}, \bibinfo{author}{{Schmidt} BP},
  \bibinfo{author}{{Tisserand} P}, \bibinfo{author}{{Walsh} JM},
  \bibinfo{author}{{Yong} D} and  \bibinfo{author}{{Yu} Q}
  (\bibinfo{year}{2015}), \bibinfo{month}{Jul.}
\bibinfo{title}{{High-Resolution Spectroscopic Study of Extremely Metal-Poor
  Star Candidates from the SkyMapper Survey}}.
\bibinfo{journal}{{\em ApJ}} \bibinfo{volume}{807}, \bibinfo{eid}{171}.
  \bibinfo{doi}{\doi{10.1088/0004-637X/807/2/171}}.
\eprint{1504.03344}.

\bibtype{Article}%
\bibitem[{Ji} et al.(2016{\natexlab{a}})]{ji16a}
\bibinfo{author}{{Ji} AP}, \bibinfo{author}{{Frebel} A},
  \bibinfo{author}{{Chiti} A} and  \bibinfo{author}{{Simon} JD}
  (\bibinfo{year}{2016}{\natexlab{a}}), \bibinfo{month}{Mar.}
\bibinfo{title}{{R-process enrichment from a single event in an ancient dwarf
  galaxy}}.
\bibinfo{journal}{{\em Nature}} \bibinfo{volume}{531}:
  \bibinfo{pages}{610--613}. \bibinfo{doi}{\doi{10.1038/nature17425}}.
\eprint{1512.01558}.

\bibtype{Article}%
\bibitem[{Ji} et al.(2016{\natexlab{b}})]{ji16b}
\bibinfo{author}{{Ji} AP}, \bibinfo{author}{{Frebel} A},
  \bibinfo{author}{{Simon} JD} and  \bibinfo{author}{{Chiti} A}
  (\bibinfo{year}{2016}{\natexlab{b}}), \bibinfo{month}{Oct.}
\bibinfo{title}{{Complete Element Abundances of Nine Stars in the r-process
  Galaxy Reticulum II}}.
\bibinfo{journal}{{\em ApJ}} \bibinfo{volume}{830}, \bibinfo{eid}{93}.
  \bibinfo{doi}{\doi{10.3847/0004-637X/830/2/93}}.
\eprint{1607.07447}.

\bibtype{Article}%
\bibitem[{Ji} et al.(2020)]{ji20_car}
\bibinfo{author}{{Ji} AP}, \bibinfo{author}{{Li} TS}, \bibinfo{author}{{Simon}
  JD}, \bibinfo{author}{{Marshall} J}, \bibinfo{author}{{Vivas} AK},
  \bibinfo{author}{{Pace} AB}, \bibinfo{author}{{Bechtol} K},
  \bibinfo{author}{{Drlica-Wagner} A}, \bibinfo{author}{{Koposov} SE},
  \bibinfo{author}{{Hansen} TT}, \bibinfo{author}{{Allam} S},
  \bibinfo{author}{{Gruendl} RA}, \bibinfo{author}{{Johnson} MD},
  \bibinfo{author}{{McNanna} M}, \bibinfo{author}{{No{\"e}l} NED},
  \bibinfo{author}{{Tucker} DL} and  \bibinfo{author}{{Walker} AR}
  (\bibinfo{year}{2020}), \bibinfo{month}{Jan.}
\bibinfo{title}{{Detailed Abundances in the Ultra-faint Magellanic Satellites
  Carina II and III}}.
\bibinfo{journal}{{\em ApJ}} \bibinfo{volume}{889} (\bibinfo{number}{1}),
  \bibinfo{eid}{27}. \bibinfo{doi}{\doi{10.3847/1538-4357/ab6213}}.
\eprint{1912.04963}.

\bibtype{Article}%
\bibitem[{Keenan}(1942)]{keenan42}
\bibinfo{author}{{Keenan} PC} (\bibinfo{year}{1942}), \bibinfo{month}{Jul.}
\bibinfo{title}{{The Spectra of CH Stars}}.
\bibinfo{journal}{{\em ApJ}} \bibinfo{volume}{96}: \bibinfo{pages}{101}.
  \bibinfo{doi}{\doi{10.1086/144435}}.

\bibtype{Article}%
\bibitem[{Keller} et al.(2014)]{keller14}
\bibinfo{author}{{Keller} SC}, \bibinfo{author}{{Bessell} MS},
  \bibinfo{author}{{Frebel} A}, \bibinfo{author}{{Casey} AR},
  \bibinfo{author}{{Asplund} M}, \bibinfo{author}{{Jacobson} HR},
  \bibinfo{author}{{Lind} K}, \bibinfo{author}{{Norris} JE},
  \bibinfo{author}{{Yong} D}, \bibinfo{author}{{Heger} A},
  \bibinfo{author}{{Magic} Z}, \bibinfo{author}{{da Costa} GS},
  \bibinfo{author}{{Schmidt} BP} and  \bibinfo{author}{{Tisserand} P}
  (\bibinfo{year}{2014}), \bibinfo{month}{Feb.}
\bibinfo{title}{{A single low-energy, iron-poor supernova as the source of
  metals in the star SMSS J031300.36-670839.3}}.
\bibinfo{journal}{{\em Nature}} \bibinfo{volume}{506}:
  \bibinfo{pages}{463--466}. \bibinfo{doi}{\doi{10.1038/nature12990}}.
\eprint{1402.1517}.

\bibtype{Article}%
\bibitem[{Kilpatrick} et al.(2017)]{kilpatrick17}
\bibinfo{author}{{Kilpatrick} CD}, \bibinfo{author}{{Foley} RJ},
  \bibinfo{author}{{Kasen} D}, \bibinfo{author}{{Murguia-Berthier} A},
  \bibinfo{author}{{Ramirez-Ruiz} E}, \bibinfo{author}{{Coulter} DA},
  \bibinfo{author}{{Drout} MR}, \bibinfo{author}{{Piro} AL},
  \bibinfo{author}{{Shappee} BJ}, \bibinfo{author}{{Boutsia} K},
  \bibinfo{author}{{Contreras} C}, \bibinfo{author}{{Di Mille} F},
  \bibinfo{author}{{Madore} BF}, \bibinfo{author}{{Morrell} N},
  \bibinfo{author}{{Pan} YC}, \bibinfo{author}{{Prochaska} JX},
  \bibinfo{author}{{Rest} A}, \bibinfo{author}{{Rojas-Bravo} C},
  \bibinfo{author}{{Siebert} MR}, \bibinfo{author}{{Simon} JD} and
  \bibinfo{author}{{Ulloa} N} (\bibinfo{year}{2017}), \bibinfo{month}{Dec.}
\bibinfo{title}{{Electromagnetic evidence that SSS17a is the result of a binary
  neutron star merger}}.
\bibinfo{journal}{{\em Science}} \bibinfo{volume}{358}:
  \bibinfo{pages}{1583--1587}. \bibinfo{doi}{\doi{10.1126/science.aaq0073}}.
\eprint{1710.05434}.

\bibtype{Article}%
\bibitem[{Kirby} et al.(2013)]{kirby13}
\bibinfo{author}{{Kirby} EN}, \bibinfo{author}{{Cohen} JG},
  \bibinfo{author}{{Guhathakurta} P}, \bibinfo{author}{{Cheng} L},
  \bibinfo{author}{{Bullock} JS} and  \bibinfo{author}{{Gallazzi} A}
  (\bibinfo{year}{2013}), \bibinfo{month}{Oct.}
\bibinfo{title}{{The Universal Stellar Mass-Stellar Metallicity Relation for
  Dwarf Galaxies}}.
\bibinfo{journal}{{\em ArXiv e-prints}} \eprint{1310.0814}.

\bibtype{Article}%
\bibitem[{Kirby} et al.(2020)]{kirby20}
\bibinfo{author}{{Kirby} EN}, \bibinfo{author}{{Duggan} G},
  \bibinfo{author}{{Ramirez-Ruiz} E} and  \bibinfo{author}{{Macias} P}
  (\bibinfo{year}{2020}), \bibinfo{month}{Mar.}
\bibinfo{title}{{The Stars in M15 Were Born with the r-process}}.
\bibinfo{journal}{{\em ApJL}} \bibinfo{volume}{891} (\bibinfo{number}{1}),
  \bibinfo{eid}{L13}. \bibinfo{doi}{\doi{10.3847/2041-8213/ab78a1}}.
\eprint{2002.09495}.

\bibtype{Article}%
\bibitem[{Kirby} et al.(2023)]{kirby23}
\bibinfo{author}{{Kirby} EN}, \bibinfo{author}{{Ji} AP} and
  \bibinfo{author}{{Kovalev} M} (\bibinfo{year}{2023}), \bibinfo{month}{Nov.}
\bibinfo{title}{{r-process Abundance Patterns in the Globular Cluster M92}}.
\bibinfo{journal}{{\em ApJ}} \bibinfo{volume}{958} (\bibinfo{number}{1}),
  \bibinfo{eid}{45}. \bibinfo{doi}{\doi{10.3847/1538-4357/acf309}}.

\bibtype{Article}%
\bibitem[{Klessen} and {Glover}(2023)]{klessen23}
\bibinfo{author}{{Klessen} RS} and  \bibinfo{author}{{Glover} SCO}
  (\bibinfo{year}{2023}), \bibinfo{month}{Aug.}
\bibinfo{title}{{The First Stars: Formation, Properties, and Impact}}.
\bibinfo{journal}{{\em ARA\&A}} \bibinfo{volume}{61}: \bibinfo{pages}{65--130}.
  \bibinfo{doi}{\doi{10.1146/annurev-astro-071221-053453}}.
\eprint{2303.12500}.

\bibtype{Article}%
\bibitem[{Kobayashi} et al.(2020)]{kobayashi20}
\bibinfo{author}{{Kobayashi} C}, \bibinfo{author}{{Karakas} AI} and
  \bibinfo{author}{{Lugaro} M} (\bibinfo{year}{2020}), \bibinfo{month}{Sep.}
\bibinfo{title}{{The Origin of Elements from Carbon to Uranium}}.
\bibinfo{journal}{{\em ApJ}} \bibinfo{volume}{900} (\bibinfo{number}{2}),
  \bibinfo{eid}{179}. \bibinfo{doi}{\doi{10.3847/1538-4357/abae65}}.

\bibtype{Article}%
\bibitem[{Koch} et al.(2004)]{koch04}
\bibinfo{author}{{Koch} A}, \bibinfo{author}{{Grebel} EK},
  \bibinfo{author}{{Odenkirchen} M}, \bibinfo{author}{{Mart{\'\i}nez-Delgado}
  D} and  \bibinfo{author}{{Caldwell} JAR} (\bibinfo{year}{2004}),
  \bibinfo{month}{Nov.}
\bibinfo{title}{{Mass Segregation in the Globular Cluster Palomar 5 and its
  Tidal Tails}}.
\bibinfo{journal}{{\em AJ}} \bibinfo{volume}{128} (\bibinfo{number}{5}):
  \bibinfo{pages}{2274--2287}. \bibinfo{doi}{\doi{10.1086/425046}}.
\eprint{astro-ph/0408208}.

\bibtype{Article}%
\bibitem[{Kollmeier} et al.(2017)]{Kollmeier17}
\bibinfo{author}{{Kollmeier} JA}, \bibinfo{author}{{Zasowski} G},
  \bibinfo{author}{{Rix} HW}, \bibinfo{author}{{Johns} M},
  \bibinfo{author}{{Anderson} SF}, \bibinfo{author}{{Drory} N},
  \bibinfo{author}{{Johnson} JA}, \bibinfo{author}{{Pogge} RW},
  \bibinfo{author}{{Bird} JC}, \bibinfo{author}{{Blanc} GA},
  \bibinfo{author}{{Brownstein} JR}, \bibinfo{author}{{Crane} JD},
  \bibinfo{author}{{De Lee} NM}, \bibinfo{author}{{Klaene} MA},
  \bibinfo{author}{{Kreckel} K}, \bibinfo{author}{{MacDonald} N},
  \bibinfo{author}{{Merloni} A}, \bibinfo{author}{{Ness} MK},
  \bibinfo{author}{{O'Brien} T}, \bibinfo{author}{{Sanchez-Gallego} JR},
  \bibinfo{author}{{Sayres} CC}, \bibinfo{author}{{Shen} Y},
  \bibinfo{author}{{Thakar} AR}, \bibinfo{author}{{Tkachenko} A},
  \bibinfo{author}{{Aerts} C}, \bibinfo{author}{{Blanton} MR},
  \bibinfo{author}{{Eisenstein} DJ}, \bibinfo{author}{{Holtzman} JA},
  \bibinfo{author}{{Maoz} D}, \bibinfo{author}{{Nandra} K},
  \bibinfo{author}{{Rockosi} C}, \bibinfo{author}{{Weinberg} DH},
  \bibinfo{author}{{Bovy} J}, \bibinfo{author}{{Casey} AR},
  \bibinfo{author}{{Chaname} J}, \bibinfo{author}{{Clerc} N},
  \bibinfo{author}{{Conroy} C}, \bibinfo{author}{{Eracleous} M},
  \bibinfo{author}{{G{\"a}nsicke} BT}, \bibinfo{author}{{Hekker} S},
  \bibinfo{author}{{Horne} K}, \bibinfo{author}{{Kauffmann} J},
  \bibinfo{author}{{McQuinn} KBW}, \bibinfo{author}{{Pellegrini} EW},
  \bibinfo{author}{{Schinnerer} E}, \bibinfo{author}{{Schlafly} EF},
  \bibinfo{author}{{Schwope} AD}, \bibinfo{author}{{Seibert} M},
  \bibinfo{author}{{Teske} JK} and  \bibinfo{author}{{van Saders} JL}
  (\bibinfo{year}{2017}), \bibinfo{month}{Nov.}
\bibinfo{title}{{SDSS-V: Pioneering Panoptic Spectroscopy}}.
\bibinfo{journal}{{\em arXiv e-prints}} ,
  \bibinfo{eid}{arXiv:1711.03234}\eprint{1711.03234}.

\bibtype{Article}%
\bibitem[{Koposov} et al.(2023)]{koposov23}
\bibinfo{author}{{Koposov} SE}, \bibinfo{author}{{Erkal} D},
  \bibinfo{author}{{Li} TS}, \bibinfo{author}{{Da Costa} GS},
  \bibinfo{author}{{Cullinane} LR}, \bibinfo{author}{{Ji} AP},
  \bibinfo{author}{{Kuehn} K}, \bibinfo{author}{{Lewis} GF},
  \bibinfo{author}{{Pace} AB}, \bibinfo{author}{{Shipp} N},
  \bibinfo{author}{{Zucker} DB}, \bibinfo{author}{{Bland-Hawthorn} J},
  \bibinfo{author}{{Lilleengen} S}, \bibinfo{author}{{Martell} SL} and
  \bibinfo{author}{{S5 Collaboration}} (\bibinfo{year}{2023}),
  \bibinfo{month}{Jun.}
\bibinfo{title}{{S $^{5}$: Probing the Milky Way and Magellanic Clouds
  potentials with the 6D map of the Orphan-Chenab stream}}.
\bibinfo{journal}{{\em MNRAS}} \bibinfo{volume}{521} (\bibinfo{number}{4}):
  \bibinfo{pages}{4936--4962}. \bibinfo{doi}{\doi{10.1093/mnras/stad551}}.
\eprint{2211.04495}.

\bibtype{Article}%
\bibitem[{Larsen} et al.(2020)]{larsen20}
\bibinfo{author}{{Larsen} SS}, \bibinfo{author}{{Romanowsky} AJ},
  \bibinfo{author}{{Brodie} JP} and  \bibinfo{author}{{Wasserman} A}
  (\bibinfo{year}{2020}), \bibinfo{month}{Nov.}
\bibinfo{title}{{An extremely metal-deficient globular cluster in the Andromeda
  Galaxy}}.
\bibinfo{journal}{{\em Science}} \bibinfo{volume}{370}
  (\bibinfo{number}{6519}): \bibinfo{pages}{970--973}.
  \bibinfo{doi}{\doi{10.1126/science.abb1970}}.

\bibtype{Article}%
\bibitem[{Lemasle} et al.(2014)]{lemasle14}
\bibinfo{author}{{Lemasle} B}, \bibinfo{author}{{de Boer} TJL},
  \bibinfo{author}{{Hill} V}, \bibinfo{author}{{Tolstoy} E},
  \bibinfo{author}{{Irwin} MJ}, \bibinfo{author}{{Jablonka} P},
  \bibinfo{author}{{Venn} K}, \bibinfo{author}{{Battaglia} G},
  \bibinfo{author}{{Starkenburg} E}, \bibinfo{author}{{Shetrone} M},
  \bibinfo{author}{{Letarte} B}, \bibinfo{author}{{Fran{\c{c}}ois} P},
  \bibinfo{author}{{Helmi} A}, \bibinfo{author}{{Primas} F},
  \bibinfo{author}{{Kaufer} A} and  \bibinfo{author}{{Szeifert} T}
  (\bibinfo{year}{2014}), \bibinfo{month}{Dec.}
\bibinfo{title}{{VLT/FLAMES spectroscopy of red giant branch stars in the
  Fornax dwarf spheroidal galaxy}}.
\bibinfo{journal}{{\em A\&A}} \bibinfo{volume}{572}, \bibinfo{eid}{A88}.
  \bibinfo{doi}{\doi{10.1051/0004-6361/201423919}}.

\bibtype{Article}%
\bibitem[{Li} and {Zhao}(2017)]{Li2017}
\bibinfo{author}{{Li} C} and  \bibinfo{author}{{Zhao} G}
  (\bibinfo{year}{2017}), \bibinfo{month}{Nov.}
\bibinfo{title}{{The Evolution of the Galactic Thick Disk with the LAMOST
  Survey}}.
\bibinfo{journal}{{\em ApJ}} \bibinfo{volume}{850} (\bibinfo{number}{1}),
  \bibinfo{eid}{25}. \bibinfo{doi}{\doi{10.3847/1538-4357/aa93f4}}.

\bibtype{Article}%
\bibitem[{Lind} and {Amarsi}(2024)]{lind24}
\bibinfo{author}{{Lind} K} and  \bibinfo{author}{{Amarsi} AM}
  (\bibinfo{year}{2024}), \bibinfo{month}{Sep.}
\bibinfo{title}{{Three-Dimensional Nonlocal Thermodynamic Equilibrium Abundance
  Analyses of Late-Type Stars}}.
\bibinfo{journal}{{\em ARA\&A}} \bibinfo{volume}{62} (\bibinfo{number}{1}):
  \bibinfo{pages}{475--527}.
  \bibinfo{doi}{\doi{10.1146/annurev-astro-052722-103557}}.

\bibtype{Article}%
\bibitem[{Lind} et al.(2012)]{lind12}
\bibinfo{author}{{Lind} K}, \bibinfo{author}{{Bergemann} M} and
  \bibinfo{author}{{Asplund} M} (\bibinfo{year}{2012}), \bibinfo{month}{Nov.}
\bibinfo{title}{{Non-LTE line formation of Fe in late-type stars - II. 1D
  spectroscopic stellar parameters}}.
\bibinfo{journal}{{\em MNRAS}} \bibinfo{volume}{427}: \bibinfo{pages}{50--60}.
  \bibinfo{doi}{\doi{10.1111/j.1365-2966.2012.21686.x}}.
\eprint{1207.2454}.

\bibtype{Article}%
\bibitem[{Majewski} et al.(2003)]{majewski03}
\bibinfo{author}{{Majewski} SR}, \bibinfo{author}{{Skrutskie} MF},
  \bibinfo{author}{{Weinberg} MD} and  \bibinfo{author}{{Ostheimer} JC}
  (\bibinfo{year}{2003}), \bibinfo{month}{Dec.}
\bibinfo{title}{{A Two Micron All Sky Survey View of the Sagittarius Dwarf
  Galaxy. I. Morphology of the Sagittarius Core and Tidal Arms}}.
\bibinfo{journal}{{\em ApJ}} \bibinfo{volume}{599} (\bibinfo{number}{2}):
  \bibinfo{pages}{1082--1115}. \bibinfo{doi}{\doi{10.1086/379504}}.
\eprint{astro-ph/0304198}.

\bibtype{Article}%
\bibitem[{Majewski} et al.(2017)]{Majewski17}
\bibinfo{author}{{Majewski} SR}, \bibinfo{author}{{Schiavon} RP},
  \bibinfo{author}{{Frinchaboy} PM}, \bibinfo{author}{{Allende Prieto} C},
  \bibinfo{author}{{Barkhouser} R}, \bibinfo{author}{{Bizyaev} D},
  \bibinfo{author}{{Blank} B}, \bibinfo{author}{{Brunner} S},
  \bibinfo{author}{{Burton} A}, \bibinfo{author}{{Carrera} R},
  \bibinfo{author}{{Chojnowski} SD}, \bibinfo{author}{{Cunha} K},
  \bibinfo{author}{{Epstein} C}, \bibinfo{author}{{Fitzgerald} G},
  \bibinfo{author}{{Garc{\'\i}a P{\'e}rez} AE}, \bibinfo{author}{{Hearty} FR},
  \bibinfo{author}{{Henderson} C}, \bibinfo{author}{{Holtzman} JA},
  \bibinfo{author}{{Johnson} JA}, \bibinfo{author}{{Lam} CR},
  \bibinfo{author}{{Lawler} JE}, \bibinfo{author}{{Maseman} P},
  \bibinfo{author}{{M{\'e}sz{\'a}ros} S}, \bibinfo{author}{{Nelson} M},
  \bibinfo{author}{{Nguyen} DC}, \bibinfo{author}{{Nidever} DL},
  \bibinfo{author}{{Pinsonneault} M}, \bibinfo{author}{{Shetrone} M},
  \bibinfo{author}{{Smee} S}, \bibinfo{author}{{Smith} VV},
  \bibinfo{author}{{Stolberg} T}, \bibinfo{author}{{Skrutskie} MF},
  \bibinfo{author}{{Walker} E}, \bibinfo{author}{{Wilson} JC},
  \bibinfo{author}{{Zasowski} G}, \bibinfo{author}{{Anders} F},
  \bibinfo{author}{{Basu} S}, \bibinfo{author}{{Beland} S},
  \bibinfo{author}{{Blanton} MR}, \bibinfo{author}{{Bovy} J},
  \bibinfo{author}{{Brownstein} JR}, \bibinfo{author}{{Carlberg} J},
  \bibinfo{author}{{Chaplin} W}, \bibinfo{author}{{Chiappini} C},
  \bibinfo{author}{{Eisenstein} DJ}, \bibinfo{author}{{Elsworth} Y},
  \bibinfo{author}{{Feuillet} D}, \bibinfo{author}{{Fleming} SW},
  \bibinfo{author}{{Galbraith-Frew} J}, \bibinfo{author}{{Garc{\'\i}a} RA},
  \bibinfo{author}{{Garc{\'\i}a-Hern{\'a}ndez} DA},
  \bibinfo{author}{{Gillespie} BA}, \bibinfo{author}{{Girardi} L},
  \bibinfo{author}{{Gunn} JE}, \bibinfo{author}{{Hasselquist} S},
  \bibinfo{author}{{Hayden} MR}, \bibinfo{author}{{Hekker} S},
  \bibinfo{author}{{Ivans} I}, \bibinfo{author}{{Kinemuchi} K},
  \bibinfo{author}{{Klaene} M}, \bibinfo{author}{{Mahadevan} S},
  \bibinfo{author}{{Mathur} S}, \bibinfo{author}{{Mosser} B},
  \bibinfo{author}{{Muna} D}, \bibinfo{author}{{Munn} JA},
  \bibinfo{author}{{Nichol} RC}, \bibinfo{author}{{O'Connell} RW},
  \bibinfo{author}{{Parejko} JK}, \bibinfo{author}{{Robin} AC},
  \bibinfo{author}{{Rocha-Pinto} H}, \bibinfo{author}{{Schultheis} M},
  \bibinfo{author}{{Serenelli} AM}, \bibinfo{author}{{Shane} N},
  \bibinfo{author}{{Silva Aguirre} V}, \bibinfo{author}{{Sobeck} JS},
  \bibinfo{author}{{Thompson} B}, \bibinfo{author}{{Troup} NW},
  \bibinfo{author}{{Weinberg} DH} and  \bibinfo{author}{{Zamora} O}
  (\bibinfo{year}{2017}), \bibinfo{month}{Sep.}
\bibinfo{title}{{The Apache Point Observatory Galactic Evolution Experiment
  (APOGEE)}}.
\bibinfo{journal}{{\em AJ}} \bibinfo{volume}{154} (\bibinfo{number}{3}),
  \bibinfo{eid}{94}. \bibinfo{doi}{\doi{10.3847/1538-3881/aa784d}}.
\eprint{1509.05420}.

\bibtype{Article}%
\bibitem[{Maoz} and {Nakar}(2024)]{maoz24}
\bibinfo{author}{{Maoz} D} and  \bibinfo{author}{{Nakar} E}
  (\bibinfo{year}{2024}), \bibinfo{month}{Jun.}
\bibinfo{title}{{The neutron-star merger delay-time distribution, r-process
  ``knees'', and the metal budget of the Galaxy}}.
\bibinfo{journal}{{\em arXiv e-prints}} ,
  \bibinfo{eid}{arXiv:2406.08630}\bibinfo{doi}{\doi{10.48550/arXiv.2406.08630}}.
\eprint{2406.08630}.

\bibtype{Article}%
\bibitem[{Mardini} et al.(2022{\natexlab{a}})]{mardini22_atari}
\bibinfo{author}{{Mardini} MK}, \bibinfo{author}{{Frebel} A},
  \bibinfo{author}{{Chiti} A}, \bibinfo{author}{{Meiron} Y},
  \bibinfo{author}{{Brauer} KV} and  \bibinfo{author}{{Ou} X}
  (\bibinfo{year}{2022}{\natexlab{a}}), \bibinfo{month}{Sep.}
\bibinfo{title}{{The Atari Disk, a Metal-poor Stellar Population in the Disk
  System of the Milky Way}}.
\bibinfo{journal}{{\em ApJ}} \bibinfo{volume}{936} (\bibinfo{number}{1}),
  \bibinfo{eid}{78}. \bibinfo{doi}{\doi{10.3847/1538-4357/ac8102}}.

\bibtype{Article}%
\bibitem[{Mardini} et al.(2022{\natexlab{b}})]{mardini22_thin}
\bibinfo{author}{{Mardini} MK}, \bibinfo{author}{{Frebel} A},
  \bibinfo{author}{{Ezzeddine} R}, \bibinfo{author}{{Chiti} A},
  \bibinfo{author}{{Meiron} Y}, \bibinfo{author}{{Ji} AP},
  \bibinfo{author}{{Placco} VM}, \bibinfo{author}{{Roederer} IU} and
  \bibinfo{author}{{Mel{\'e}ndez} J} (\bibinfo{year}{2022}{\natexlab{b}}),
  \bibinfo{month}{Dec.}
\bibinfo{title}{{The chemical abundance pattern of the extremely metal-poor
  thin disc star 2MASS J1808-5104 and its origins}}.
\bibinfo{journal}{{\em MNRAS}} \bibinfo{volume}{517} (\bibinfo{number}{3}):
  \bibinfo{pages}{3993--4004}. \bibinfo{doi}{\doi{10.1093/mnras/stac2783}}.

\bibtype{Article}%
\bibitem[{Mardini} et al.(2024)]{mardini24_atari}
\bibinfo{author}{{Mardini} MK}, \bibinfo{author}{{Frebel} A} and
  \bibinfo{author}{{Chiti} A} (\bibinfo{year}{2024}), \bibinfo{month}{Mar.}
\bibinfo{title}{{A strontium-rich ultra-metal-poor star in the Atari disc
  component}} \bibinfo{volume}{529} (\bibinfo{number}{1}):
  \bibinfo{pages}{L60--L66}. \bibinfo{doi}{\doi{10.1093/mnrasl/slad197}}.
\eprint{2402.07046}.

\bibtype{Article}%
\bibitem[{Mardini} et al.(2025)]{mardini_ustar}
\bibinfo{author}{{Mardini} MK}, \bibinfo{author}{{Ezzeddine} R},
  \bibinfo{author}{{Frebel} A}, \bibinfo{author}{{Hansen} TT},
  \bibinfo{author}{{Holmbeck} EM}, \bibinfo{author}{{Beers} TC},
  \bibinfo{author}{{Placco} VM}, \bibinfo{author}{{Roederer} IU} and
  \bibinfo{author}{{Sakari} CM} (\bibinfo{year}{2025}).
\bibinfo{title}{{The $R$-Process Alliance: U\,II Detection in the $R$-Process
  Enhanced and Actinide-Deficient Star J1453$+$0040}}.
\bibinfo{journal}{{\em subm. to ApJ}} .

\bibtype{Article}%
\bibitem[{Martin} et al.(2022)]{martin22}
\bibinfo{author}{{Martin} NF}, \bibinfo{author}{{Venn} KA},
  \bibinfo{author}{{Aguado} DS}, \bibinfo{author}{{Starkenburg} E},
  \bibinfo{author}{{Gonz{\'a}lez Hern{\'a}ndez} JI}, \bibinfo{author}{{Ibata}
  RA}, \bibinfo{author}{{Bonifacio} P}, \bibinfo{author}{{Caffau} E},
  \bibinfo{author}{{Sestito} F}, \bibinfo{author}{{Arentsen} A},
  \bibinfo{author}{{Allende Prieto} C}, \bibinfo{author}{{Carlberg} RG},
  \bibinfo{author}{{Fabbro} S}, \bibinfo{author}{{Fouesneau} M},
  \bibinfo{author}{{Hill} V}, \bibinfo{author}{{Jablonka} P},
  \bibinfo{author}{{Kordopatis} G}, \bibinfo{author}{{Lardo} C},
  \bibinfo{author}{{Malhan} K}, \bibinfo{author}{{Mashonkina} LI},
  \bibinfo{author}{{McConnachie} AW}, \bibinfo{author}{{Navarro} JF},
  \bibinfo{author}{{S{\'a}nchez-Janssen} R}, \bibinfo{author}{{Thomas} GF},
  \bibinfo{author}{{Yuan} Z} and  \bibinfo{author}{{Mucciarelli} A}
  (\bibinfo{year}{2022}), \bibinfo{month}{Jan.}
\bibinfo{title}{{A stellar stream remnant of a globular cluster below the
  metallicity floor}}.
\bibinfo{journal}{{\em Nature}} \bibinfo{volume}{601} (\bibinfo{number}{7891}):
  \bibinfo{pages}{45--48}. \bibinfo{doi}{\doi{10.1038/s41586-021-04162-2}}.

\bibtype{Article}%
\bibitem[{Merrill}(1952)]{merrill52}
\bibinfo{author}{{Merrill} PW} (\bibinfo{year}{1952}), \bibinfo{month}{Jan.}
\bibinfo{title}{{Technetium in the stars}}.
\bibinfo{journal}{{\em Science}} \bibinfo{volume}{115}
  (\bibinfo{number}{2992}): \bibinfo{pages}{484}.
  \bibinfo{doi}{\doi{10.1126/science.115.2992.479}}.

\bibtype{Article}%
\bibitem[{Milone} et al.(2012)]{milone12}
\bibinfo{author}{{Milone} AP}, \bibinfo{author}{{Piotto} G},
  \bibinfo{author}{{Bedin} LR}, \bibinfo{author}{{King} IR},
  \bibinfo{author}{{Anderson} J}, \bibinfo{author}{{Marino} AF},
  \bibinfo{author}{{Bellini} A}, \bibinfo{author}{{Gratton} R},
  \bibinfo{author}{{Renzini} A}, \bibinfo{author}{{Stetson} PB},
  \bibinfo{author}{{Cassisi} S}, \bibinfo{author}{{Aparicio} A},
  \bibinfo{author}{{Bragaglia} A}, \bibinfo{author}{{Carretta} E},
  \bibinfo{author}{{D'Antona} F}, \bibinfo{author}{{Di Criscienzo} M},
  \bibinfo{author}{{Lucatello} S}, \bibinfo{author}{{Monelli} M} and
  \bibinfo{author}{{Pietrinferni} A} (\bibinfo{year}{2012}),
  \bibinfo{month}{Jan.}
\bibinfo{title}{{Multiple Stellar Populations in 47 Tucanae}}.
\bibinfo{journal}{{\em ApJ}} \bibinfo{volume}{744} (\bibinfo{number}{1}),
  \bibinfo{eid}{58}.
  \bibinfo{doi}{\doi{10.1088/0004-637X/744/1/5810.1086/141918}}.
\eprint{1109.0900}.

\bibtype{Article}%
\bibitem[{Milone} et al.(2017)]{milone17}
\bibinfo{author}{{Milone} AP}, \bibinfo{author}{{Marino} AF},
  \bibinfo{author}{{Bedin} LR}, \bibinfo{author}{{Anderson} J},
  \bibinfo{author}{{Apai} D}, \bibinfo{author}{{Bellini} A},
  \bibinfo{author}{{Bergeron} P}, \bibinfo{author}{{Burgasser} AJ},
  \bibinfo{author}{{Dotter} A} and  \bibinfo{author}{{Rees} JM}
  (\bibinfo{year}{2017}), \bibinfo{month}{Jul.}
\bibinfo{title}{{The HST large programme on {\ensuremath{\omega}} Centauri - I.
  Multiple stellar populations at the bottom of the main sequence probed in
  NIR-Optical}}.
\bibinfo{journal}{{\em MNRAS}} \bibinfo{volume}{469} (\bibinfo{number}{1}):
  \bibinfo{pages}{800--812}. \bibinfo{doi}{\doi{10.1093/mnras/stx836}}.
\eprint{1704.00418}.

\bibtype{Article}%
\bibitem[{M{\"o}sta} et al.(2018)]{moesta18}
\bibinfo{author}{{M{\"o}sta} P}, \bibinfo{author}{{Roberts} LF},
  \bibinfo{author}{{Halevi} G}, \bibinfo{author}{{Ott} CD},
  \bibinfo{author}{{Lippuner} J}, \bibinfo{author}{{Haas} R} and
  \bibinfo{author}{{Schnetter} E} (\bibinfo{year}{2018}), \bibinfo{month}{Sep.}
\bibinfo{title}{{r-process Nucleosynthesis from Three-dimensional
  Magnetorotational Core-collapse Supernovae}}.
\bibinfo{journal}{{\em ApJ}} \bibinfo{volume}{864} (\bibinfo{number}{2}),
  \bibinfo{eid}{171}. \bibinfo{doi}{\doi{10.3847/1538-4357/aad6ec}}.

\bibtype{Article}%
\bibitem[{Nordlander} et al.(2017)]{Nordlander17}
\bibinfo{author}{{Nordlander} T}, \bibinfo{author}{{Amarsi} AM},
  \bibinfo{author}{{Lind} K}, \bibinfo{author}{{Asplund} M},
  \bibinfo{author}{{Barklem} PS}, \bibinfo{author}{{Casey} AR},
  \bibinfo{author}{{Collet} R} and  \bibinfo{author}{{Leenaarts} J}
  (\bibinfo{year}{2017}), \bibinfo{month}{Jan.}
\bibinfo{title}{{3D NLTE analysis of the most iron-deficient star,
  SMSS0313-6708}}.
\bibinfo{journal}{{\em A\&A}} \bibinfo{volume}{597}, \bibinfo{eid}{A6}.
  \bibinfo{doi}{\doi{10.1051/0004-6361/201629202}}.
\eprint{1609.07416}.

\bibtype{Article}%
\bibitem[{Norris} et al.(1985)]{norris85}
\bibinfo{author}{{Norris} J}, \bibinfo{author}{{Bessell} MS} and
  \bibinfo{author}{{Pickles} AJ} (\bibinfo{year}{1985}), \bibinfo{month}{Jul.}
\bibinfo{title}{{Population studies. I. The Bidelman-MacConnell ``weak-metal''
  stars.}}
\bibinfo{journal}{{\em ApJs}} \bibinfo{volume}{58}: \bibinfo{pages}{463--492}.
  \bibinfo{doi}{\doi{10.1086/191049}}.

\bibtype{Article}%
\bibitem[{Norris} et al.(1997)]{norris97_carbon}
\bibinfo{author}{{Norris} JE}, \bibinfo{author}{{Ryan} SG} and
  \bibinfo{author}{{Beers} TC} (\bibinfo{year}{1997}), \bibinfo{month}{Oct.}
\bibinfo{title}{{Extremely Metal-poor Stars. IV. The Carbon-rich Objects}}.
\bibinfo{journal}{{\em ApJ}} \bibinfo{volume}{488}: \bibinfo{pages}{350}.

\bibtype{Article}%
\bibitem[{Norris} et al.(2010)]{norris10a}
\bibinfo{author}{{Norris} JE}, \bibinfo{author}{{Wyse} RFG},
  \bibinfo{author}{{Gilmore} G}, \bibinfo{author}{{Yong} D},
  \bibinfo{author}{{Frebel} A}, \bibinfo{author}{{Wilkinson} MI},
  \bibinfo{author}{{Belokurov} V} and  \bibinfo{author}{{Zucker} DB}
  (\bibinfo{year}{2010}), \bibinfo{month}{Nov.}
\bibinfo{title}{{Chemical Enrichment in the Faintest Galaxies: The Carbon and
  Iron Abundance Spreads in the Bo{\"o}tes I Dwarf Spheroidal Galaxy and the
  Segue 1 System}}.
\bibinfo{journal}{{\em ApJ}} \bibinfo{volume}{723}:
  \bibinfo{pages}{1632--1650}.
  \bibinfo{doi}{\doi{10.1088/0004-637X/723/2/1632}}.

\bibtype{Article}%
\bibitem[{Norris} et al.(2013)]{norris13_I}
\bibinfo{author}{{Norris} JE}, \bibinfo{author}{{Bessell} MS},
  \bibinfo{author}{{Yong} D}, \bibinfo{author}{{Christlieb} N},
  \bibinfo{author}{{Barklem} PS}, \bibinfo{author}{{Asplund} M},
  \bibinfo{author}{{Murphy} SJ}, \bibinfo{author}{{Beers} TC},
  \bibinfo{author}{{Frebel} A} and  \bibinfo{author}{{Ryan} SG}
  (\bibinfo{year}{2013}), \bibinfo{month}{Jan.}
\bibinfo{title}{{The Most Metal-poor Stars. I. Discovery, Data, and Atmospheric
  Parameters}}.
\bibinfo{journal}{{\em ApJ}} \bibinfo{volume}{762}, \bibinfo{eid}{25}.
  \bibinfo{doi}{\doi{10.1088/0004-637X/762/1/25}}.
\eprint{1208.2999}.

\bibtype{Article}%
\bibitem[{Piotto} et al.(2007)]{piotto07}
\bibinfo{author}{{Piotto} G}, \bibinfo{author}{{Bedin} LR},
  \bibinfo{author}{{Anderson} J}, \bibinfo{author}{{King} IR},
  \bibinfo{author}{{Cassisi} S}, \bibinfo{author}{{Milone} AP},
  \bibinfo{author}{{Villanova} S}, \bibinfo{author}{{Pietrinferni} A} and
  \bibinfo{author}{{Renzini} A} (\bibinfo{year}{2007}), \bibinfo{month}{May}.
\bibinfo{title}{{A Triple Main Sequence in the Globular Cluster NGC 2808}}.
\bibinfo{journal}{{\em ApJl}} \bibinfo{volume}{661} (\bibinfo{number}{1}):
  \bibinfo{pages}{L53--L56}. \bibinfo{doi}{\doi{10.1086/518503}}.
\eprint{astro-ph/0703767}.

\bibtype{Article}%
\bibitem[{Placco} et al.(2014)]{placco14}
\bibinfo{author}{{Placco} VM}, \bibinfo{author}{{Frebel} A},
  \bibinfo{author}{{Beers} TC} and  \bibinfo{author}{{Stancliffe} RJ}
  (\bibinfo{year}{2014}), \bibinfo{month}{Dec.}
\bibinfo{title}{{Carbon-enhanced Metal-poor Star Frequencies in the Galaxy:
  Corrections for the Effect of Evolutionary Status on Carbon Abundances}}.
\bibinfo{journal}{{\em ApJ}} \bibinfo{volume}{797}, \bibinfo{eid}{21}.
  \bibinfo{doi}{\doi{10.1088/0004-637X/797/1/21}}.
\eprint{1410.2223}.

\bibtype{Article}%
\bibitem[{Przybylski}(1966)]{przybylski66}
\bibinfo{author}{{Przybylski} A} (\bibinfo{year}{1966}), \bibinfo{month}{Apr.}
\bibinfo{title}{{Abundance Analysis of the Peculiar Star HD 101065}}.
\bibinfo{journal}{{\em Nature}} \bibinfo{volume}{210} (\bibinfo{number}{5031}):
  \bibinfo{pages}{20--22}. \bibinfo{doi}{\doi{10.1038/210020a0}}.

\bibtype{Article}%
\bibitem[{Reggiani} et al.(2021)]{reggiani21}
\bibinfo{author}{{Reggiani} H}, \bibinfo{author}{{Schlaufman} KC},
  \bibinfo{author}{{Casey} AR}, \bibinfo{author}{{Simon} JD} and
  \bibinfo{author}{{Ji} AP} (\bibinfo{year}{2021}), \bibinfo{month}{Dec.}
\bibinfo{title}{{The Most Metal-poor Stars in the Magellanic Clouds Are
  r-process Enhanced}}.
\bibinfo{journal}{{\em AJ}} \bibinfo{volume}{162} (\bibinfo{number}{6}),
  \bibinfo{eid}{229}. \bibinfo{doi}{\doi{10.3847/1538-3881/ac1f9a}}.
\eprint{2108.10880}.

\bibtype{Article}%
\bibitem[{Roederer} and {Gnedin}(2019)]{roederer19_sylger}
\bibinfo{author}{{Roederer} IU} and  \bibinfo{author}{{Gnedin} OY}
  (\bibinfo{year}{2019}), \bibinfo{month}{Sep.}
\bibinfo{title}{{High-resolution Optical Spectroscopy of Stars in the Sylgr
  Stellar Stream}}.
\bibinfo{journal}{{\em ApJ}} \bibinfo{volume}{883} (\bibinfo{number}{1}),
  \bibinfo{eid}{84}. \bibinfo{doi}{\doi{10.3847/1538-4357/ab365c}}.

\bibtype{Article}%
\bibitem[{Roederer} et al.(2012)]{roederer12_hubble}
\bibinfo{author}{{Roederer} IU}, \bibinfo{author}{{Lawler} JE},
  \bibinfo{author}{{Sobeck} JS}, \bibinfo{author}{{Beers} TC},
  \bibinfo{author}{{Cowan} JJ}, \bibinfo{author}{{Frebel} A},
  \bibinfo{author}{{Ivans} II}, \bibinfo{author}{{Schatz} H},
  \bibinfo{author}{{Sneden} C} and  \bibinfo{author}{{Thompson} IB}
  (\bibinfo{year}{2012}), \bibinfo{month}{Dec.}
\bibinfo{title}{{New Hubble Space Telescope Observations of Heavy Elements in
  Four Metal-Poor Stars}}.
\bibinfo{journal}{{\em ApJs}} \bibinfo{volume}{203} (\bibinfo{number}{2}),
  \bibinfo{eid}{27}. \bibinfo{doi}{\doi{10.1088/0067-0049/203/2/27}}.

\bibtype{Article}%
\bibitem[{Roederer} et al.(2016)]{Roederer16b}
\bibinfo{author}{{Roederer} IU}, \bibinfo{author}{{Mateo} M},
  \bibinfo{author}{{Bailey} III JI}, \bibinfo{author}{{Song} Y},
  \bibinfo{author}{{Bell} EF}, \bibinfo{author}{{Crane} JD},
  \bibinfo{author}{{Loebman} S}, \bibinfo{author}{{Nidever} DL},
  \bibinfo{author}{{Olszewski} EW}, \bibinfo{author}{{Shectman} SA},
  \bibinfo{author}{{Thompson} IB}, \bibinfo{author}{{Valluri} M} and
  \bibinfo{author}{{Walker} MG} (\bibinfo{year}{2016}), \bibinfo{month}{Mar.}
\bibinfo{title}{{Detailed Chemical Abundances in the r-process-rich Ultra-faint
  Dwarf Galaxy Reticulum 2}}.
\bibinfo{journal}{{\em AJ}} \bibinfo{volume}{151}, \bibinfo{eid}{82}.
  \bibinfo{doi}{\doi{10.3847/0004-6256/151/3/82}}.
\eprint{1601.04070}.

\bibtype{Article}%
\bibitem[{Roederer} et al.(2018)]{roederer18_kin}
\bibinfo{author}{{Roederer} IU}, \bibinfo{author}{{Hattori} K} and
  \bibinfo{author}{{Valluri} M} (\bibinfo{year}{2018}), \bibinfo{month}{Oct.}
\bibinfo{title}{{Kinematics of Highly r-process-enhanced Field Stars: Evidence
  for an Accretion Origin and Detection of Several Groups from Disrupted
  Satellites}}.
\bibinfo{journal}{{\em AJ}} \bibinfo{volume}{156} (\bibinfo{number}{4}),
  \bibinfo{eid}{179}. \bibinfo{doi}{\doi{10.3847/1538-3881/aadd9c}}.

\bibtype{Article}%
\bibitem[{Roederer} et al.(2020)]{roederer20_pb}
\bibinfo{author}{{Roederer} IU}, \bibinfo{author}{{Lawler} JE},
  \bibinfo{author}{{Holmbeck} EM}, \bibinfo{author}{{Beers} TC},
  \bibinfo{author}{{Ezzeddine} R}, \bibinfo{author}{{Frebel} A},
  \bibinfo{author}{{Hansen} TT}, \bibinfo{author}{{Ivans} II},
  \bibinfo{author}{{Karakas} AI}, \bibinfo{author}{{Placco} VM} and
  \bibinfo{author}{{Sakari} CM} (\bibinfo{year}{2020}), \bibinfo{month}{Oct.}
\bibinfo{title}{{Detection of Pb II in the Ultraviolet Spectra of Three
  Metal-poor Stars}}.
\bibinfo{journal}{{\em ApJl}} \bibinfo{volume}{902} (\bibinfo{number}{1}),
  \bibinfo{eid}{L24}. \bibinfo{doi}{\doi{10.3847/2041-8213/abbc21}}.

\bibtype{Article}%
\bibitem[{Roederer} et al.(2022)]{roederer22}
\bibinfo{author}{{Roederer} IU}, \bibinfo{author}{{Lawler} JE},
  \bibinfo{author}{{Den Hartog} EA}, \bibinfo{author}{{Placco} VM},
  \bibinfo{author}{{Surman} R}, \bibinfo{author}{{Beers} TC},
  \bibinfo{author}{{Ezzeddine} R}, \bibinfo{author}{{Frebel} A},
  \bibinfo{author}{{Hansen} TT}, \bibinfo{author}{{Hattori} K},
  \bibinfo{author}{{Holmbeck} EM} and  \bibinfo{author}{{Sakari} CM}
  (\bibinfo{year}{2022}), \bibinfo{month}{Jun.}
\bibinfo{title}{{The R-process Alliance: A Nearly Complete R-process Abundance
  Template Derived from Ultraviolet Spectroscopy of the R-process-enhanced
  Metal-poor Star HD 222925}}.
\bibinfo{journal}{{\em ApJS}} \bibinfo{volume}{260} (\bibinfo{number}{2}),
  \bibinfo{eid}{27}. \bibinfo{doi}{\doi{10.3847/1538-4365/ac5cbc}}.
\eprint{2205.03426}.

\bibtype{Article}%
\bibitem[{Roederer} et al.(2023)]{roederer23_fission}
\bibinfo{author}{{Roederer} IU}, \bibinfo{author}{{Vassh} N},
  \bibinfo{author}{{Holmbeck} EM}, \bibinfo{author}{{Mumpower} MR},
  \bibinfo{author}{{Surman} R}, \bibinfo{author}{{Cowan} JJ},
  \bibinfo{author}{{Beers} TC}, \bibinfo{author}{{Ezzeddine} R},
  \bibinfo{author}{{Frebel} A}, \bibinfo{author}{{Hansen} TT},
  \bibinfo{author}{{Placco} VM} and  \bibinfo{author}{{Sakari} CM}
  (\bibinfo{year}{2023}), \bibinfo{month}{Dec.}
\bibinfo{title}{{Element abundance patterns in stars indicate fission of nuclei
  heavier than uranium}}.
\bibinfo{journal}{{\em Science}} \bibinfo{volume}{382}
  (\bibinfo{number}{6675}): \bibinfo{pages}{1177--1180}.
  \bibinfo{doi}{\doi{10.1126/science.adf1341}}.
\eprint{2312.06844}.

\bibtype{Article}%
\bibitem[{Romanyuk}(2007)]{romanyuk07}
\bibinfo{author}{{Romanyuk} II} (\bibinfo{year}{2007}), \bibinfo{month}{Mar.}
\bibinfo{title}{{Main-sequence magnetic CP stars: II. Physical parameters and
  chemical composition of the atmosphere}}.
\bibinfo{journal}{{\em Astrophysical Bulletin}} \bibinfo{volume}{62}
  (\bibinfo{number}{1}): \bibinfo{pages}{62--89}.
  \bibinfo{doi}{\doi{10.1134/S1990341307010063}}.

\bibtype{Inproceedings}%
\bibitem[{Rossi} et al.(1999)]{1999rossicarbon}
\bibinfo{author}{{Rossi} S}, \bibinfo{author}{{Beers} TC} and
  \bibinfo{author}{{Sneden} C} (\bibinfo{year}{1999}), \bibinfo{title}{{Carbon
  Abundances for Metal-Poor Stars Based on Medium-Resolution Spectra}},
  \bibinfo{booktitle}{ASP Conf. Ser. 165: The Third Stromlo Symposium: The
  Galactic Halo}, pp. \bibinfo{pages}{264}.

\bibtype{Article}%
\bibitem[{Schuler} et al.(2007)]{schuler07}
\bibinfo{author}{{Schuler} SC}, \bibinfo{author}{{Cunha} K},
  \bibinfo{author}{{Smith} VV}, \bibinfo{author}{{Sivarani} T},
  \bibinfo{author}{{Beers} TC} and  \bibinfo{author}{{Lee} YS}
  (\bibinfo{year}{2007}), \bibinfo{month}{Sep.}
\bibinfo{title}{{Fluorine in a Carbon-enhanced Metal-poor Star}}.
\bibinfo{journal}{{\em ApJl}} \bibinfo{volume}{667} (\bibinfo{number}{1}):
  \bibinfo{pages}{L81--L84}. \bibinfo{doi}{\doi{10.1086/521951}}.

\bibtype{Article}%
\bibitem[{Shappee} et al.(2017)]{shappee17}
\bibinfo{author}{{Shappee} BJ}, \bibinfo{author}{{Simon} JD},
  \bibinfo{author}{{Drout} MR}, \bibinfo{author}{{Piro} AL},
  \bibinfo{author}{{Morrell} N}, \bibinfo{author}{{Prieto} JL},
  \bibinfo{author}{{Kasen} D}, \bibinfo{author}{{Holoien} TWS},
  \bibinfo{author}{{Kollmeier} JA}, \bibinfo{author}{{Kelson} DD},
  \bibinfo{author}{{Coulter} DA}, \bibinfo{author}{{Foley} RJ},
  \bibinfo{author}{{Kilpatrick} CD}, \bibinfo{author}{{Siebert} MR},
  \bibinfo{author}{{Madore} BF}, \bibinfo{author}{{Murguia-Berthier} A},
  \bibinfo{author}{{Pan} YC}, \bibinfo{author}{{Prochaska} JX},
  \bibinfo{author}{{Ramirez-Ruiz} E}, \bibinfo{author}{{Rest} A},
  \bibinfo{author}{{Adams} C}, \bibinfo{author}{{Alatalo} K},
  \bibinfo{author}{{Ba{\~n}ados} E}, \bibinfo{author}{{Baughman} J},
  \bibinfo{author}{{Bernstein} RA}, \bibinfo{author}{{Bitsakis} T},
  \bibinfo{author}{{Boutsia} K}, \bibinfo{author}{{Bravo} JR},
  \bibinfo{author}{{Di Mille} F}, \bibinfo{author}{{Higgs} CR},
  \bibinfo{author}{{Ji} AP}, \bibinfo{author}{{Maravelias} G},
  \bibinfo{author}{{Marshall} JL}, \bibinfo{author}{{Placco} VM},
  \bibinfo{author}{{Prieto} G} and  \bibinfo{author}{{Wan} Z}
  (\bibinfo{year}{2017}), \bibinfo{month}{Dec.}
\bibinfo{title}{{Early spectra of the gravitational wave source GW170817:
  Evolution of a neutron star merger}}.
\bibinfo{journal}{{\em Science}} \bibinfo{volume}{358}:
  \bibinfo{pages}{1574--1578}. \bibinfo{doi}{\doi{10.1126/science.aaq0186}}.
\eprint{1710.05432}.

\bibtype{Article}%
\bibitem[{Shulyak} et al.(2010)]{shulyak10}
\bibinfo{author}{{Shulyak} D}, \bibinfo{author}{{Ryabchikova} T},
  \bibinfo{author}{{Kildiyarova} R} and  \bibinfo{author}{{Kochukhov} O}
  (\bibinfo{year}{2010}), \bibinfo{month}{Sep.}
\bibinfo{title}{{Realistic model atmosphere and revised abundances of the
  coolest Ap star HD 101065}}.
\bibinfo{journal}{{\em A\&A}} \bibinfo{volume}{520}, \bibinfo{eid}{A88}.
  \bibinfo{doi}{\doi{10.1051/0004-6361/200913750}}.
\eprint{1004.0246}.

\bibtype{Article}%
\bibitem[{Simmerer} et al.(2004)]{simmerer04}
\bibinfo{author}{{Simmerer} J}, \bibinfo{author}{{Sneden} C},
  \bibinfo{author}{{Cowan} JJ}, \bibinfo{author}{{Collier} J},
  \bibinfo{author}{{Woolf} VM} and  \bibinfo{author}{{Lawler} JE}
  (\bibinfo{year}{2004}), \bibinfo{month}{Dec.}
\bibinfo{title}{{The Rise of the s-Process in the Galaxy}}.
\bibinfo{journal}{{\em ApJ}} \bibinfo{volume}{617}:
  \bibinfo{pages}{1091--1114}. \bibinfo{doi}{\doi{10.1086/424504}}.
\eprint{astro-ph/0410396}.

\bibtype{Article}%
\bibitem[{Simon}(2019)]{simon19}
\bibinfo{author}{{Simon} JD} (\bibinfo{year}{2019}), \bibinfo{month}{Aug.}
\bibinfo{title}{{The Faintest Dwarf Galaxies}}.
\bibinfo{journal}{{\em ARA\&A}} \bibinfo{volume}{57}:
  \bibinfo{pages}{375--415}.
  \bibinfo{doi}{\doi{10.1146/annurev-astro-091918-104453}}.

\bibtype{Article}%
\bibitem[{Simon} et al.(2017)]{simon17}
\bibinfo{author}{{Simon} JD}, \bibinfo{author}{{Li} TS},
  \bibinfo{author}{{Drlica-Wagner} A}, \bibinfo{author}{{Bechtol} K},
  \bibinfo{author}{{Marshall} JL}, \bibinfo{author}{{James} DJ},
  \bibinfo{author}{{Wang} MY}, \bibinfo{author}{{Strigari} L},
  \bibinfo{author}{{Balbinot} E}, \bibinfo{author}{{Kuehn} K},
  \bibinfo{author}{{Walker} AR}, \bibinfo{author}{{Abbott} TMC},
  \bibinfo{author}{{Allam} S}, \bibinfo{author}{{Annis} J},
  \bibinfo{author}{{Benoit-L{\'e}vy} A}, \bibinfo{author}{{Brooks} D},
  \bibinfo{author}{{Buckley-Geer} E}, \bibinfo{author}{{Burke} DL},
  \bibinfo{author}{{Carnero Rosell} A}, \bibinfo{author}{{Carrasco Kind} M},
  \bibinfo{author}{{Carretero} J}, \bibinfo{author}{{Cunha} CE},
  \bibinfo{author}{{D'Andrea} CB}, \bibinfo{author}{{da Costa} LN},
  \bibinfo{author}{{DePoy} DL}, \bibinfo{author}{{Desai} S},
  \bibinfo{author}{{Doel} P}, \bibinfo{author}{{Fernandez} E},
  \bibinfo{author}{{Flaugher} B}, \bibinfo{author}{{Frieman} J},
  \bibinfo{author}{{Garc{\'{\i}}a-Bellido} J}, \bibinfo{author}{{Gaztanaga} E},
  \bibinfo{author}{{Goldstein} DA}, \bibinfo{author}{{Gruen} D},
  \bibinfo{author}{{Gutierrez} G}, \bibinfo{author}{{Kuropatkin} N},
  \bibinfo{author}{{Maia} MAG}, \bibinfo{author}{{Martini} P},
  \bibinfo{author}{{Menanteau} F}, \bibinfo{author}{{Miller} CJ},
  \bibinfo{author}{{Miquel} R}, \bibinfo{author}{{Neilsen} E},
  \bibinfo{author}{{Nord} B}, \bibinfo{author}{{Ogando} R},
  \bibinfo{author}{{Plazas} AA}, \bibinfo{author}{{Romer} AK},
  \bibinfo{author}{{Rykoff} ES}, \bibinfo{author}{{Sanchez} E},
  \bibinfo{author}{{Santiago} B}, \bibinfo{author}{{Scarpine} V},
  \bibinfo{author}{{Schubnell} M}, \bibinfo{author}{{Sevilla-Noarbe} I},
  \bibinfo{author}{{Smith} RC}, \bibinfo{author}{{Sobreira} F},
  \bibinfo{author}{{Suchyta} E}, \bibinfo{author}{{Swanson} MEC},
  \bibinfo{author}{{Tarle} G}, \bibinfo{author}{{Whiteway} L},
  \bibinfo{author}{{Yanny} B} and  \bibinfo{author}{{DES Collaboration}}
  (\bibinfo{year}{2017}), \bibinfo{month}{Mar.}
\bibinfo{title}{{Nearest Neighbor: The Low-mass Milky Way Satellite Tucana
  III}}.
\bibinfo{journal}{{\em ApJ}} \bibinfo{volume}{838}, \bibinfo{eid}{11}.
  \bibinfo{doi}{\doi{10.3847/1538-4357/aa5be7}}.
\eprint{1610.05301}.

\bibtype{Article}%
\bibitem[{Sk{\'u}lad{\'o}ttir} et al.(2015)]{skuladottir15}
\bibinfo{author}{{Sk{\'u}lad{\'o}ttir} {\'A}}, \bibinfo{author}{{Tolstoy} E},
  \bibinfo{author}{{Salvadori} S}, \bibinfo{author}{{Hill} V},
  \bibinfo{author}{{Pettini} M}, \bibinfo{author}{{Shetrone} MD} and
  \bibinfo{author}{{Starkenburg} E} (\bibinfo{year}{2015}),
  \bibinfo{month}{Feb.}
\bibinfo{title}{{The first carbon-enhanced metal-poor star found in the
  Sculptor dwarf spheroidal}}.
\bibinfo{journal}{{\em A\&A}} \bibinfo{volume}{574}, \bibinfo{eid}{A129}.
  \bibinfo{doi}{\doi{10.1051/0004-6361/201424782}}.

\bibtype{Article}%
\bibitem[{Sk{\'u}lad{\'o}ttir} et al.(2021)]{skuladottir21}
\bibinfo{author}{{Sk{\'u}lad{\'o}ttir} {\'A}}, \bibinfo{author}{{Salvadori} S},
  \bibinfo{author}{{Amarsi} AM}, \bibinfo{author}{{Tolstoy} E},
  \bibinfo{author}{{Irwin} MJ}, \bibinfo{author}{{Hill} V},
  \bibinfo{author}{{Jablonka} P}, \bibinfo{author}{{Battaglia} G},
  \bibinfo{author}{{Starkenburg} E}, \bibinfo{author}{{Massari} D},
  \bibinfo{author}{{Helmi} A} and  \bibinfo{author}{{Posti} L}
  (\bibinfo{year}{2021}), \bibinfo{month}{Jul.}
\bibinfo{title}{{Zero-metallicity Hypernova Uncovered by an Ultra-metal-poor
  Star in the Sculptor Dwarf Spheroidal Galaxy}}.
\bibinfo{journal}{{\em ApJL}} \bibinfo{volume}{915} (\bibinfo{number}{2}),
  \bibinfo{eid}{L30}. \bibinfo{doi}{\doi{10.3847/2041-8213/ac0dc2}}.
\eprint{2106.11592}.

\bibtype{Article}%
\bibitem[{Sneden} et al.(2003)]{sneden03}
\bibinfo{author}{{Sneden} C}, \bibinfo{author}{{Cowan} JJ},
  \bibinfo{author}{{Lawler} JE}, \bibinfo{author}{{Ivans} II},
  \bibinfo{author}{{Burles} S}, \bibinfo{author}{{Beers} TC},
  \bibinfo{author}{{Primas} F}, \bibinfo{author}{{Hill} V},
  \bibinfo{author}{{Truran} JW}, \bibinfo{author}{{Fuller} GM},
  \bibinfo{author}{{Pfeiffer} B} and  \bibinfo{author}{{Kratz} KL}
  (\bibinfo{year}{2003}), \bibinfo{month}{Jul.}
\bibinfo{title}{{The Extremely Metal-poor, Neutron Capture-rich Star CS
  22892-052: A Comprehensive Abundance Analysis}}.
\bibinfo{journal}{{\em ApJ}} \bibinfo{volume}{591}: \bibinfo{pages}{936--953}.
  \bibinfo{doi}{\doi{10.1086/375491}}.
\eprint{astro-ph/0303542}.

\bibtype{Article}%
\bibitem[{Sneden} et al.(2008)]{sneden08}
\bibinfo{author}{{Sneden} C}, \bibinfo{author}{{Cowan} JJ} and
  \bibinfo{author}{{Gallino} R} (\bibinfo{year}{2008}), \bibinfo{month}{Sep.}
\bibinfo{title}{{Neutron-Capture Elements in the Early Galaxy}}.
\bibinfo{journal}{{\em ARA\&A}} \bibinfo{volume}{46}:
  \bibinfo{pages}{241--288}.
  \bibinfo{doi}{\doi{10.1146/annurev.astro.46.060407.145207}}.

\bibtype{Article}%
\bibitem[{Starkenburg} et al.(2010)]{starkenburg10}
\bibinfo{author}{{Starkenburg} E}, \bibinfo{author}{{Hill} V},
  \bibinfo{author}{{Tolstoy} E}, \bibinfo{author}{{Gonz{\'a}lez Hern{\'a}ndez}
  JI}, \bibinfo{author}{{Irwin} M}, \bibinfo{author}{{Helmi} A},
  \bibinfo{author}{{Battaglia} G}, \bibinfo{author}{{Jablonka} P},
  \bibinfo{author}{{Tafelmeyer} M}, \bibinfo{author}{{Shetrone} M},
  \bibinfo{author}{{Venn} K} and  \bibinfo{author}{{de Boer} T}
  (\bibinfo{year}{2010}), \bibinfo{month}{Apr.}
\bibinfo{title}{{The NIR Ca ii triplet at low metallicity. Searching for
  extremely low-metallicity stars in classical dwarf galaxies}}.
\bibinfo{journal}{{\em A\&A}} \bibinfo{volume}{513}: \bibinfo{pages}{A34}.
  \bibinfo{doi}{\doi{10.1051/0004-6361/200913759}}.
\eprint{1002.2963}.

\bibtype{Article}%
\bibitem[{Starkenburg} et al.(2017)]{starkenburg17}
\bibinfo{author}{{Starkenburg} E}, \bibinfo{author}{{Martin} N},
  \bibinfo{author}{{Youakim} K}, \bibinfo{author}{{Aguado} DS},
  \bibinfo{author}{{Allende Prieto} C}, \bibinfo{author}{{Arentsen} A},
  \bibinfo{author}{{Bernard} EJ}, \bibinfo{author}{{Bonifacio} P},
  \bibinfo{author}{{Caffau} E}, \bibinfo{author}{{Carlberg} RG},
  \bibinfo{author}{{C{\^o}t{\'e}} P}, \bibinfo{author}{{Fouesneau} M},
  \bibinfo{author}{{Fran{\c{c}}ois} P}, \bibinfo{author}{{Franke} O},
  \bibinfo{author}{{Gonz{\'a}lez Hern{\'a}ndez} JI}, \bibinfo{author}{{Gwyn}
  SDJ}, \bibinfo{author}{{Hill} V}, \bibinfo{author}{{Ibata} RA},
  \bibinfo{author}{{Jablonka} P}, \bibinfo{author}{{Longeard} N},
  \bibinfo{author}{{McConnachie} AW}, \bibinfo{author}{{Navarro} JF},
  \bibinfo{author}{{S{\'a}nchez-Janssen} R}, \bibinfo{author}{{Tolstoy} E} and
  \bibinfo{author}{{Venn} KA} (\bibinfo{year}{2017}), \bibinfo{month}{Nov.}
\bibinfo{title}{{The Pristine survey - I. Mining the Galaxy for the most
  metal-poor stars}}.
\bibinfo{journal}{{\em MNRAS}} \bibinfo{volume}{471} (\bibinfo{number}{3}):
  \bibinfo{pages}{2587--2604}. \bibinfo{doi}{\doi{10.1093/mnras/stx1068}}.

\bibtype{Article}%
\bibitem[{Starkenburg} et al.(2018)]{Starkenburg18}
\bibinfo{author}{{Starkenburg} E}, \bibinfo{author}{{Aguado} DS},
  \bibinfo{author}{{Bonifacio} P}, \bibinfo{author}{{Caffau} E},
  \bibinfo{author}{{Jablonka} P}, \bibinfo{author}{{Lardo} C},
  \bibinfo{author}{{Martin} N}, \bibinfo{author}{{S{\'a}nchez-Janssen} R},
  \bibinfo{author}{{Sestito} F}, \bibinfo{author}{{Venn} KA},
  \bibinfo{author}{{Youakim} K}, \bibinfo{author}{{Prieto} CA},
  \bibinfo{author}{{Arentsen} A}, \bibinfo{author}{{Gentile} M},
  \bibinfo{author}{{Hern{\'a}ndez} JIG}, \bibinfo{author}{{Kielty} C},
  \bibinfo{author}{{Koppelman} HH}, \bibinfo{author}{{Longeard} N},
  \bibinfo{author}{{Tolstoy} E}, \bibinfo{author}{{Carlberg} RG},
  \bibinfo{author}{{C{\^o}t{\'e}} P}, \bibinfo{author}{{Fouesneau} M},
  \bibinfo{author}{{Hill} V}, \bibinfo{author}{{McConnachie} AW} and
  \bibinfo{author}{{Navarro} JF} (\bibinfo{year}{2018}), \bibinfo{month}{Aug.}
\bibinfo{title}{{The Pristine Survey IV: Approaching the Galactic metallicity
  floor with the discovery of an ultra metal-poor star}}.
\bibinfo{journal}{{\em MNRAS}} \bibinfo{doi}{\doi{10.1093/mnras/sty2276}}.
\eprint{1807.04292}.

\bibtype{Article}%
\bibitem[{Strigari} et al.(2008)]{strigari08}
\bibinfo{author}{{Strigari} LE}, \bibinfo{author}{{Bullock} JS},
  \bibinfo{author}{{Kaplinghat} M}, \bibinfo{author}{{Simon} JD},
  \bibinfo{author}{{Geha} M}, \bibinfo{author}{{Willman} B} and
  \bibinfo{author}{{Walker} MG} (\bibinfo{year}{2008}), \bibinfo{month}{Aug.}
\bibinfo{title}{{A common mass scale for satellite galaxies of the Milky Way}}.
\bibinfo{journal}{{\em Nature}} \bibinfo{volume}{454}:
  \bibinfo{pages}{1096--1097}. \bibinfo{doi}{\doi{10.1038/nature07222}}.
\eprint{0808.3772}.

\bibtype{Article}%
\bibitem[{Suda} et al.(2008)]{Suda08}
\bibinfo{author}{{Suda} T}, \bibinfo{author}{{Katsuta} Y},
  \bibinfo{author}{{Yamada} S}, \bibinfo{author}{{Suwa} T},
  \bibinfo{author}{{Ishizuka} C}, \bibinfo{author}{{Komiya} Y},
  \bibinfo{author}{{Sorai} K}, \bibinfo{author}{{Aikawa} M} and
  \bibinfo{author}{{Fujimoto} MY} (\bibinfo{year}{2008}), \bibinfo{month}{Oct.}
\bibinfo{title}{{Stellar Abundances for the Galactic Archeology (SAGA) Database
  --- Compilation of the Characteristics of Known Extremely Metal-Poor Stars}}.
\bibinfo{journal}{{\em PASJ}} \bibinfo{volume}{60}:
  \bibinfo{pages}{1159--1171}.
\eprint{0806.3697}.

\bibtype{Article}%
\bibitem[{Tarumi} et al.(2021)]{tarumi21}
\bibinfo{author}{{Tarumi} Y}, \bibinfo{author}{{Yoshida} N} and
  \bibinfo{author}{{Frebel} A} (\bibinfo{year}{2021}), \bibinfo{month}{Jun.}
\bibinfo{title}{{Formation of an Extended Stellar Halo around an Ultra-faint
  Dwarf Galaxy Following One of the Earliest Mergers from Galactic Building
  Blocks}}.
\bibinfo{journal}{{\em ApJl}} \bibinfo{volume}{914} (\bibinfo{number}{1}),
  \bibinfo{eid}{L10}. \bibinfo{doi}{\doi{10.3847/2041-8213/ac024e}}.

\bibtype{Article}%
\bibitem[{Tolstoy} et al.(2004)]{tolstoy04}
\bibinfo{author}{{Tolstoy} E}, \bibinfo{author}{{Irwin} MJ},
  \bibinfo{author}{{Helmi} A}, \bibinfo{author}{{Battaglia} G},
  \bibinfo{author}{{Jablonka} P}, \bibinfo{author}{{Hill} V},
  \bibinfo{author}{{Venn} KA}, \bibinfo{author}{{Shetrone} MD},
  \bibinfo{author}{{Letarte} B}, \bibinfo{author}{{Cole} AA},
  \bibinfo{author}{{Primas} F}, \bibinfo{author}{{Francois} P},
  \bibinfo{author}{{Arimoto} N}, \bibinfo{author}{{Sadakane} K},
  \bibinfo{author}{{Kaufer} A}, \bibinfo{author}{{Szeifert} T} and
  \bibinfo{author}{{Abel} T} (\bibinfo{year}{2004}), \bibinfo{month}{Dec.}
\bibinfo{title}{{Two Distinct Ancient Components in the Sculptor Dwarf
  Spheroidal Galaxy: First Results from the Dwarf Abundances and Radial
  Velocities Team}}.
\bibinfo{journal}{{\em ApJL}} \bibinfo{volume}{617}:
  \bibinfo{pages}{L119--L122}. \bibinfo{doi}{\doi{10.1086/427388}}.
\eprint{arXiv:astro-ph/0411029}.

\bibtype{Article}%
\bibitem[{Travaglio} et al.(2004)]{Travaglio04}
\bibinfo{author}{{Travaglio} C}, \bibinfo{author}{{Gallino} R},
  \bibinfo{author}{{Arnone} E}, \bibinfo{author}{{Cowan} J},
  \bibinfo{author}{{Jordan} F} and  \bibinfo{author}{{Sneden} C}
  (\bibinfo{year}{2004}), \bibinfo{month}{Feb.}
\bibinfo{title}{{Galactic Evolution of Sr, Y, And Zr: A Multiplicity of
  Nucleosynthetic Processes}}.
\bibinfo{journal}{{\em ApJ}} \bibinfo{volume}{601}: \bibinfo{pages}{864--884}.
  \bibinfo{doi}{\doi{10.1086/380507}}.
\eprint{astro-ph/0310189}.

\bibtype{Article}%
\bibitem[{Tumlinson}(2010)]{tumlinson10}
\bibinfo{author}{{Tumlinson} J} (\bibinfo{year}{2010}), \bibinfo{month}{Jan.}
\bibinfo{title}{{Chemical Evolution in Hierarchical Models of Cosmic Structure.
  II. The Formation of the Milky Way Stellar Halo and the Distribution of the
  Oldest Stars}}.
\bibinfo{journal}{{\em ApJ}} \bibinfo{volume}{708}:
  \bibinfo{pages}{1398--1418}.
  \bibinfo{doi}{\doi{10.1088/0004-637X/708/2/1398}}.
\eprint{0911.1786}.

\bibtype{Article}%
\bibitem[{Umeda} and {Nomoto}(2002)]{UmedaNomoto:2002}
\bibinfo{author}{{Umeda} H} and  \bibinfo{author}{{Nomoto} K}
  (\bibinfo{year}{2002}), \bibinfo{month}{Jan.}
\bibinfo{title}{{Nucleosynthesis of Zinc and Iron Peak Elements in Population
  III Type II Supernovae: Comparison with Abundances of Very Metal Poor Halo
  Stars}}.
\bibinfo{journal}{{\em ApJ}} \bibinfo{volume}{565}: \bibinfo{pages}{385--404}.

\bibtype{Article}%
\bibitem[{Van Eck} et al.(2001)]{2001vaneck}
\bibinfo{author}{{Van Eck} S}, \bibinfo{author}{{Goriely} S},
  \bibinfo{author}{{Jorissen} A} and  \bibinfo{author}{{Plez} B}
  (\bibinfo{year}{2001}), \bibinfo{month}{Aug.}
\bibinfo{title}{{Discovery of three lead-rich stars}}.
\bibinfo{journal}{{\em Nature}} \bibinfo{volume}{412}:
  \bibinfo{pages}{793--795}.

\bibtype{Article}%
\bibitem[{Wallerstein} et al.(1963)]{wallerstein63}
\bibinfo{author}{{Wallerstein} G}, \bibinfo{author}{{Greenstein} JL},
  \bibinfo{author}{{Parker} R}, \bibinfo{author}{{Helfer} HL} and
  \bibinfo{author}{{Aller} LH} (\bibinfo{year}{1963}), \bibinfo{month}{Jan.}
\bibinfo{title}{{Red Giants with Extreme Metal Deficiencies.}}
\bibinfo{journal}{{\em ApJ}} \bibinfo{volume}{137}: \bibinfo{pages}{280}.
  \bibinfo{doi}{\doi{10.1086/147501}}.

\bibtype{Article}%
\bibitem[{Wan} et al.(2020)]{wan20}
\bibinfo{author}{{Wan} Z}, \bibinfo{author}{{Lewis} GF}, \bibinfo{author}{{Li}
  TS}, \bibinfo{author}{{Simpson} JD}, \bibinfo{author}{{Martell} SL},
  \bibinfo{author}{{Zucker} DB}, \bibinfo{author}{{Mould} JR},
  \bibinfo{author}{{Erkal} D}, \bibinfo{author}{{Pace} AB},
  \bibinfo{author}{{Mackey} D}, \bibinfo{author}{{Ji} AP},
  \bibinfo{author}{{Koposov} SE}, \bibinfo{author}{{Kuehn} K},
  \bibinfo{author}{{Shipp} N}, \bibinfo{author}{{Balbinot} E},
  \bibinfo{author}{{Bland-Hawthorn} J}, \bibinfo{author}{{Casey} AR},
  \bibinfo{author}{{Da Costa} GS}, \bibinfo{author}{{Kafle} P},
  \bibinfo{author}{{Sharma} S} and  \bibinfo{author}{{De Silva} GM}
  (\bibinfo{year}{2020}), \bibinfo{month}{Jul.}
\bibinfo{title}{{The tidal remnant of an unusually metal-poor globular
  cluster}}.
\bibinfo{journal}{{\em Nature}} \bibinfo{volume}{583} (\bibinfo{number}{7818}):
  \bibinfo{pages}{768--770}. \bibinfo{doi}{\doi{10.1038/s41586-020-2483-6}}.
\eprint{2007.14577}.

\bibtype{Article}%
\bibitem[{Wanajo} et al.(2001)]{wanajo01}
\bibinfo{author}{{Wanajo} S}, \bibinfo{author}{{Kajino} T},
  \bibinfo{author}{{Mathews} GJ} and  \bibinfo{author}{{Otsuki} K}
  (\bibinfo{year}{2001}), \bibinfo{month}{Jun.}
\bibinfo{title}{{The r-Process in Neutrino-driven Winds from Nascent,
  ``Compact'' Neutron Stars of Core-Collapse Supernovae}}.
\bibinfo{journal}{{\em ApJ}} \bibinfo{volume}{554}: \bibinfo{pages}{578--586}.
  \bibinfo{doi}{\doi{10.1086/321339}}.
\eprint{arXiv:astro-ph/0102261}.

\bibtype{Article}%
\bibitem[{Xylakis-Dornbusch} et al.(2024)]{xylakis24}
\bibinfo{author}{{Xylakis-Dornbusch} T}, \bibinfo{author}{{Hansen} TT},
  \bibinfo{author}{{Beers} TC}, \bibinfo{author}{{Christlieb} N},
  \bibinfo{author}{{Ezzeddine} R}, \bibinfo{author}{{Frebel} A},
  \bibinfo{author}{{Holmbeck} E}, \bibinfo{author}{{Placco} VM},
  \bibinfo{author}{{Roederer} IU}, \bibinfo{author}{{Sakari} CM} and
  \bibinfo{author}{{Sneden} C} (\bibinfo{year}{2024}), \bibinfo{month}{Apr.}
\bibinfo{title}{{The R-Process Alliance: Analysis of Limited-r Stars}}.
\bibinfo{journal}{{\em arXiv e-prints}} ,
  \bibinfo{eid}{arXiv:2404.03379}\bibinfo{doi}{\doi{10.48550/arXiv.2404.03379}}.
\eprint{2404.03379}.

\bibtype{Article}%
\bibitem[{Yong} et al.(2013)]{yong13_II}
\bibinfo{author}{{Yong} D}, \bibinfo{author}{{Norris} JE},
  \bibinfo{author}{{Bessell} MS}, \bibinfo{author}{{Christlieb} N},
  \bibinfo{author}{{Asplund} M}, \bibinfo{author}{{Beers} TC},
  \bibinfo{author}{{Barklem} PS}, \bibinfo{author}{{Frebel} A} and
  \bibinfo{author}{{Ryan} SG} (\bibinfo{year}{2013}), \bibinfo{month}{Jan.}
\bibinfo{title}{{The Most Metal-Poor Stars. II. Chemical Abundances of 190
  Metal-Poor Stars Including 10 New Stars With [Fe/H] < -3.5}}.
\bibinfo{journal}{{\em ApJ}} \bibinfo{volume}{762}, \bibinfo{eid}{26}.
  \bibinfo{doi}{\doi{10.1088/0004-637X/762/1/26}}.
\eprint{1208.3003}.

\bibtype{Article}%
\bibitem[{Yong} et al.(2014)]{yong14}
\bibinfo{author}{{Yong} D}, \bibinfo{author}{{Roederer} IU},
  \bibinfo{author}{{Grundahl} F}, \bibinfo{author}{{Da Costa} GS},
  \bibinfo{author}{{Karakas} AI}, \bibinfo{author}{{Norris} JE},
  \bibinfo{author}{{Aoki} W}, \bibinfo{author}{{Fishlock} CK},
  \bibinfo{author}{{Marino} AF}, \bibinfo{author}{{Milone} AP} and
  \bibinfo{author}{{Shingles} LJ} (\bibinfo{year}{2014}), \bibinfo{month}{Jul.}
\bibinfo{title}{{Iron and neutron-capture element abundance variations in the
  globular cluster M2 (NGC 7089)}}.
\bibinfo{journal}{{\em MNRAS}} \bibinfo{volume}{441}:
  \bibinfo{pages}{3396--3416}. \bibinfo{doi}{\doi{10.1093/mnras/stu806}}.
\eprint{1404.6873}.

\bibtype{Article}%
\bibitem[{Yong} et al.(2021)]{yong21}
\bibinfo{author}{{Yong} D}, \bibinfo{author}{{Kobayashi} C},
  \bibinfo{author}{{Da Costa} GS}, \bibinfo{author}{{Bessell} MS},
  \bibinfo{author}{{Chiti} A}, \bibinfo{author}{{Frebel} A},
  \bibinfo{author}{{Lind} K}, \bibinfo{author}{{Mackey} AD},
  \bibinfo{author}{{Nordlander} T}, \bibinfo{author}{{Asplund} M},
  \bibinfo{author}{{Casey} AR}, \bibinfo{author}{{Marino} AF},
  \bibinfo{author}{{Murphy} SJ} and  \bibinfo{author}{{Schmidt} BP}
  (\bibinfo{year}{2021}), \bibinfo{month}{Jul.}
\bibinfo{title}{{r-Process elements from magnetorotational hypernovae}}.
\bibinfo{journal}{{\em Nature}} \bibinfo{volume}{595} (\bibinfo{number}{7866}):
  \bibinfo{pages}{223--226}. \bibinfo{doi}{\doi{10.1038/s41586-021-03611-2}}.

\bibtype{Article}%
\bibitem[{Zhao} et al.(2012)]{zhao_lamost12}
\bibinfo{author}{{Zhao} G}, \bibinfo{author}{{Zhao} YH}, \bibinfo{author}{{Chu}
  YQ}, \bibinfo{author}{{Jing} YP} and  \bibinfo{author}{{Deng} LC}
  (\bibinfo{year}{2012}), \bibinfo{month}{Jul.}
\bibinfo{title}{{LAMOST spectral survey -- An overview}}.
\bibinfo{journal}{{\em Research in Astronomy and Astrophysics}}
  \bibinfo{volume}{12} (\bibinfo{number}{7}): \bibinfo{pages}{723--734}.
  \bibinfo{doi}{\doi{10.1088/1674-4527/12/7/002}}.

\end{thebibliography*}

\end{document}